\documentclass{aa}
\usepackage{graphicx}
\usepackage{caption}
\usepackage{txfonts}
\usepackage{epstopdf}
\usepackage{color}
\usepackage{multirow}
\usepackage{amsmath}
\usepackage{amssymb}
\usepackage{upgreek}
\usepackage{hyperref}
\usepackage{upgreek}

\begin{document}

\title{IZ~Tel and UW~Vir: Southern oscillating eclipsing Algol systems with active mass transfer}
     \author{A. Liakos\inst{1}
             \and
              D. J. W. Moriarty\inst{2,3}
             \and
              J. F. West\inst{3}
             \and
              A. Erdem\inst{4,5}
             }
 \institute{Institute for Astronomy, Astrophysics, Space Applications and Remote Sensing, National Observatory of Athens,\\
              Metaxa \& Vas. Pavlou St., GR-15236, Penteli, Athens, Greece \\
              \email{alliakos@noa.gr}
              \and
              School of Mathematics and Physics, The University of Queensland, Queensland 4072, Australia
              \and
              Astronomical Association of Queensland, St. Lucia, Queensland, 4067, Australia
              \and
              Astrophysics Research Center and Ulup{\i}nar Observatory, \c{C}anakkale Onsekiz Mart University, TR-17100, \c{C}anakkale, T\"{u}rkiye
              \and
              Department of Physics, Faculty of Arts and Sciences, \c{C}anakkale Onsekiz Mart University, Terzio\u{g}lu Kamp\"{u}s\"{u}, TR-17100, \\ \c{C}anakkale, T\"{u}rkiye}

           \date{Received XX XXX 2025; accepted XX XXX 2026}


\abstract
{This study is an in-depth examination of IZ~Tel and UW~Vir which are semi-detached oscillating Eclipsing Algol binary systems (oEA stars). The radial velocities of both components of each system were derived using spectra observed with the Australian National University's 2.3~m telescope. The spectral types of the IZ~Tel primary and secondary components were determined as F2V and K2IV; and those of UW~Vir were determined as A7V and K6IV, respectively. Spectroscopy revealed mass transfer in progress which was confirmed by the photometric analysis for both cases and also by Eclipse-Timing Variation analysis in the case of UW~Vir. Data from the Transiting Exoplanet Survey Satellite (TESS) and ground-based observations enabled detailed light-curve modelling and pulsation analysis. We determined component masses of $M_1=1.48$~M$_\sun$ and $M_2=0.33$~M$_\sun$ for IZ~Tel, and $M_1=2.39$~M$_\sun$ and $M_2=0.67$~M$_\sun$ for UW~Vir from the simultaneous solution of the light and radial velocity curves. Spectra during total eclipses of the primary components revealed H$\alpha$ emission was present. Both primary components are $\delta$~Sct stars. That of IZ~Tel pulsates with a dominant frequency of 13.56~d$^{-1}$, which is revealed as non-radial pressure mode, as well as in another 16 combination frequencies. The primary component of UW~Vir oscillates in three main frequencies within the range 34.9-43.3~d$^{-1}$ and in more than 50 another combination frequencies. Mode coupling was detected in the three main frequencies, which showed amplitude and phase modulations within a time span of approximately four years.  The physical and pulsational properties of the $\delta$~Sct stars of both systems were compared with other members of oEA stars.}

\keywords{binaries: eclipsing -- stars: fundamental parameters -– binaries: close -– stars: oscillations –- stars: variables: delta Scuti -- stars: individual (IZ~Tel) -- stars: individual (UW~Vir)}

\maketitle


\section{Introduction}
\label{Sec:Intro}
$\delta$~Sct stars display rapid, multi-periodic pulsations spanning roughly 4–80~d$^{-1}$ \citep{AER10}. Their variability is mainly driven by radial and low-order non-radial pressure (p)~modes excited through the $\kappa$-mechanism \citep[see][]{ZHE63, BRE00, AER10, BAL15}. They may also exhibit high-order non-radial oscillations influenced by turbulent pressure within the hydrogen convection zone \citep{ANT14}. These stars generally have masses of about 1.5–2.5~M$_{\sun}$ and correspond to spectral types A to early F. They are found primarily within the classical instability strip, which spans from the main sequence to the giant branch \citep{AER10}.

Eclipsing binaries (EBs) are essential tools for determining the physical properties of their stellar components—such as masses, radii, luminosities, and evolutionary states. When light curves (LCs) and radial velocity (RV) curves are analyzed together, those properties can be derived with high precision. Furthermore, eclipse timing variations (ETVs) enable the detection of processes that influence changes in the orbital period. Systems containing a pulsating component are especially valuable because they shed light on the structure and evolution of pulsating stars while also providing a unique framework to examine how binarity—through effects such as tidal interactions or mass transfer—shapes pulsation behaviour.

Oscillating eclipsing Algols (oEA stars) are a subclass of semi-detached Algol-type binary systems in which the mass-accreting primary component exhibits $\delta$~Sct type pulsations \citep{MKR02}. Their geometry enables not only precise determination of fundamental stellar properties, but also how active mass transfer and tidal interactions influence pulsational behaviour \citep{MKR18}. Therefore, oEA stars have important roles for investigating the interplay between accretion processes, stellar structure, and oscillation properties. Systematic studies and cataloguing of these systems began twenty years ago \citep{SOY06, LIA09, LIAN12, LIAN17, KAH17, LIA25}. To date, 107 oEA stars are known \citep[][and online\footnote{https://alexiosliakos.weebly.com/catalogue.html}]{LIA25}, but the physical properties (e.g.~masses, radii) have been accurately determined only for 41 of them, which are members of double-lined spectroscopic and eclipsing systems. For another 41 oEAs, the absolute parameters were estimates based on assumptions, as they were single-lined spectroscopic and eclipsing systems or completely lacked spectroscopic information.

This work is a thorough examination of the binaries IZ~Tel and UW~Vir, which have not been studied in detail previously (i.e.~they lack accurate physical parameters and detailed pulsational modelling). The oscillating behaviour of IZ~Tel (TIC~360877907) was discovered by \citet{PIG07}, who found a dominant pulsational frequency of $\sim13.6$~d$^{-1}$. For UW~Vir (TIC~347809861), \citet{QIA00} and \citet{ZHAJ09} suggested that orbital period changes of the system could be caused by the light-time effect due to a third component and/or mass transfer as the most likely modulation mechanisms. \citet{MKR17} announced that the system exhibits pulsations and detected two modes ($f_1\sim28.8$~d$^{-1}$ and $f_2\sim46.9$~d$^{-1}$).

In our study, using space telescopes and ground-based photometry, spectroscopy, and historical data of times of eclipses, our aim has been to determine accurate physical parameters and the pulsation models of their oscillating components. Orbital models were calculated using RVs of both components and the LCs of each system; LC residuals were used for asteroseismic analyses. In addition, we analysed eclipse timing variations of UW~Vir and we present the mechanisms that modify its orbital period. Furthermore, we have compared the systems' properties with other similar ones. Based on the present results, this work increases the sample of oEA stars with accurate absolute parameters by $\sim5$\%.

Throughout this paper, the designations primary (S1) and secondary (S2) correspond, respectively, to the more luminous and to the less luminous, component of each system. The primary eclipse is at phase 0.0, during which the more luminous star is eclipsed by its companion. Uncertainties listed in all tables throughout this work are provided in parentheses alongside the values and correspond to the last digit(s).

\section{Spectroscopy}
\label{Sec:Spectroscopy}

Spectra were observed with the stellar aperture on the wide field spectrograph (WiFeS) mounted at the Nasmyth focal position on the Australian National University's 2.3~m telescope at Siding Spring Observatory (SSO), Australia. For details of the equipment, observing and reduction procedures, see \citet{DOP07, DOP10}, and \citet{CHI14}. The B3000 grating was used for spectral typing, as it has a larger wavelength range extending into the far blue region that includes the Ca~II~H and K lines \citep{MOR19}. In 2023, the telescope was switched to automatic operations; data and calibration frames are taken automatically without human intervention \citep{PRI24}. On 4 March 2026, three 10-min exposures of UW~Vir were recorded during a primary eclipse, which has duration of just 35 minutes.

The detailed observations log is given in Table~\ref{table:SPLOG}. For comparisons of spectral line strengths of UW~Vir, equivalent widths were calculated (Table~\ref{table:UWVir_ew}, Fig.~\ref{Fig:uwvir_ew}) on normalised spectra with the IRAF {\sc splot} task \citep{TOD93}. Uncertainties were $\pm0.02$ for He~I and Na~I~D lines and $\pm0.04$ for H$\alpha$; these were calculated using the $m$ command in {\sc splot} on sections of the spectrum that were clear of large absorption lines.

\subsection{Spectral classification}
\label{Sec:SpectrClass}

Spectral types were determined using the {\sc Xclass} and {\sc Mkclass}\footnote{http://www.appstate.edu/~grayro/mkclass} programs as described in \citet{MOR19}. The MK classification of A–F type stars relies primarily on the blue spectral region between $\lambda\lambda3800$ and 4600~{\AA}. In late K stars, magnesium hydride and the Mg~I triplicate $\lambda\lambda5167$, 5172, 5183~{\AA} are important; those lines are marked in Fig.~\ref{Fig:PhSerSpec}, with an expanded wavelength scale. In Algol-type systems, like those described in this paper, the H$\alpha$ and Na~I~D lines are affected by mass transfer events, making it difficult to assign spectral types by comparison with spectra of single stars.

\begin{figure}
\centering
\includegraphics[width=\columnwidth]{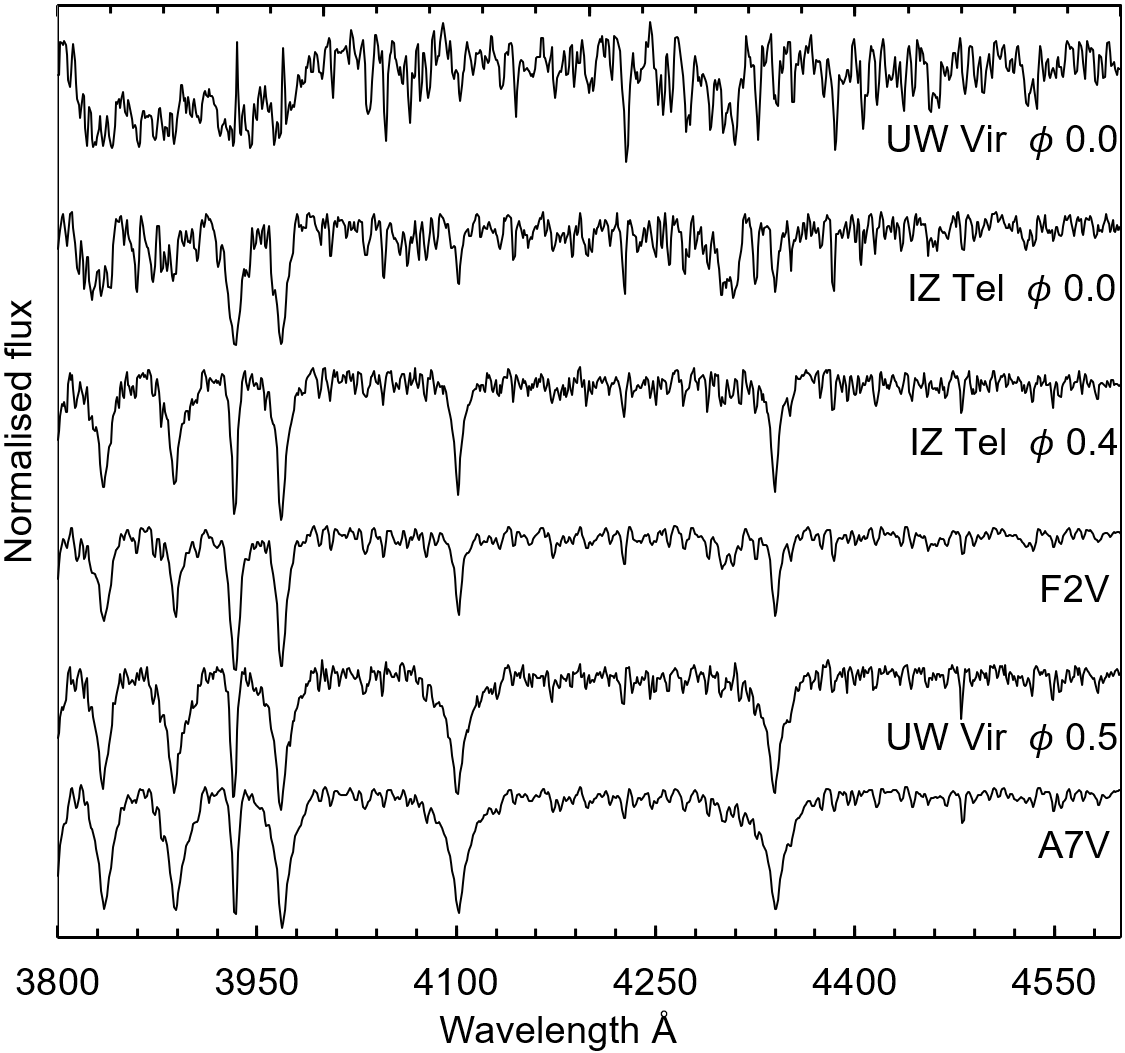}
\caption{Rectified spectra of the primary and secondary components of IZ~Tel and UW~Vir and spectra for comparison: HD~33256 (F2V) and HD~27819 (A7V).}
\label{Fig:spectra}
\end{figure}

The best fit for the primary stars was as follows, IZ~Tel: F2V, UW~Vir: A7V/IV, and the spectral types of secondary stars were determined as, IZ~Tel: K2IV and for UW~Vir: K6IV. The spectra were rectified to deliver a linear continuum baseline using the `Autorectify' function in the XMK25 display routine in {\sc Xclass} (Fig.~\ref{Fig:spectra}).

The luminosity classification of  A7 stars is difficult \citep{GRA09}. The ionised blend of Fe~II and Ti~II lines at $\lambda\lambda4172-4179~{\AA}$ and at $\lambda\lambda4500-4600~{\AA}$, which are stronger in more luminous stars, are stronger in UW~Vir than in HD~27819 (Fig.~\ref{Fig:spectra}). These and other features explain why the {\sc Xclass} and {\sc Mkclass} programs indicate that UW~Vir's spectral type is between A7V and A7IV. As discussed in Sections~\ref{Sec:AbsPar}, \ref{Sec:ETV}, and \ref{Sec:Disc}, the primary component has received a high proportion of the mass of the evolved secondary component, which would include a greater abundance of metals than found in single A7V stars.

\subsection{Radial velocities}
\label{Sec:RVs}

Spectra for RV determinations were captured with the B7000 grating, which has a velocity resolution of 45~km~s$^{-1}$ and wavelength range from 4180 to 5300~{\AA}. Exposure times were long enough to allow detection of the faint secondary stars' metal lines with the broadening function in the {\sc Ravespan} graphical application written by \citet{PIL17} \citep[see also][]{RUC92, RUC02} and are listed in Table~\ref{Tab:RVs}.

The RV values were checked with velocities determined from Na~I~D and H$\alpha$ spectra observed with the R7000 grating; the full wavelength range of that grating includes many telluric lines, which are difficult to remove. The procedures that we used are described in general by \citet{MOR19} and \citet{LIA22, LIA24}. The {\sc Ravespan} settings chosen specifically for the systems in this paper included a preference setting of velocity range $1.6 + 90$, which gave the best fit of broadening function (BF) velocities to velocities determined from the Na~I~D and H$\alpha$ spectral lines. Preference settings in {\sc Ravespan} are given in Table \ref{table:RAVESPAN}. Pulsations were selected in Preferences for both systems. For UW~Vir, these settings excluded the Balmer lines and far blue spectral region to emphasise the lines of the faint secondary component of this system. 

The RVs of the secondary components were clearly evident (see Fig.~\ref{Fig:BFs}). There was no evidence for a third component in these systems, as the broadening function velocity peaks of the primary components were symmetrical, in contrast to the asymmetrical velocity peaks due to tertiary components, as shown, for example in the BF~Vel and RR~Lep systems \citep{LIA24}.

\subsection{Spectral evidence for mass transfer}
\label{Sec:evidence_for_mt}

Evidence for mass transfer from the secondary (donor) components to the primary (accretor) components was apparent in both systems as emission in the H$\alpha$ line during total eclipses of their primary stars and also as broadening or infilling in the H$\alpha$ and Na~I~D doublet absorption spectra of the primary components of each system. Episodes of gas streams impacting on the primary components of both systems were indicated by increases in strength of the He~I~5876 and 6678~{\AA} absorption lines. The spectra indicate that the secondary component of UW Vir was in a very active stage of mass transfer to the primary component in February and March, and which continued to a lesser extent into April and May. It was also active in 2026 when its primary eclipse was observed.

\subsubsection{IZ~Tel}
\label{Sec:iztel_for_mt}

\begin{figure}
\centering
\begin{tabular}{cc}
\includegraphics[height=4.5cm]{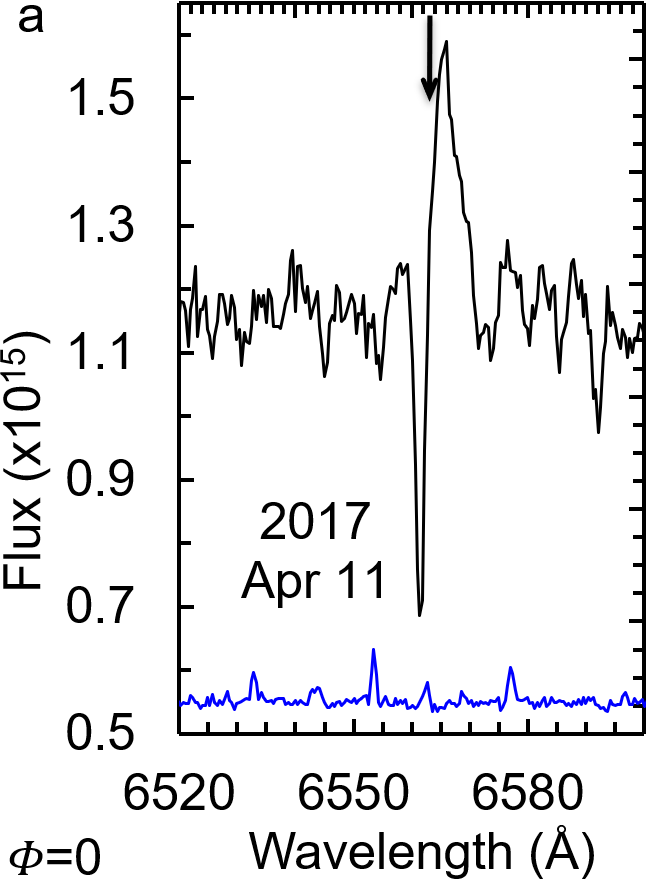}&\includegraphics[height=4.5cm]{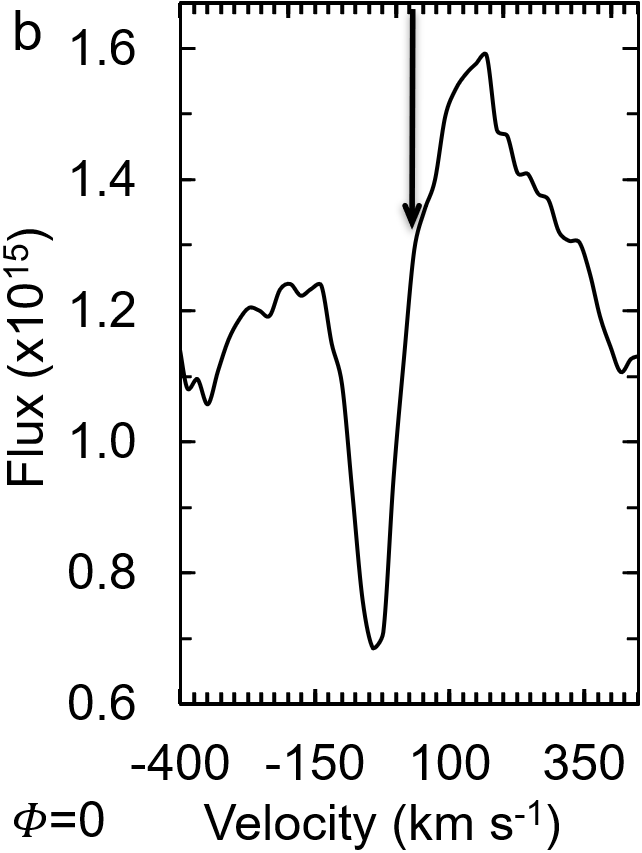}\\
\multicolumn{2}{c}{\includegraphics[height=4.5cm]{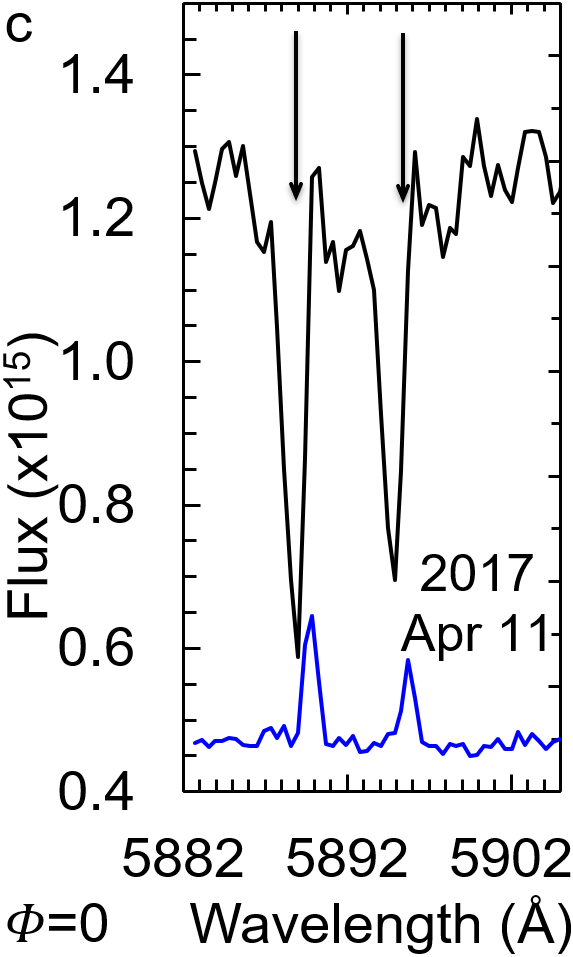}}\\
\end{tabular}
\caption{Spectra of IZ~Tel secondary component during a primary eclipse; (a, b)~the H$\alpha$ spectrum and velocity with systemic velocity and background sky flux subtracted; (c)~Na~I~D spectra and sky background. Downward arrows mark the calculated positions of the line centres. Exp.~time:~600~s.}
\label{Fig:iztel_ecl}
\end{figure}

The spectrum of the IZ~Tel secondary component was observed in a series of six exposures during the total eclipse of its primary star in April 2017. In each observation, the H$\alpha$ line displayed an absorption component with a blue shift from the line centre and a strong emission component to the red; the velocity ranged from $-100$ to 300~km~s$^{-1}$, in appearance not dissimilar to a P~Cygni profile (Fig.~\ref{Fig:iztel_ecl}a, b). The Na~I~D lines of the secondary component were broadened to the blue (Fig.~\ref{Fig:iztel_ecl}c). The H$\alpha$ spectrum appears to be a blend of the secondary component’s normal absorption line, blue-shifted by its atmosphere outflowing in the direction of the observer, and emission possibly generated from a localised hot spot on the primary or within its atmosphere, caused by the impact of gas streaming from the secondary onto the primary, or emission within the gas flow itself, stimulated by the radiation from the primary star. Despite being considered, there was no clear evidence that an accretion disk was present — no double- or single-peaked emission spectra were detected at any phase, as might be expected from the work of \citet{RIC99}.

The velocity of the H$\alpha$ absorption line on the blue side, ranging from rest to about $-$100~km~s$^{-1}$, indicates the velocity of the gas leaving the secondary in the direction of the observer. Some possibly leaves the system and the remainder moves around the secondary within the systemic Roche equipotential boundary and towards the primary component. This P~Cygni-like profile was observed only in the H$\alpha$ spectrum and not in other spectral lines that we observed between $\lambda$3800~{\AA} and $\lambda$7000~{\AA}. It is likely that either a very small accretion disk or a heated primary star surface or hot spot acts as a source of photons that excite hydrogen atoms in the gas flowing around and from the secondary component.  This would yield further emission as these atoms de-excite, some of which would be directed towards the observer. Given that the gas flow is generally away from the observer, such emission would be red-shifted by the gas velocity, as indeed was observed. If the velocity of gas flow as it accelerated towards the primary component ranged from $-100$~km~s$^{-1}$ to 300~km~s$^{-1}$, it would yield an aggregate emission peak similar to that observed.

\begin{figure}
\centering
\begin{tabular}{cc}
\includegraphics[height=4cm]{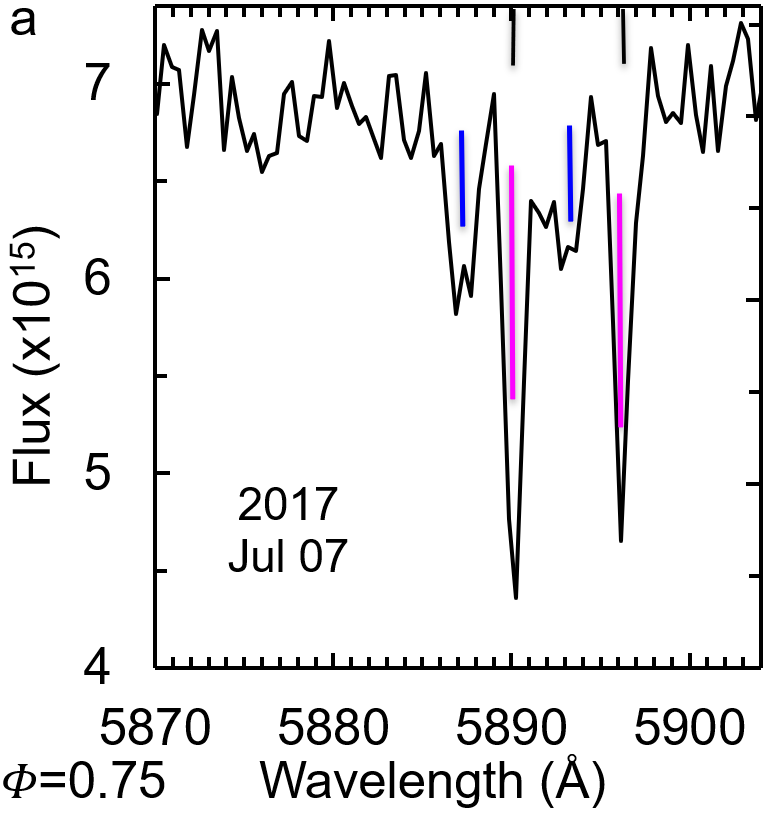}&\includegraphics[height=4cm]{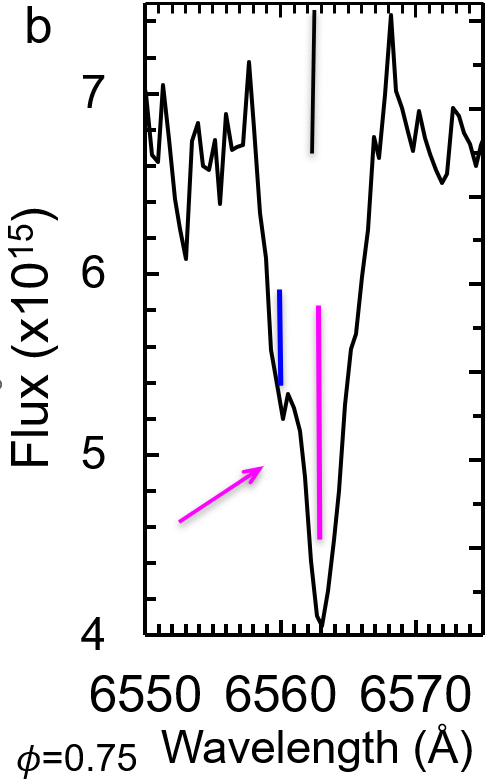}\\
\includegraphics[height=4cm]{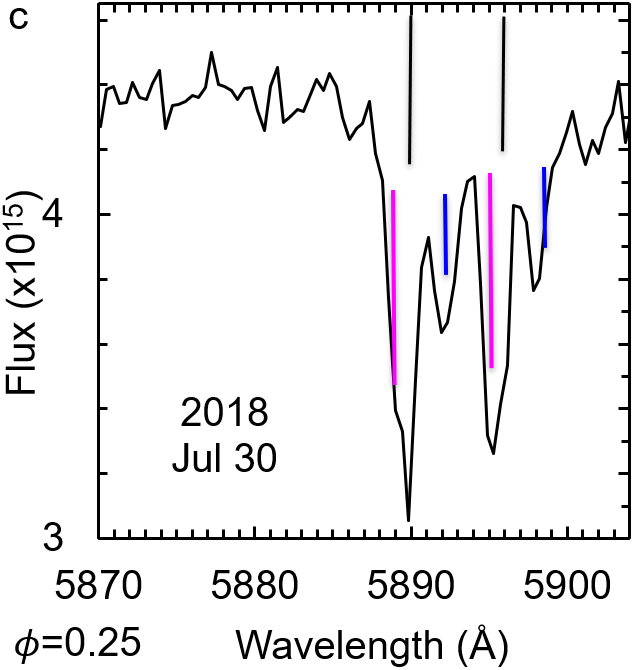}&\includegraphics[height=4cm]{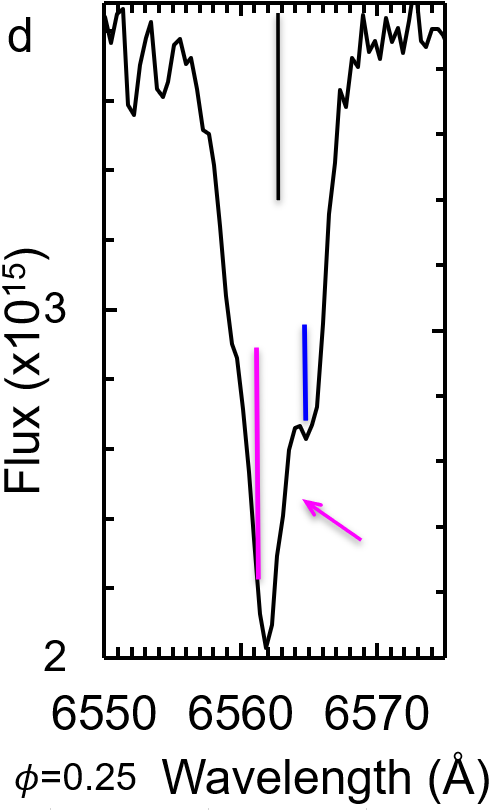}\\
\end{tabular}
\caption{Spectra of IZ~Tel in the wavelength ranges of the Na~I~D and H$\alpha$ lines at the first and second quadrature phases ($\phi$). The positions of the primary and secondary components' lines are marked with magenta and blue lines respectively. Magenta (angled) arrows indicate emission infilling in the H$\alpha$ absorption lines. The rest wavelengths are indicated with black lines. Exp.~times:~(a,~b)~180~s; (c,~d)~600~s.}
\label{Fig:iztel_quad}
\end{figure}

At phase 0.75, the Na~I~D and H$\alpha$ line centres of the primary component were close to the wavelength positions calculated from the sum of the orbital and systemic velocities (27 and $-24$~km~s$^{-1}$ respectively), thus confirming the RV values determined with {\sc Ravespan} (Fig.~\ref{Fig:iztel_quad}a,~b). With the relatively short exposure time of 180~s, the absorption lines were only slightly broadened by mass transfer from the secondary component. The primary component's H$\alpha$ line was asymmetrical. It was broadened to the blue by the velocity the secondary component's H$\alpha$ line, which was $-151~$km~s$^{-1}$ at phase 0.75, and by infilling due to emission in gas from the secondary component flowing around the primary component (cf.~Fig.~\ref{Fig:iztel_ecl}).

At phase 0.25 with a longer exposure time and the orbital and systemic velocities contributing to a blue shift of 51~km~s$^{-1}$, both the sodium and H$\alpha$ lines were strongly broadened to the red by the gas flowing towards and around the primary component (Fig.~\ref{Fig:iztel_quad}c,~d). The primary component's H$\alpha$ line was asymmetrical with infilling caused by emission in gas flowing from the secondary component. Between phases 0.3 and 0.67, the He~I $\lambda5876~{\AA}$ line was moderately strong, indicating impacts of mass on the primary star streaming from the secondary (Fig.~\ref{Fig:PhSerSpec}).

\subsubsection{UW~Vir}
\label{Sec:uwvir_for_mt}

\begin{figure}[h!]
\centering
\begin{tabular}{cc}

\includegraphics[height=3.5cm]{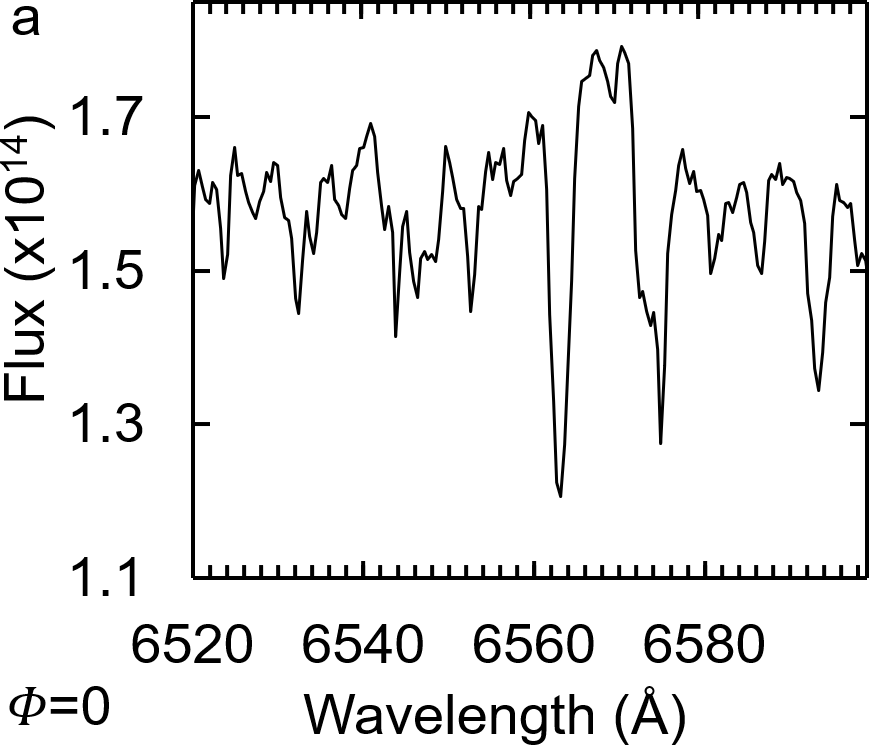}&\includegraphics[height=3.5cm]{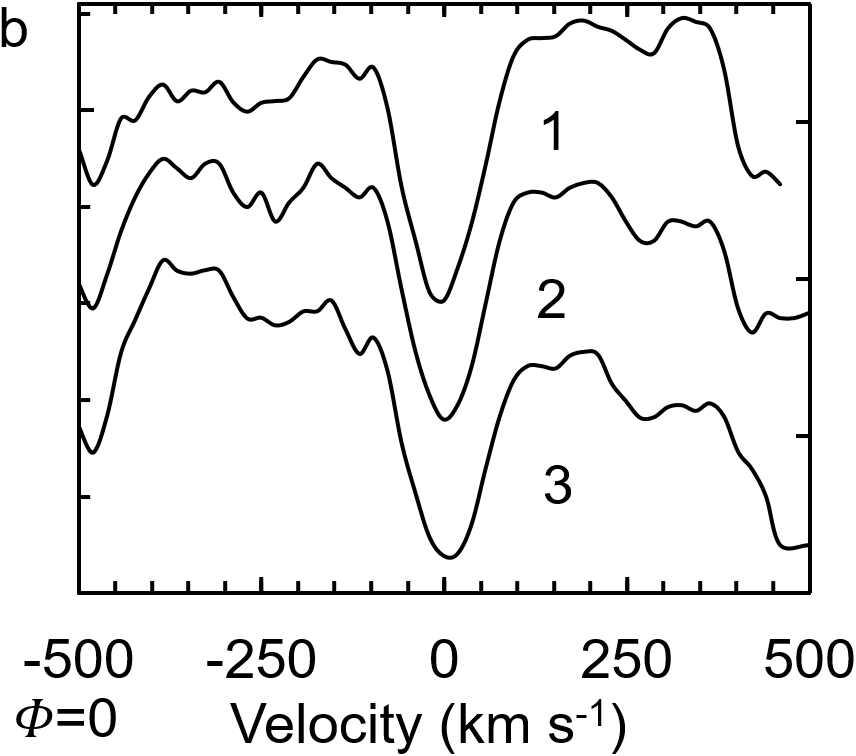}\\
\includegraphics[height=3.5cm]{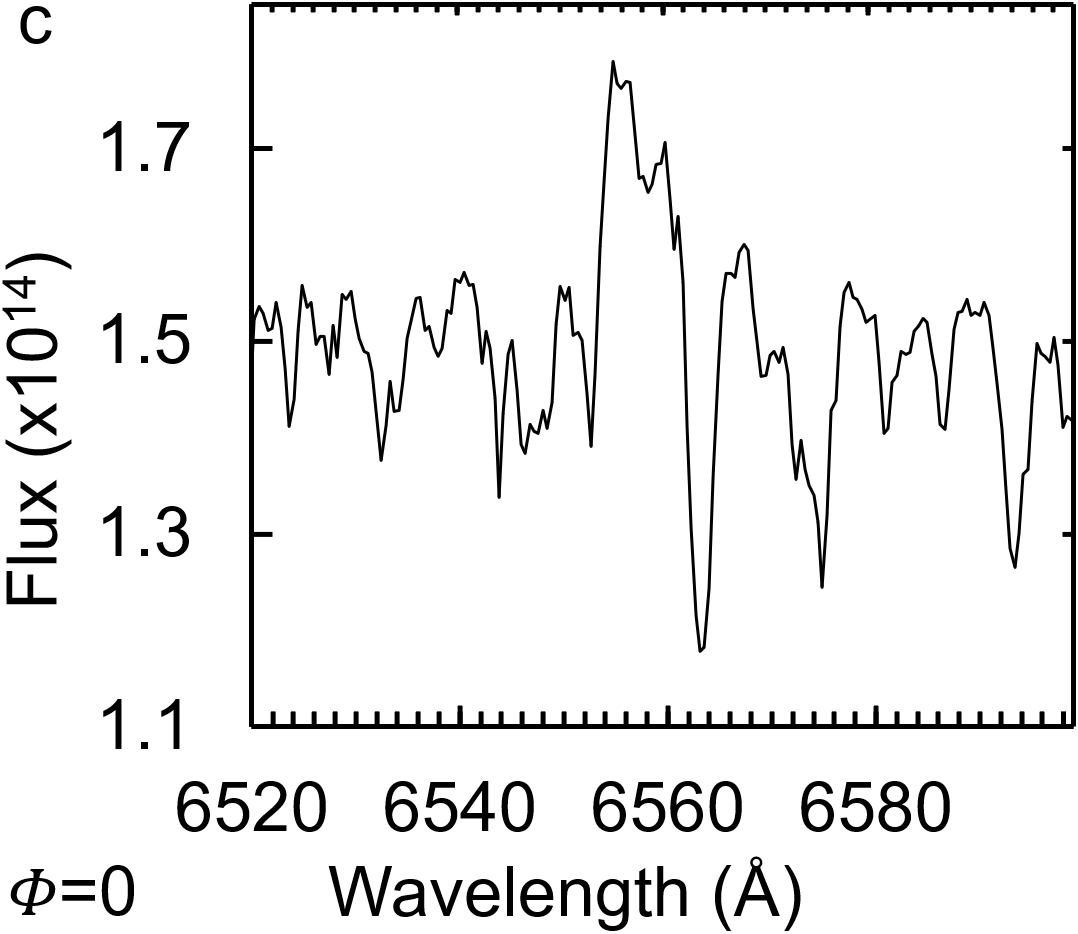}&\includegraphics[height=3.5cm]{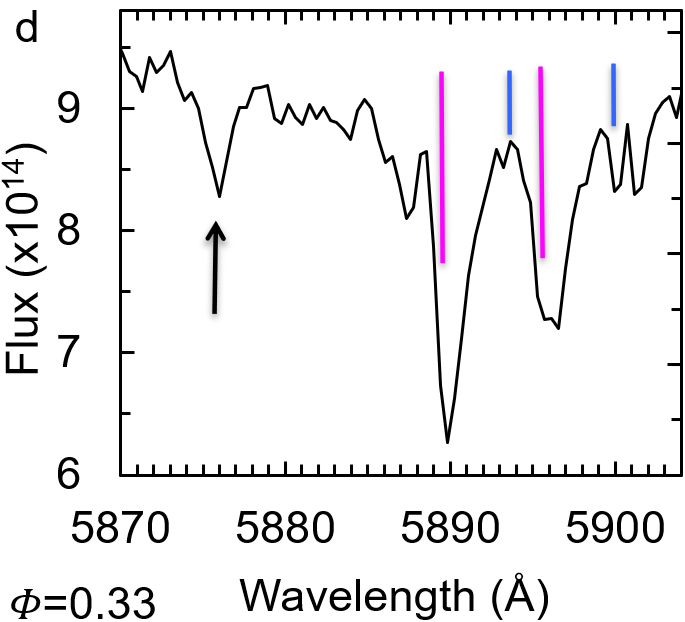}\\
\includegraphics[height=3.5cm]{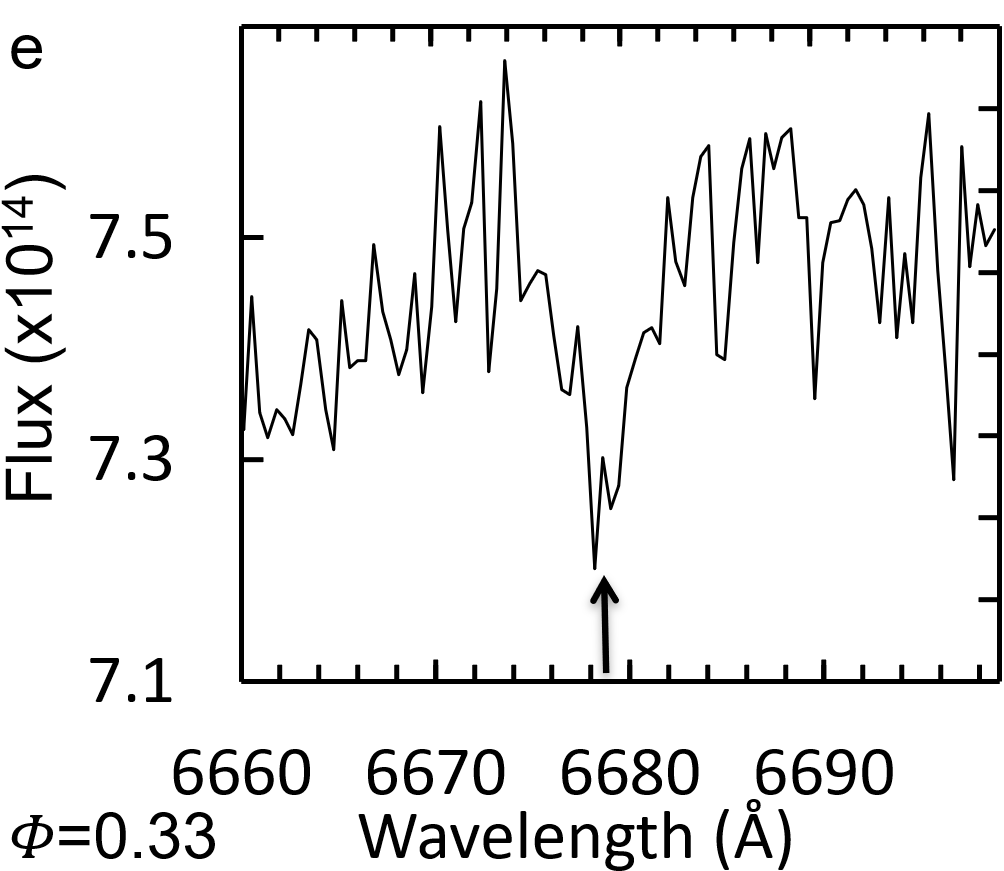}&\includegraphics[height=3.5cm]{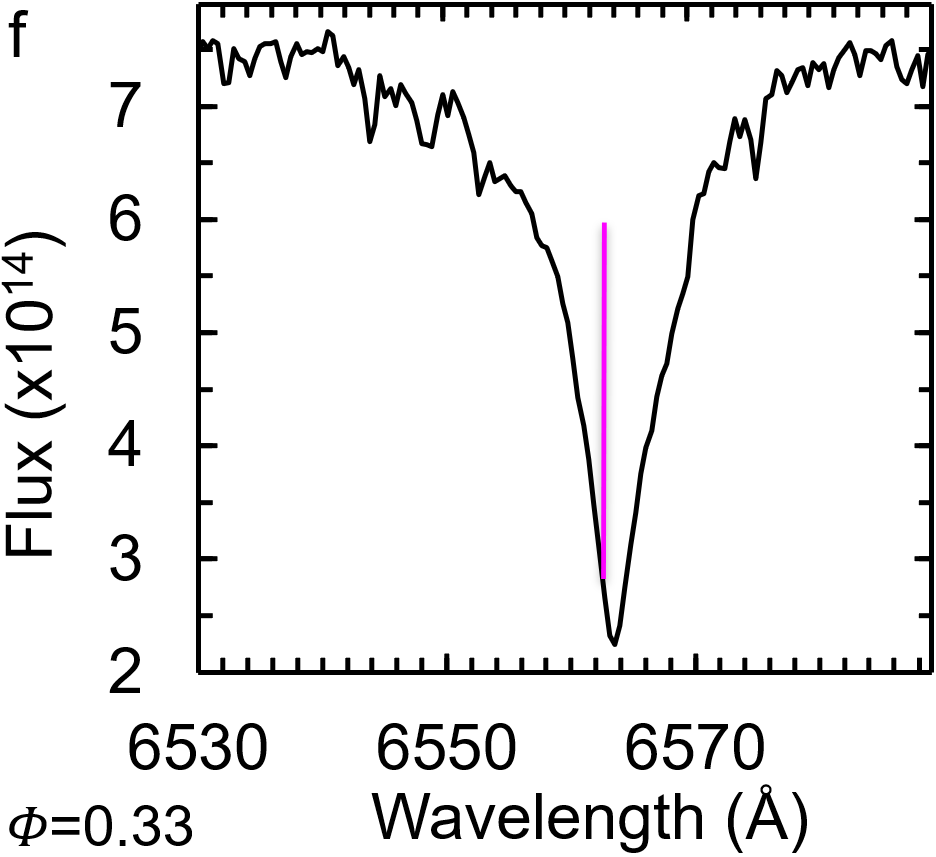}\\
\includegraphics[height=3.5cm]{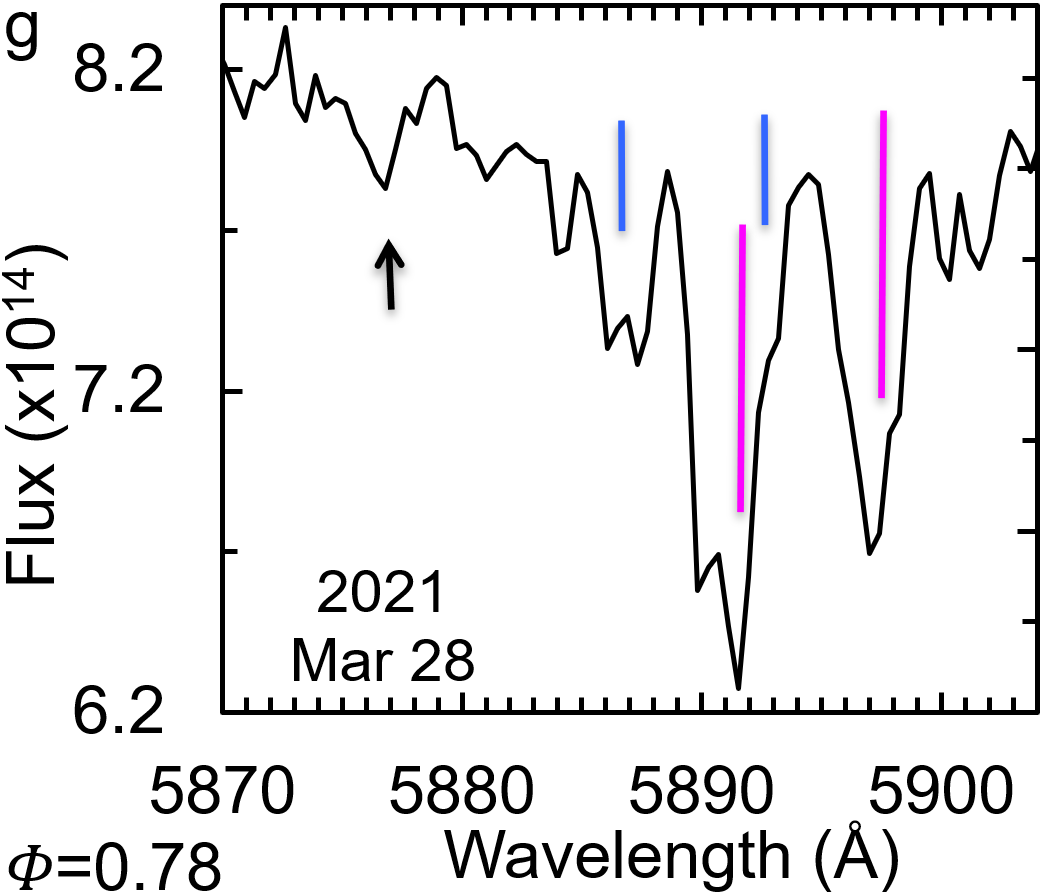}&\includegraphics[height=3.5cm]{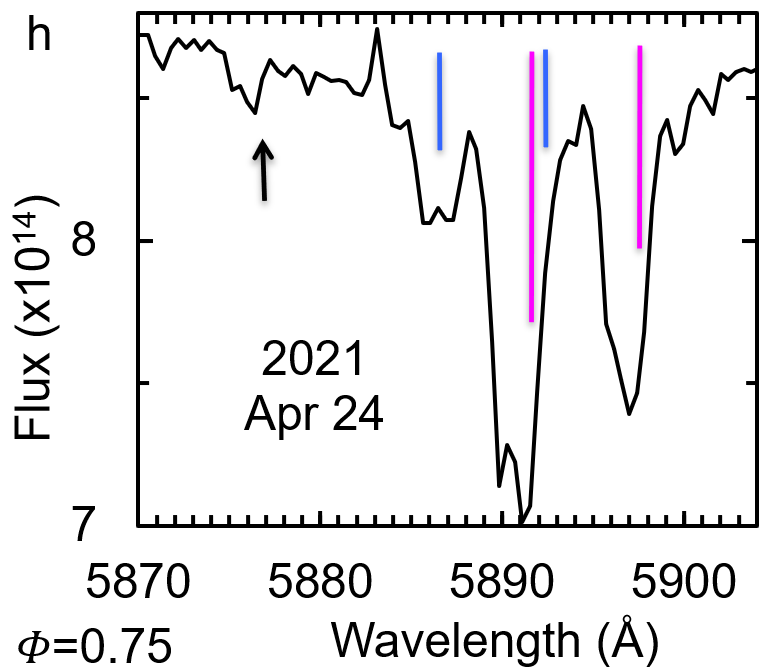}\\
\multicolumn{2}{c}{\includegraphics[height=3.5cm]{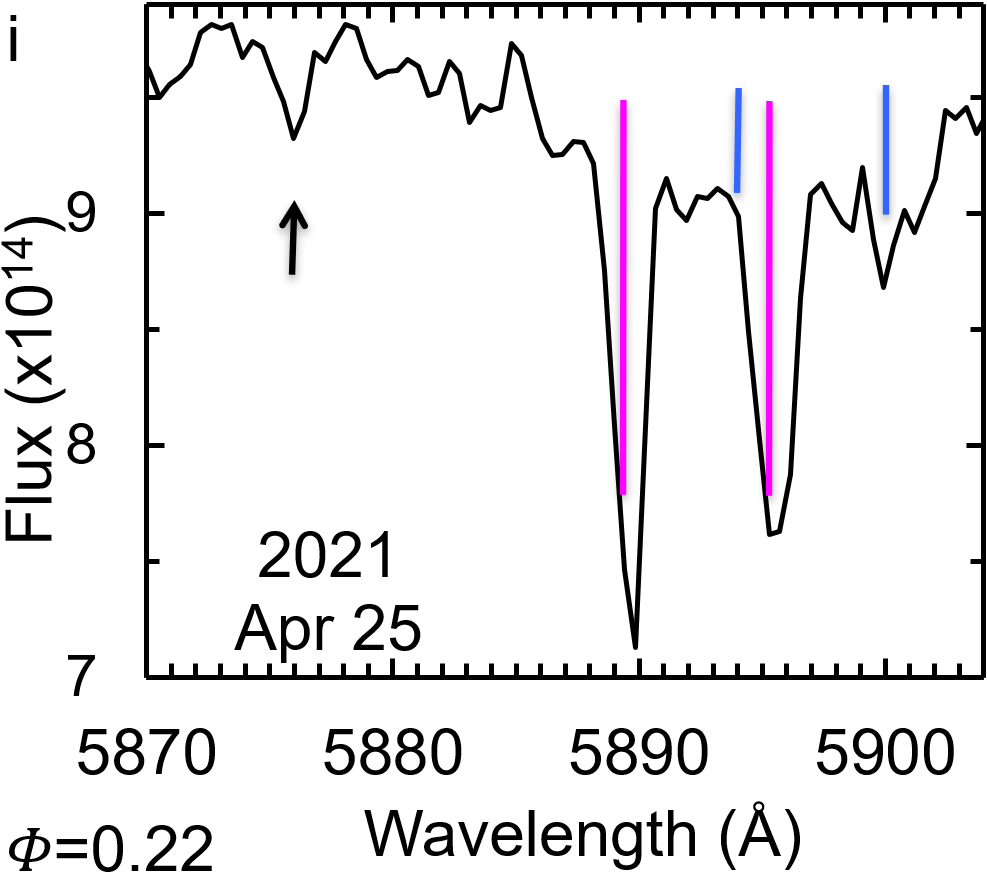}}\\
\end{tabular}
\caption{Spectra of UW~Vir's components. (a)~Emission in the H$\alpha$ spectrum (R=7000) of the secondary component for the first exposure, (b)~velocity profiles of the H$\alpha$ line (Flux values of the velocity profiles offset for clarity), and (c)~the same as (a) but for the third exposure. (d-i)~Examples of the Na~I~D, He~I~5876 and 6678~{\AA} and H$\alpha$ lines of the primary star. Magenta lines mark the calculated positions of the Na~I~D and H$\alpha$ lines for the primary star; blue lines mark the calculated positions of the Na~I~D secondary component's lines. Upward arrows mark the positions of the He~I lines. Exp.~times: (d,~e,~f)~60~s, (g,~h)~240~s, (i)~120~s.}
\label{Fig:uwvir_mt}
\end{figure}

During a primary eclipse of UW~Vir observed on 4~March 2026, the presence of emission in the Ca~II~H and K lines of the secondary component indicates that it is chromospherically active (Fig.~\ref{Fig:uwvir_2rysp}). Emission in the H$\alpha$ line was evident, and it was detectable only during the primary eclipse, which has a duration of approx.~35~min. In contrast to the spectra of IZ~Tel during a primary eclipse, the H$\alpha$ emission, which was present in each of the three exposures, varied considerably in strength and velocity. In the first exposure, its spectrum was similar to that of IZ~Tel with an absorption component on the blue side of the H$\alpha$ line centre and gas with a velocity of 325~km~s$^{-1}$ to the red (Fig.~\ref{Fig:uwvir_mt}a,~b). In the second exposure, ten minutes later, the gas velocity was lower on the red side and slightly higher to the blue (Fig.~\ref{Fig:uwvir_mt}b). In the third exposure the gas velocity was 385~km~s$^{-1}$ on the blue side of the H$\alpha$ line centre (Fig.~\ref{Fig:uwvir_mt}b,~c). Those rapid variations indicate the gas originated in the very active chromosphere of the secondary star, which has an equatorial velocity tidally locked to it orbital velocity of 195~km~s$^{-1}$ (Table~\ref{Tab:Models_LCs}).

The UW~Vir primary component’s He~I~5876 and 6678~{\AA} lines were very strong and the Na~I~D and H$\alpha$ lines were broadened considerably to the red at phase 0.33 on 2021 February 27 (Fig.~\ref{Fig:uwvir_mt}d,~e,~f). The equivalent width of the Na~I~D$_1$ line in the gas stream from the secondary star was similar to that of the primary component, and the equivalent width of the Na~I~D$_2$ line in the gas stream was about a third stronger than that of the primary component (Table~\ref{table:UWVir_ew}). The large mass ejection from the secondary component in February 2021 is also evident in the equivalent width of the H$\alpha$ line (5.57), which was much larger than in May~2021 (Table~\ref{table:UWVir_ew}).

The equivalent width of He~I~5876 at phase 0.33 was twice the strength of that at phase 0.78 one month later; the difference may be due to the  phase at which impacts on the primary star occurred in addition to a decrease in mass transfer (Figs.~\ref{Fig:uwvir_mt}d,~g and \ref{Fig:PhSerSpec}). Both helium lines decreased further in strength in April and May, and varied in strength with phase (Table~\ref{table:UWVir_ew}, Figs.~\ref{Fig:uwvir_mt}i and \ref{Fig:PhSerSpec}).

The Na~I~D and H$\alpha$ spectra were broadened to the red in the first quadrature and to the blue in the second, indicating the presence of a gas stream or transient disc around the primary component (Figs.~\ref{Fig:uwvir_mt}d,~f,~g,~h,~i and \ref{Fig:PhSerSpec}). In the first quadrature, the gas was travelling significantly more slowly towards the observer than the primary star, indicating that the gas was moving towards or onto the primary star, having passed around it after leaving the secondary component. The primary component's D$_2$ line was broadened not only by the gas stream, but also by the secondary component's weak D$_1$ line in the second quadrature phases (Figs.~\ref{Fig:uwvir_mt}g,~h and \ref{Fig:PhSerSpec}). The presence of low velocity gas around the primary component is also shown by the partial split in the Na~I~D$_2$ line at phase 0.75, where the primary component's velocity is 72~km~s$^{-1}$ (a combination of the systemic velocity of 15~km~s$^{-1}$ and orbital velocity of 57~km~s$^{-1}$). Such a distinct split indicates that at those instances, the gas velocity still has a significant component away from the primary component and towards the observer, prior to being drawn around it.

At phase 0.25 the primary component's velocity of 42~km~s$^{-1}$ results in a greater degree of blending of its lines with those of the lower velocity gas. Although the He~I lines were weaker in April and May, mass transfer from the secondary component was continuing as shown by the low to moderate strength of the helium lines and broadening of the sodium~D and H$\alpha$ lines (Table~\ref{table:UWVir_ew} and Figs.~\ref{Fig:uwvir_mt}i and \ref{Fig:PhSerSpec}).

\section{Ground-based photometry}
\label{Sec:G-B_Photometry}

The LCs of IZ~Tel were observed on a 356~mm Schmidt~Cassegrain telescope fitted with a Moravian G3-6303 CCD camera with Johnson $B$, $V$, and Bessel $I$ filters in 14 nights between July~2018 and July~2020. The Johnson $B$ and $V$ and Sloan $I$ magnitudes of the comparison star TYC~8798~1480-1 were taken from the AAVSO Photometric All-Sky Survey (APASS) data release~9 \citep{HEN15}. See \citet{MOR19} for details of the observatory and photometric reduction procedures. Outside of the primary eclipse exposure times were mostly 180-200~s for $B$ and 90~s for the $V$ and $I$ bands. Between phases 0.98–0.03 exposure times were 300~s for $B$, 150~s for $V$, and 120~s for the $I$ band.

\section{Spaceborne photometry}
\label{Sec:Space_Photometry}

UW~Vir was observed in three sectors with a 2-min cadence by TESS: Sector~10 (March~26 to April~22, 2019), sector~37 (2-28~April, 2021) and sector~67 (1-28~July, 2023). IZ~Tel was also observed in three sectors by TESS: With a 30-min cadence in sector~13 (June~19 to July~17, 2019), with a 10-min cadence in sector~27 (5-30~July, 2020) and with a 200-s cadence in sector~67 (1-28~July, 2023). We downloaded the TESS data from the Mikulski Archive for Space Telescopes (MAST)\footnote{https://mast.stsci.edu/portal/Mashup/Clients/Mast/Portal.html} \citep[cf.][]{JEN16}. The straight Simple Aperture Photometry (SAP) data were used, since the Pre-search Data Conditioning Simple Aperture Photometry (PDCSAP) detrending produces artificial side-effects associated with the search for planetary transits. Of the photometric data of IZ~Tel and UW~Vir in the Hipparcos catalogue\footnote{https://www.cosmos.esa.int/web/hipparcos/search-facility}, only those of the latter were sufficient for analysis.

\section{Modelling light and radial velocity curves}
\label{Sec:Models}

The numerical integration method of \citet[][WD]{WIL71} was applied to solve the LC and RV curves of IZ~Tel and UW~Vir simultaneously. This method models the LC and (or) RV curve of an eclipsing binary by taking into account the ellipticity and proximity effects of the components' Roche equipotential surfaces. The WD program combined with a `Monte Carlo' (MC) search procedure, as discussed in \citet{ZOL04}, was also used to find starting values and uncertainties for simultaneous solution of LC and RV curves.

The effective temperatures of 6800~K and 7750~K, corresponding to the spectral types of the IZ~Tel and UW~Vir primary stars (i.e.~F2V and A7V, respectively) were taken from \citet{PEC13} and \citet{EKE18}. As preliminary analysis of TESS LCs of UW~Vir indicated that the primary component may be a main sequence star, the effective temperature of a spectral class of A7V star was used, rather than A7IV (cf.~Section~\ref{Sec:SpectrClass}). An uncertainty of 300~K was assumed, based on the subtype uncertainty in the spectral classification and from the intrinsic scatter of the calibration relations. The effective temperatures of the secondary components ($T_2$) were adjusted in the WD iterations.

A quadratic limb-darkening law was used, with coefficients from \citet{CLA17} applied to the TESS LCs. For the ground-based LCs, the limb darkening coefficients were taken from the tables of \citet{CLA11}, according to the effective temperature of the components and the filters used. Bolometric gravity-darkening exponents of 0.32 and 1.0 were adopted from \citet{LUC67} and \citet{ZEI24} for convective ($T < 7200$~K) and radiative ($T>7200$~K) stellar atmospheres, respectively. The bolometric albedos of the components were set to 0.5 for convective atmospheres and to 1.0 for radiative atmospheres \citep{RUC69}. Due to reflection effect in the secondary component of UW~Vir, its albedo value was adjusted and was found $\sim0.6$.

\begin{figure*}
\centering
\begin{tabular}{cc}
\includegraphics[width=8.4cm]{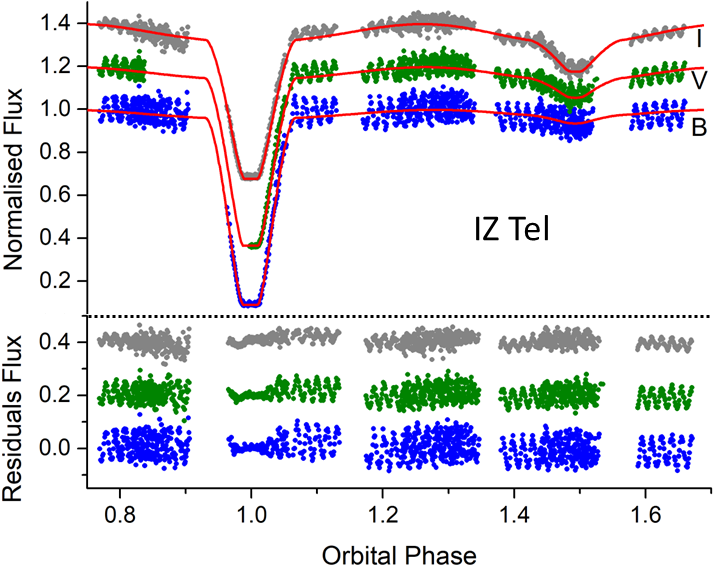}&\includegraphics[width=8.4cm]{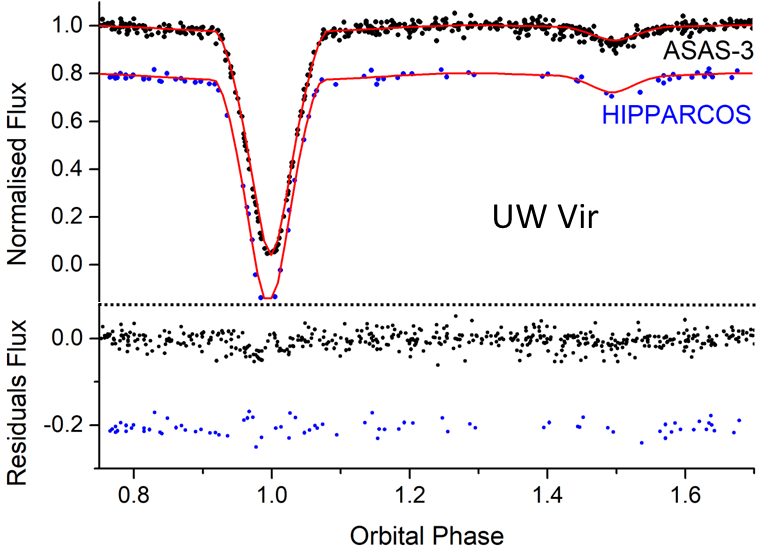}\\
\includegraphics[width=8.4cm]{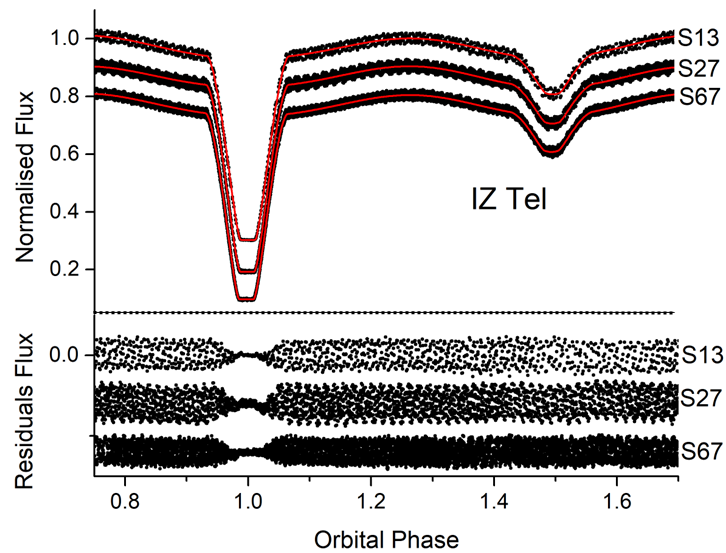}&\includegraphics[width=8.4cm]{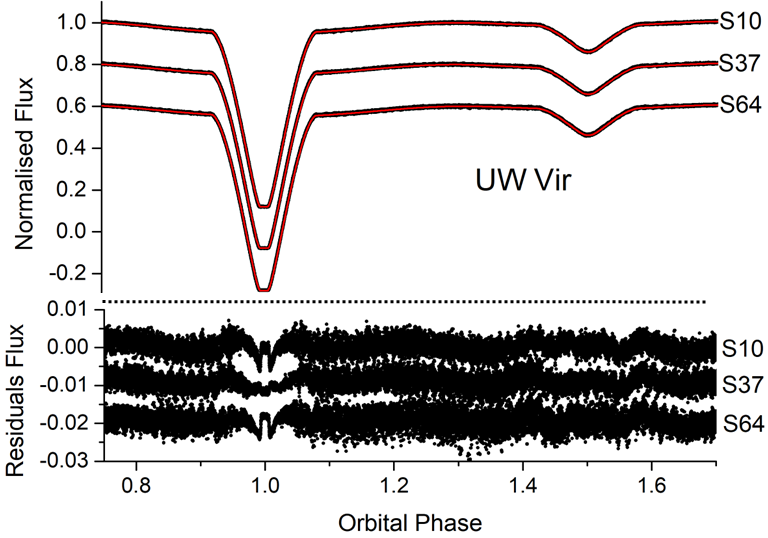}\\
\includegraphics[width=8.4cm]{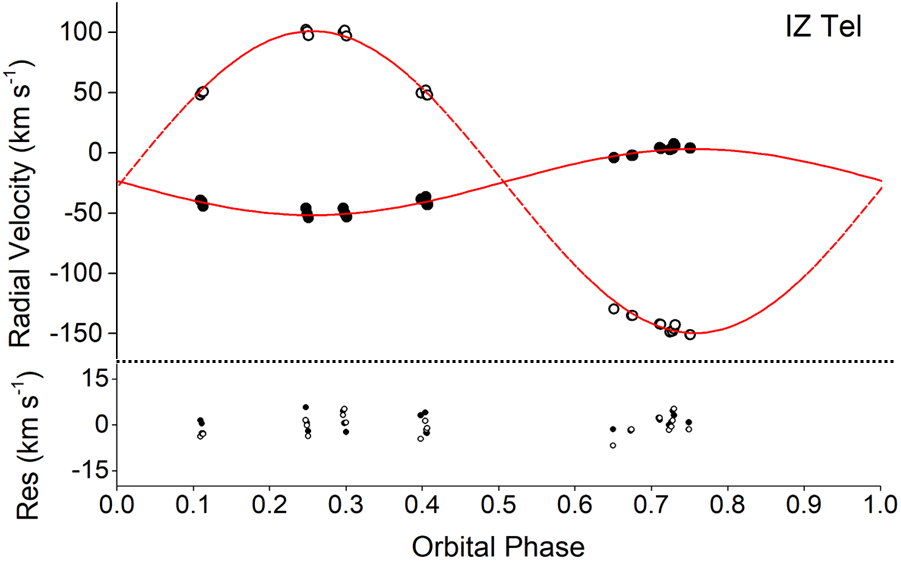}&\includegraphics[width=8.4cm]{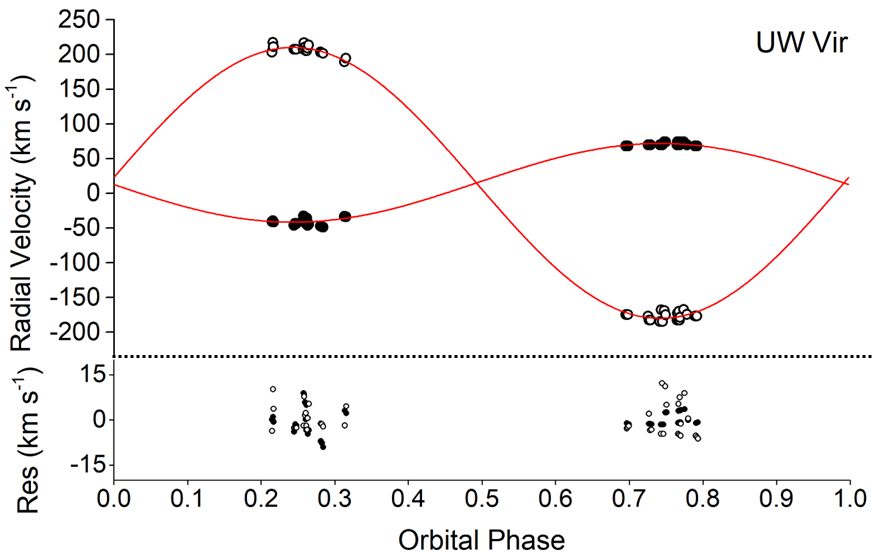}\\
\end{tabular}
\caption{Light and radial velocity curves (symbols) with the WD fittings (lines) and residuals of the solutions (bottom plots) of IZ~Tel (left panels) and UW~Vir (right panels). Left upper and middle panels show the $BVI$ and TESS sectors 13, 27, and 67 LCs along with the respective models for IZ~Tel. Right upper and middle panels illustrate the ASAS-3 and Hipparcos, and three TESS sector observed and synthetic light curves for UW~Vir. Both lower panels show the radial velocities of the primary (filled symbols) and secondary (hollow symbols) components. Residuals in all panels except the bottom ones are shifted to enhance visibility.}
\label{fig:LCRV}
\end{figure*}

\begin{table*}
\begin{center}
\caption{Results of the WD fitting to the radial velocity and light curves of IZ~Tel and UW~Vir.
\label{Tab:Models_LCs}}
\begin{tabular}{lcccc| ccc}
\hline\hline															
System:	&	\multicolumn{4}{c}{IZ~Tel}							&	\multicolumn{3}{c}{UW~Vir}					\\
\hline															
Parameter   	&	 $B$ 	&	 $V$ 	&	 $I$ 	&	 TESS 	&	ASAS-3 	&	 HIPPARCOS 	&	 TESS	\\
\hline															
$T_0$ (2400000+)	&	\multicolumn{3}{c}{58655.6677$^a$}					&	58655.6685$^b$	&	52501.10820$^a$	&	48501.1400$^a$	&	59315.04261$^b$	\\
$P$ (d)	&	 \multicolumn{3}{c}{4.8803178}					&	4.8803178	&	1.8107666	&	1.81078	&	1.8107817	\\
\hline															
$a$~(R$_{\sun}$)	& 14.6(2)	& 14.6(1)	& 14.7(1)	& 14.7(1)	& 9.01(4)	& 9.22(6)	& 9.01(4) 	\\
$V_\gamma$		& $-24(1)$	& $-24(1)$	& $-24(1)$	& $-24(1)$	& 15(1)	& 15(2)	& 15(1)  \\
$K_1$ (km~s$^{-1}$)	& 26.2(1.9)	& 26.3(1.8)	& 26.4(1.8)	& 27.6(1.6)	& 56.6(2.1)	& 57.9(2.4)	& 56.6(1.98)    	\\
$K_2$ (km~s$^{-1}$)	& 124.9(3.2)	& 125.1(2.4)	& 125.6(2.5)	& 125.2(1.8)	& 195.0(3.2)	& 199.6(3.6)	& 195.1(3.1)   	\\
$\Delta \phi$	&	0.0014(1)	&	0.0015(1)	&	0.0016(1)	&	0.0002(1)	&	$-0.0007$(1) 	&	$-0.0041$(2) 	&	0.0001(1)	\\
$i$ ($\degr$)	&	88.4(8)	&	88.2(6)	&	 87.1(4)	&	 88.3(5)	&	88.7(5) 	&	89.7(8) 	&	88.9(2)	\\
$T_1$ (K)	&	6800 (fixed)	&	6800 (fixed)	&	6800 (fixed)	&	6800 (fixed) 	&	7750 (fixed) 	&	7750 (fixed) 	&	7750 (fixed)	\\
$T_2$ (K)	&	4105(33)	&	4263(28)	&	4336(27)	&	4327(17)	&	4063(34)	&	4144(45)	&	4130(30)	\\
$\Omega_1$	&	6.31(13)	&	6.39(10)	&	 6.52(11)	&	 6.40(10)	&	4.96(4)	&	4.99(10)	&	4.923(8)	\\

$\Omega_2$	&	2.258	&	2.258	&	 2.258	&	 2.299(3)	&	2.485(12)	&	2.481(23)	&	2.469(2)	\\

$q=M_2/M_1$	&	0.21(1)	& 0.21(1)	& 0.21(1)	& 0.22(1) 	&	0.29(1)	&	0.29(1) 	&	0.29(1)	\\

$fof_1$	&	$-0.64$	&	$-0.62$	&	$-0.63$	&	$-0.64$ 	&	$-0.502$	&	$-0.507$	&	$-0.500$ 	\\

$fof_2$	&	0	&	0	&	0	&	$-0.007$ 	&	$-0.017$	&	$-0.015$	&	$-0.010$  	\\

$r_1$ (volume)	&	0.16(3)	&	0.16(2)	&	0.16(2)	&	  0.162(6)	&	0.21(2)	&	0.21(3)	&	0.216(6)	\\

$r_2$ (volume)	&	0.25(3)	&	0.25(2)	&	0.25(2)	&	0.251(8)	&	0.26(3)	&	0.27(3)	&	0.269(8)	\\

$L_1/(L_1+L_2)$	&	 0.89(5)	&	0.80(3)	&	 0.65(3)	&	0.679(19)	&	0.94(3)	&	0.93(5)	&	0.853(10)	\\
$L_2/(L_1+L_2)$	&	 0.11(1)	&	0.20(2)	&	0.35(2)	&	0.321(13)	&	0.06(1)	&	0.07(1)	&	0.147(5)	\\
$l_3$			&	0.015(3)	&	0.016(2)	&	0.010(2)	&	0.048(6)	&	0	&	0	&	0	\\
\hline				  											
\end{tabular}
\\
\textbf{Notes.} $^a$in HJD, $^b$in BJD										
\end{center}
\end{table*}

In the TESS SAP data, the CROWDSAP parameter, which estimates the ratio of target flux to total flux in the photometric aperture, indicated that almost all of the flux originated from the UW~Vir system. For IZ~Tel, 3–6\% of the observed flux in the SAP aperture does not come from the binary system. As our spectra showed no evidence for a tertiary component (Sect.~\ref{Sec:RVs}) the third light ($l_3$) would be due the H$\alpha$ emission in the gas ring around the primary star (Sect.~\ref{Sec:iztel_for_mt}). In the TESS LCs of these two systems there was a slight asymmetry between maximum light intensity (Fig.~\ref{fig:LCRV}). In the case of UW~Vir, the episodes of mass impacts on the primary star would create spots around its equatorial region (Sect.~\ref{Sec:uwvir_for_mt}). An equatorial hot spot on the primary component fitted the LCs with a longitude $\sim100\degr$ with an angular dimension of up to $20\degr$, and a temperature factor of approximately 1.05. Although no evidence of third light was detected in the spectroscopic analysis of this EB, its Eclipse Timing Variation diagram (see Sect.~\ref{Sec:ETV}) suggests the existence of a tertiary component. Thus, although we explored a WD solution that accounted for the $l_3$ parameter, it converged to an essentially zero value and was therefore excluded from further consideration.

The $BVI$ LCs of IZ~Tel did not show any significant asymmetry, probably due to the large amplitude of the pulsations. Equatorial hot spots with an angular dimension of $\sim20\degr$, temperature factor of $\sim1.06$ and longitudes of $105\degr$ and $77\degr$ fitted the LCs for the TESS sectors~13 and 67, respectively.

The fill-out factor for IZ~Tel indicates that the primary component filled approximately 40\% of its Roche lobe (RL). In $BVI$~LCs solutions, the secondary component was found to fill its RL completely and in TESS LCs solutions by 99.3\%. The fill-out factor for UW~Vir in TESS LCs solutions indicates that the primary component filled $\sim50$\% of its RL, while the secondary component filled $\sim99$\% of its RL. Considering the errors in the surface potential, $\Omega_2$, parameter for both systems and the widespread convective outer atmosphere in their secondary components, we conclude that both IZ~Tel and UW~Vir are semi-detached systems with each secondary component filling its RL.

The values of the WD modelling results are listed in Table~\ref{Tab:Models_LCs}. This table contains: The ephemeris elements used for phasing the LCs (i.e.~reference time of minimum $T_0$ and orbital period $P$), semi-major axis ($a$), systemic velocity ($V_\gamma$), amplitudes of the RVs ($K_1$ and $K_2$), phase-shift ($\Delta \phi$), orbital inclination ($i$) with respect to the line-of-sight, effective temperature ($T$), surface potential ($\Omega$), mass ratio ($q$), fill-out factor ($fof$), relative radius ($r$), relative luminosity ($L$), and third light contribution ($l_3$). As the amplitudes of the RVs, K1 and K2, are not directly given by the WD method, they were calculated using the relevant parameters in the RV curve fitting. The weighted averages of the values for each TESS sector are shown in this table. The comparisons of the RV measurements and photometric observations with the WD models for both systems are shown in Fig.~\ref{fig:LCRV}.

\section{Physical parameters, distances and evolutionary status}
\label{Sec:AbsPar}

The physical parameters of the components of both systems were derived based on the simultaneous RV+LC solutions (Table~\ref{Tab:Models_LCs}) and are listed in Table~\ref{table:abs_par}. In these calculations, the simultaneous RV + TESS LC solutions were generally preferred, given the relatively high precision of TESS data. The first six rows in Table~\ref{table:abs_par} contain the physical parameters directly provided by the WD program. The next five rows contain the magnitudes necessary for calculating the distances to the binary systems.

The following method was used to calculate the distances to these systems: (i)~The bolometric magnitudes ($M_{\rm bol,1,2}$) of the components derived by WD were converted to TESS-band absolute magnitudes ($M_{\rm TESS, 1, 2}$) using the bolometric correction formula: $M_{\rm TESS,1,2}= M_{\rm bol,1,2} - BC_{1,2}$.  Bolometric corrections for the components were taken from the study of \citet{EKE23}, according to their effective temperatures. (ii)~The TESS-band absolute magnitudes of the system was also computed from following equation:
\begin{equation}
M_{\rm TESS,~system}=M_{\rm TESS,2} -2.5 \log \left(1+10^{-0.4\left(M_{\rm TESS,~1}-M_{\rm TESS,~2}\right)}\right).
\end{equation}
(iii)~The distances to the systems were then calculated from the distance modulus ($d = 10^{m_{\rm TESS} - M_{\rm TESS} + 5 - A_{\rm TESS}}$). Here, $m_{\rm TESS}$ is the apparent magnitude and $A_{\rm TESS}$ is the interstellar extinction in the given band. The TESS-band apparent magnitude $m_{\rm TESS}$ was taken from the MAST Portal, and the parameter $A_{\rm TESS}$ was calculated using the $A_{\rm TESS}=1.940~E(B-V)$ relation given by \citet{EKE23}. The colour excess $E(B-V)$ was determined as described in \citet{LIA24}.

As a way to check the accuracy of the absolute parameters in Table \ref{table:abs_par}, the photometric parallax ($\pi_{\rm phot}=1/d$) was compared with the trigonometric parallax ($\pi$) in Gaia~DR3 \citep{GAIA23}. We derived $\log \pi_{\rm phot} = -3.10$ and $\log \pi_{\rm phot} = -2.41$ for IZ~Tel and UW~Vir, respectively. These parallaxes are in close agreement with those in the Gaia~DR3 catalogue ($\log \pi = -3.07$ and $\log \pi = -2.39$).

\begin{table}
\centering
\caption{Absolute parameters of IZ~Tel and UW~Vir.}
\label{table:abs_par}
\begin{tabular}{lcccc}
\hline\hline									
System: 	&	\multicolumn{2}{c}{IZ~Tel}			&	\multicolumn{2}{c}{UW~Vir}			\\
\hline									
Component:	&	Prim.	&	Sec.	&	Prim.	&	Sec.	\\
\hline									
$M$~(M$_{\sun}$)	&	1.48(9)	&	0.33(3)	&	2.33(9)	&	0.68(5)	\\
$R$~(R$_{\sun}$)	&	2.39(10)	&	3.70(14)	&	1.95(6)	&	2.42(8)	\\
$\log g$~(cm~s$^{-2}$) 	&	3.85(1)	&	2.82(1)	&	4.23(1)	&	3.50(3)	\\
$L$~(L$_{\sun}$)	&	11(3)	&	4.3(1.1)	&	12(3)	 &	1.54(40)	\\
$a$~(R$_{\sun}$)	&	2.66(6)	&	12.09(8)	&	2.03(3)	&	6.98(4)	\\
$M_{\rm bol}$ (mag)	&	2.14(28)	&	3.15(28)	&	2.01(24)	&	4.27(28)	\\
\hline									
$BC_{\rm TESS}$ (mag)	&	0.33(1)	&	0.67(1)	&	0.14(1)	&	0.67(1)	\\
$M_{\rm TESS}$ (mag)	&	1.81(28)	&	2.48(28)	&	1.87(24)	&	3.60(28)	\\
$m_{TESS}$ (mag)	&	\multicolumn{2}{c}{11.93(1)}			&	\multicolumn{2}{c}{8.82(1)}			\\
$A_{\rm TESS}$ (mag)	&	\multicolumn{2}{c}{0.11(1)}			&	\multicolumn{2}{c}{0.11(1)}			\\
$M_{\rm TESS}$ (mag)	&	\multicolumn{2}{c}{1.34(30)}			&	\multicolumn{2}{c}{1.67(30)}			\\
$d$ (pc)	&	\multicolumn{2}{c}{1247(150)}			&	\multicolumn{2}{c}{256(34)}			\\
\hline			
\end{tabular}		
\textbf{Notes.} Prim.=Primary, Sec.=Secondary	
\end{table}		

\begin{figure}
\centering
\includegraphics[width=\columnwidth]{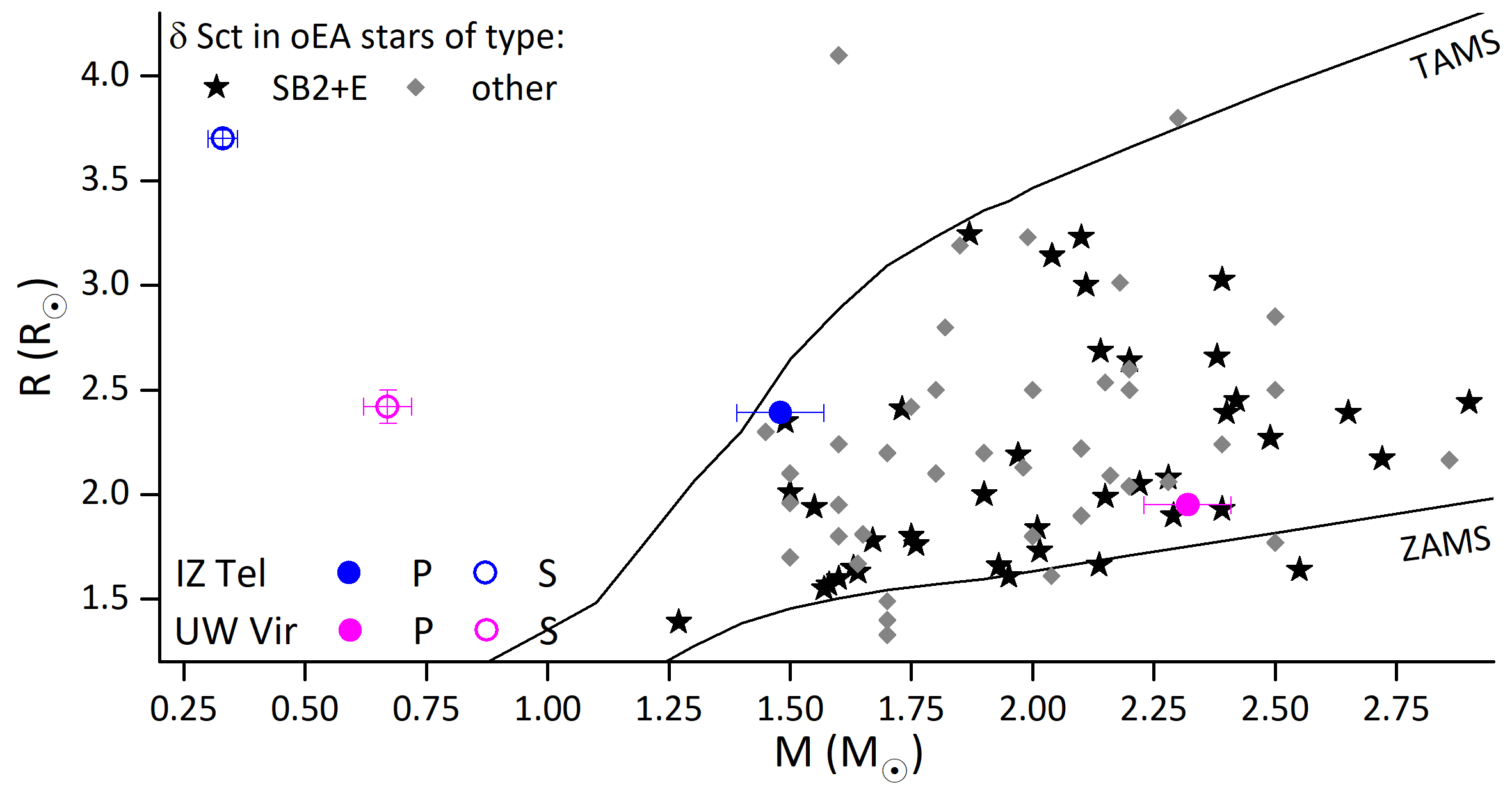}  \\
\includegraphics[width=\columnwidth]{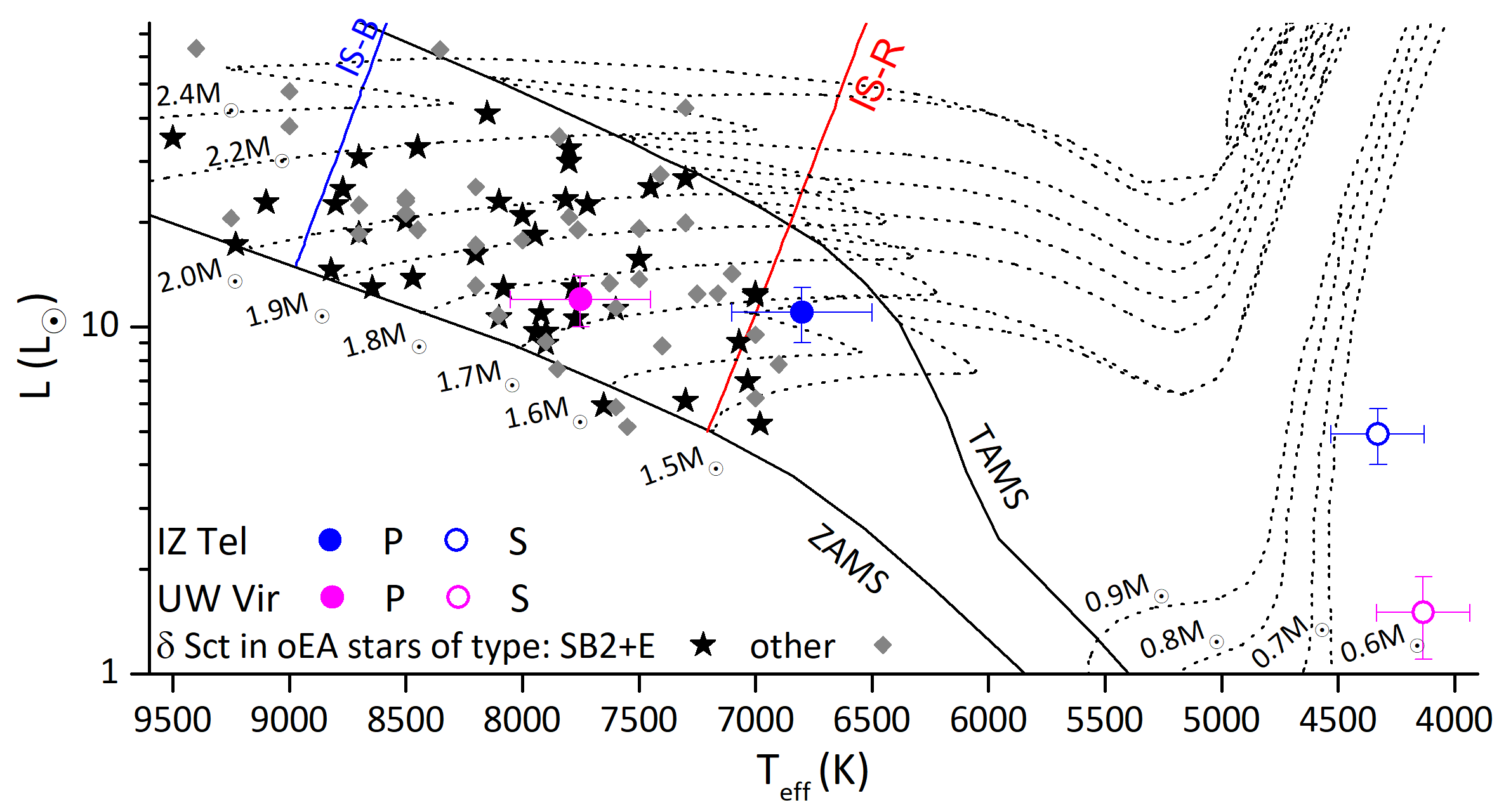}
\caption{Mass--Radius (top panel) and Hertzsprung--Russell (bottom panel) evolutionary diagrams for oEA stars. In both diagrams, black symbols denote the $\delta$~Scuti members of oEA stars that belong to double-lined and eclipsing (SB2+E) systems and the gray symbols those that belong either to spectroscopic single-lined and eclipsing systems or only in eclipsing systems. Blue and magenta symbols refer to the primary (filled) and secondary (open) components of IZ~Tel and UW~Virl, respectively. The sample was mainly gathered from \citet{LIAN17} and \citet{LIA25}. Solid black lines indicate the ZAMS and TAMS limits and the coloured solid lines (B = blue, R = red) the boundaries of the instability strip \citep[IS;][]{SOY06}}.
\label{fig:evol}
\end{figure}

The primary components of both systems are main sequence stars (Fig.~\ref{fig:evol}). The primary of IZ~Tel is located close to Terminal-Age main sequence (TAMS) and slightly outside of the classical instability strip, whereas that of UW~Vir is close to the Zero-Age main sequence (ZAMS). This situation indicates mass transfer in the past from the original primary component of IZ~Tel to the original secondary component, leading to inversion of the mass ratio. The original primary component, and now current secondary component, has evolved to the subgiant stage, filled its Roche lobe and now is transferring mass to the current primary component, as shown by the H$\alpha$ spectrum at phase 0. The $\sim$2.3~M$_{\sun}$ mass value of UW~Vir's primary component and its position near Zero-Age main sequence (ZAMS) in the $H-R$ diagram indicates mass transfer from the current secondary to the primary component, as shown in the spectra (Sect.~\ref{Sec:uwvir_for_mt}). The secondary of UW~Vir appears much less luminous and cooler than expected for its mass. The evolution of Algol binary systems, in which mass ratio reversal occurs, is complex and is not a topic of discussion in this paper as their positions cannot be compared directly with the evolutionary tracks of single stars (Fig.~\ref{fig:evol}).

\section{Eclipse Timing Variation analysis}
\label{Sec:ETV}	

Of the two systems studied here, for IZ~Tel there are only two historical minima timings and 25 more from our $BVI$ and TESS LCs, a time span of five years, and hence these are not sufficient for an ETV analysis. For future use, we list these timings in Table~\ref{Tab:ToM1}. There were enough past times of minima (ToM) for UW~Vir to perform an orbital period analysis; there are more than 80 past ToM for UW~Vir ranging between early 1900's to date. Historical times of minima were compiled from online databases\footnote{https://www.variablestarssouth.org/}$^,$\footnote{http://var2.astro.cz/ocgate/}$^,$\footnote{http://www.oa.uj.edu.pl/ktt/krttk\_dn.html}$^,$\footnote{https://www.bav-astro.eu/index.php/}$^,$\footnote{https://www.as.up.krakow.pl/ephem/}$^,$\footnote{https://binaries.boulder.swri.edu/binaries/omc/}
and supplemented with values derived from the TESS data. The ToM used in the analysis are listed in Table~\ref{Tab:ToM2}.

\begin{figure}
\centering
\includegraphics[width=\columnwidth]{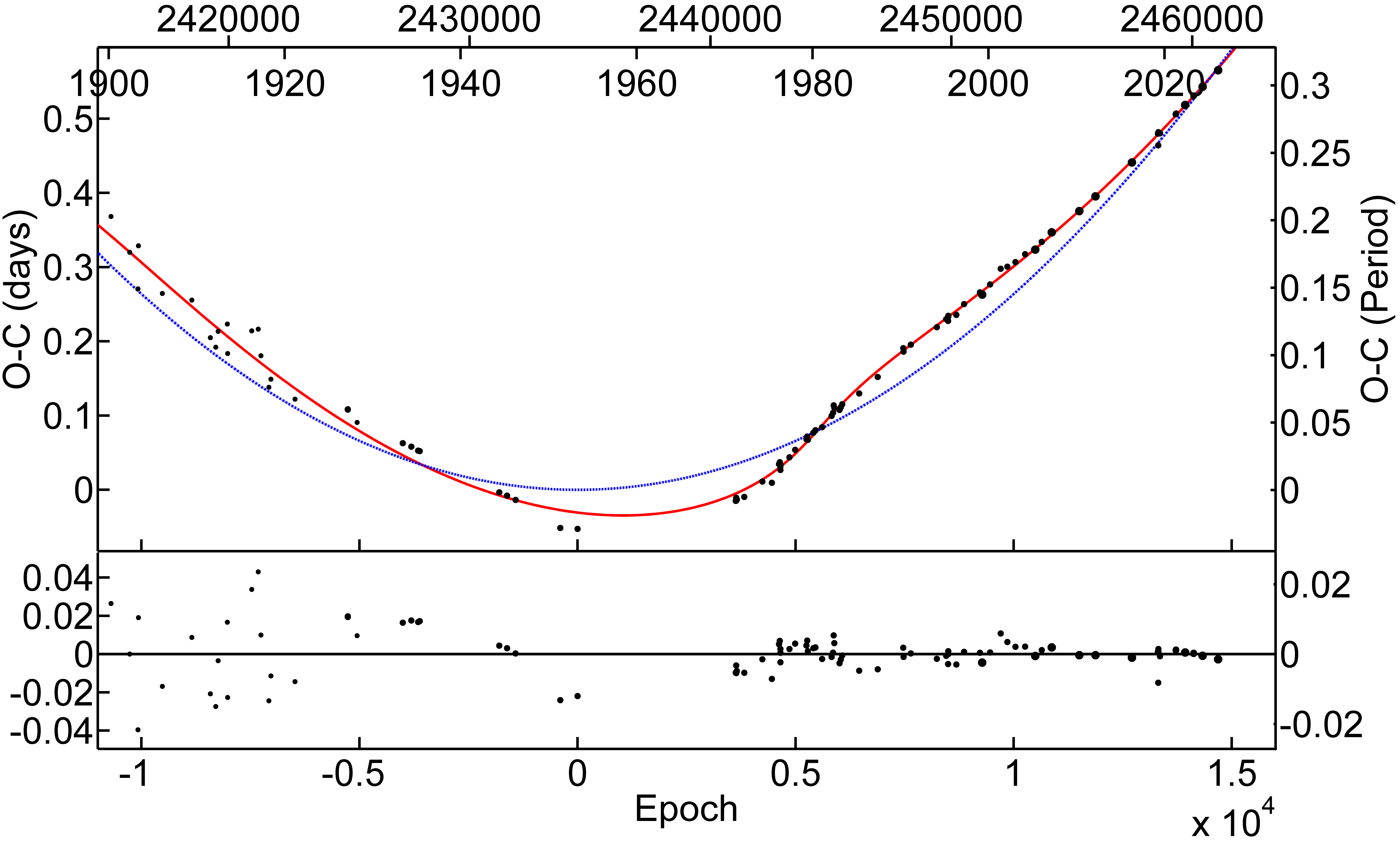}\\
\includegraphics[width=\columnwidth]{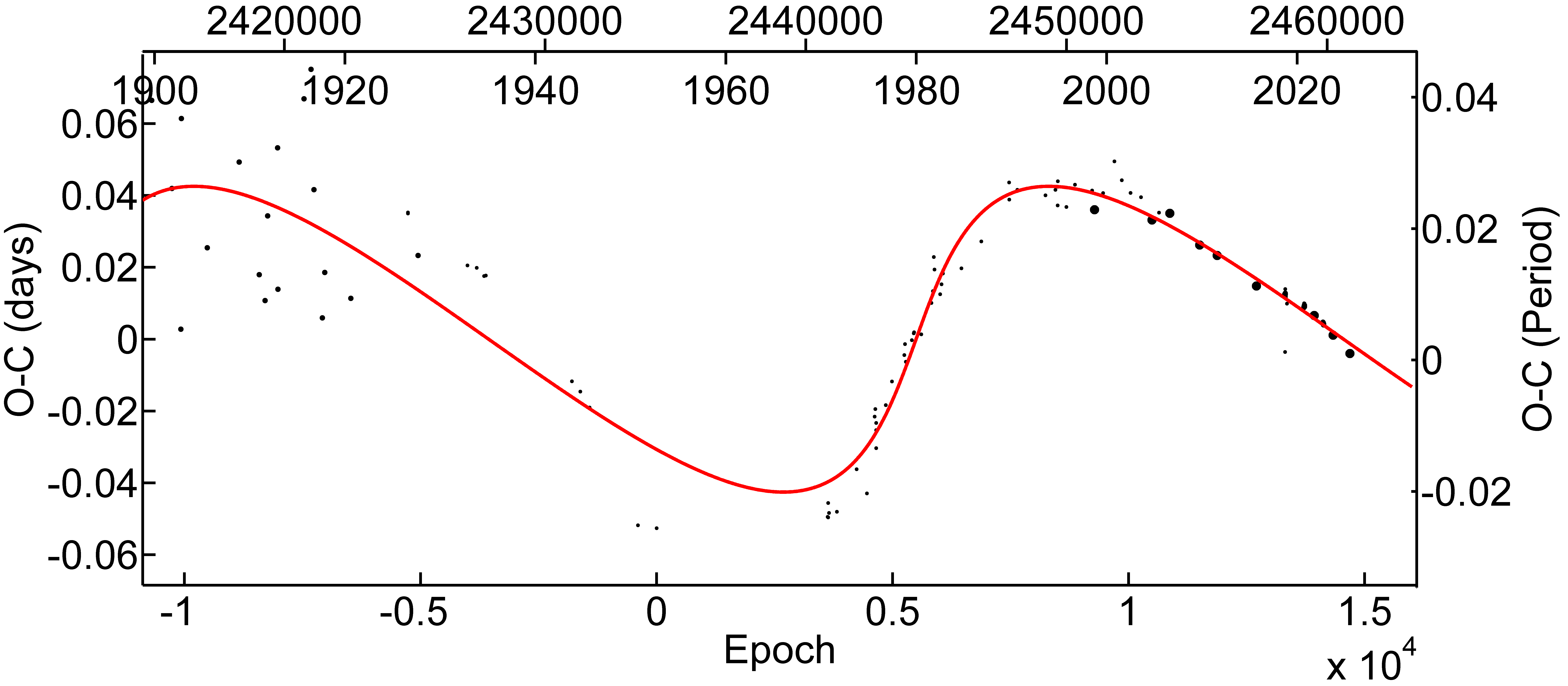}\\
\caption{Upper plot: Fitting of a LITE curve and a parabola on the minima timings of UW~Vir and the residuals of the solution. The red line is the sum of the two fitted curves, while the blue line represents only the parabola. Lower plot: Fitting of the LITE curve (red line) after the subtraction of parabola. The bigger the symbol the bigger the statistical weight.}
\label{Fig:uwvir_ETV}
\end{figure}

Orbital period variations were analysed using the LITE code of \citet{ZAS09}, which applies statistical weights ($w$) to the times of minima based on the observational technique. Visual and photographic minima were assigned $w=1$, photoelectric $w=7$, and CCD $w=10$. The TESS minima timings were assigned $w=1$ because they are too many and concentrated into a very narrow time range of four years. The code fits the observed data using both Light-Time Effect \citep[LITE;][]{IRW59} and parabolic models. The LITE model includes seven free parameters: The binary ephemeris ($T_0$, $P$), the amplitude of variation ($A_{\rm ETV}$), the orbital period of the third body ($P_3$), the time of periastron passage ($JD_0$), the argument of periastron ($\Omega$), and the orbital eccentricity ($e_3$). The parabolic model involves three free parameters: the binary ephemeris and the quadratic term ($C_2$). A modified ephemeris from \citet{KRE04} was adopted and allowed to vary during the fitting process. The fitting is illustrated in Fig.~\ref{Fig:uwvir_ETV} and the results are listed in Table~\ref{Tab:ETV}. The resulting ETV parameters were then used as inputs in the InPeVeb software \citep{LIA15B} to derive the characteristics of the most probable orbital modulation mechanisms.

The results indicate that the mass transfer direction is from the less to more massive component of the system, which is in good agreement with the conventional semi-detached configuration resulting from the LC+RV analysis. The period variation rate, and hence the mass transfer rate, is one order of magnitude higher than in similar systems \citep[e.g.~][]{LIA22, LIA24}. This relatively high mass flow rate concurs with the spectroscopic findings (see Sect.~\ref{Sec:uwvir_for_mt}).

In contrast, the periodic modulation of an EB's period cannot be explained at present. Taking into account the $f(m_3)$ that is based on the total mass of the components of the EB (i.e.~$M_1=2.32$~M$_{\sun}$ and $M_2=0.67$~M$_{\sun}$), the resulting minimum mass (i.e.~for coplanar orbit) of the hypothetical third component is $\sim1.19$~M$_{\sun}$. If this component were a main sequence star, we would expect a third light contribution of at least 11\%. Using the distance of the system (Table~\ref{table:abs_par}), we found that the orbit of the tertiary should have a semi-major axis of $\sim29$~AU and $\sim104$~mas angular separation from the EB. However, no third light was detected in either the LC or RV analyses; therefore, its nature cannot be determined. We speculate that, if present, this component might be an exotic object (e.g.~a black hole, neutron star, or white dwarf) or even an additional binary system. For the latter scenario, it must consist of two low-mass stars whose total luminosity is much less than that of a $\sim1.19$~M$_{\sun}$ star. The latter might explain the absence from our LC+RV analyses. As an alternative explanation for the periodic shift of the EB's orbital period, we examined Applegate's mechanism \citep{APP92}. Assuming the secondary component is the potential magnetically active star, the quadrupole moment variation ($\Delta$Q) would be of the order of $10^{50}$~gr~cm$^2$, which is well inside the range set by \citet{LAN02}, and might explain the observed periodic period changes. However, the ETV model revealed a relatively high eccentricity value that contradicts the latter possible explanation. Therefore, we conclude that the clarification regarding the cyclic changes UW Vir's orbital period remains open, and any further examination is beyond the scope of this work.

\begin{table}			
\begin{center}			
\caption{ETV parameters of UW~Vir.}			
\label{Tab:ETV}			
\begin{tabular}{lc}			
\hline\hline			
Parameter	&	Eclipsing binary	\\
\hline			
$T_0$ (HJD)	&	2434478.732(4)	\\
$P$ (d)	&	1.8107178(3)	\\
\hline			
	&	Light-time effect	\\
\hline			
$P_3$ (yr)	&	89.8(1.4)	\\
$JD_0$ (HJD)	&	2444432(176)	\\
$A_{\rm ETV}$ (d)	&	0.043(2)	\\
$\Omega$ ($\degr$)	&	0(2)	\\
e	&	0.59(5)\\
$f(m_3)$ (M$_{\sun}$)	&	0.0956(6)	\\
\hline			
	&	Quadruple moment variation	\\
\hline			
$\Delta$Q (gr~cm$^2$)	&	4.8(3)~$\times 10^{50}$	\\
\hline			
	&	Mass transfer	\\
\hline			
$c_2$ (d cycle$^{-1}$)	&	2.6442(1)~$\times 10^{-9}$	\\
$\dot{P}$ (d~yr$^{-1}$)	&	1.070(4)~$\times 10^{-7}$	\\
$\dot{M}_{\rm tr}$ (M$_{\sun}$~yr$^{-1}$)	&	1.9(2)~$\times 10^{-7}$	\\
\hline			
$\sum res^2$	&	0.032	\\
\hline			
\end{tabular}
\end{center}			
\end{table}		

\section{Pulsation Analyses}
\label{Sec:Puls}

The frequency analysis was carried out using the software \textsc{PERIOD04} v.1.2 \citep{LEN05}, which employs classical Fourier techniques. Although $\delta$~Sct stars typically show pulsations within the 4–80~d$^{-1}$ range \citep{BRE00, BOW18}, we extended the search interval to 0–80~d$^{-1}$ because $\delta$~Sct stars in binary systems may also exhibit $g$-mode pulsations linked to their orbital periods or display hybrid $\gamma$~Doradus–$\delta$~Scuti behaviour. For these analyses, we used the residuals of each system's LC (Fig.~\ref{fig:LCRV}) of the out-of-eclipse data in order the sample to be homogeneous (i.e.~constant light contribution from the system's components). The analyses were made in the LCs of each systems which are the best in terms of time resolution, observations time span, and photometric quality. To ensure the detection of reliable frequencies—especially in cases where many closely spaced frequencies increase the local background—we adopted the signal-to-noise ratio ($S/N$) approach proposed by \citet{LIA17} \citep[cf.][]{LIA25b}. In brief, the background noise was calculated within a frequency region free of signals, using a box size of 2 within a 2~d$^{-1}$ window. The $S/N$ of each detected frequency was then obtained by dividing its amplitude by this background value. After identifying each frequency, we pre-whitened the residuals before proceeding to the next one. The search was terminated once the $S/N$ of a detected peak fell to approximately 5 as proposed for long time-series data \citep{BAR15, BOW21}. For both systems we calculated the orbital frequency ($f_{\rm orb} = 1 / P_{\rm orb}$) and searched for possible sidelobes. Moreover, we calculated all possible combinations of frequencies ($f_{\rm combo}$) based on the independent ones and the $f_{\rm orb}$ using the following formula:
\begin{equation}
f_{\rm combo}= a f_{\rm i}+b f_{\rm j} + c f_{\rm k} + d f_{\rm orb},
\end{equation}
where $a,~b,~c=0, \pm 1, \pm2, \pm3, \pm4$, $d=0, \pm 1, \pm2...\pm 20$, and $f_{\rm i}$, $f_{\rm j}$ and $f_{\rm k}$ the independent frequencies. Every calculated $f_{\rm combo}$ was checked against the detected frequencies, and if a match was found within the frequency resolution \citep[$\delta f = 1.5 / \delta t$, where $\delta t$ is the total observational time span in days; cf.][]{LOU78}, the detected frequency was identified as arising from that combination.

According to the spectroscopy and physical parameters (Sections~\ref{Sec:SpectrClass} and \ref{Sec:AbsPar}) only the primary components of the studied systems are consistent with the characteristic mass and temperature profiles of $\delta$~Sct stars. Therefore, we conclude that the observed pulsations originate from these stars. Results of the analyses are presented in Table~\ref{tab:Indfreqs}, which lists the independent frequencies found for the $\delta$~Sct components of each system in each band and TESS sectors. This table includes: the frequency index ($i$), the frequency value ($f_{\rm i}$), the amplitude ($A$), the phase ($\Phi$), the pulsation constant $Q$, and the spherical $l$-degrees. The other (combination) frequencies for each  system are given in Tables~\ref{Tab:Vircombo1}-\ref{Tab:Telcombo}. In the following subsections, we present our analysis strategy and comments on the results for each system.

\begin{table}			
\centering		
\caption{Independent oscillation frequencies for IZ~Tel and UW~Vir.}			
\label{tab:Indfreqs}			
\scalebox{0.9}{			
\begin{tabular}{cccc ccc}		
\hline\hline													
Band	&	$i$	&	  $f_{\rm i}$	&	$A$	&	  $\Phi$	&	$Q$	&	$l$	\\
	&		&	     (d$^{-1}$)	&	(mmag)	&	(2$\pi$~rad)	&	(d)	&		\\
\hline													
\multicolumn{7}{c}{IZ~Tel}													\\
\hline													
$B$	&	\multirow{4}{*}{1} 	&	13.560(1)	&	60.2(7)	&	0.646(2)	&	\multirow{4}{*}{0.024(2)} 	&	\multirow{4}{*}{1-p2} 	\\
$V$	&		&	13.559(1)	&	41.4(4)	&	0.648(2)	&		&		\\
$I$	&		&	13.558(3)	&	20.5(5)	&	0.639(4)	&		&		\\
T s67	&		&	13.5579(1)	&	15.49(4)	&	0.8093(4)	&		&		\\
\hline													
\multicolumn{7}{c}{UW~Vir}													\\
\hline													
\multirow{3}{*}{T s10} 	&	2	&	43.2500(3)	&	1.04(1)	&	0.554(2)	&	0.013(1)	&	2-p5	\\
	&	3	&	36.9257(3)	&	0.92(1)	&	0.475(2)	&	0.015(1)	&	2-p4	\\
	&	6	&	34.9217(4)	&	0.63(1)	&	0.282(3)	&	0.016(1)	&	1-p4	\\
\hline													
\multirow{3}{*}{T s37} 	&	2	&	43.2495(2)	&	0.96(1)	&	0.667(2)	&	0.013(1)	&	2-p5	\\
	&	4	&	36.9250(3)	&	0.84(1)	&	0.306(2)	&	0.015(1)	&	2-p4	\\
	&	6	&	34.9191(4)	&	0.62(1)	&	0.508(3)	&	0.016(1)	&	1-p4	\\
\hline													
\multirow{3}{*}{T s64} 	&	2	&	36.9253(3)	&	0.82(1)	&	0.379(2)	&	0.015(1)	&	2-p4	\\
	&	4	&	34.9188(3)	&	0.73(1)	&	0.935(2)	&	0.016(1)	&	1-p4	\\
	&	5	&	43.2506(3)	&	0.69(1)	&	0.796(2)	&	0.013(1)	&	2-p5	\\
\hline																																																	
\end{tabular}}				
\textbf{Notes.} T=TESS, s=sector, p=pressure. Values next to `p' denote the radial order $n$.	
\end{table}

\subsection{IZ~Tel}
\label{Sec:Puls_IZTel}

The $BVI$ LCs of IZ~Tel (Fig.~\ref{fig:LCRV}) have a time resolution of $\sim7$~min and expand within almost two years. However, the first six observing nights were made in July~2018 in a time span of a week. These multi-wavelength observations were analysed in order to detect only the strongest pulsation frequencies and to calculate their spherical degrees, $l$, based on the amplitude and phase ratios. In the analyses of these LCs, we used the $S/N=4$ limit for the detection of frequencies as suggested by \citet{LEN05}. Of the three available TESS LCs of IZ~Tel, only that of sector~67 was observed in short-cadence mode (i.e.~time resolution 3.3~min) and it expands in $\sim24$~d, hence, this is the only set we used for the detailed frequency analysis.

\begin{figure}
\centering
\includegraphics[width=\columnwidth]{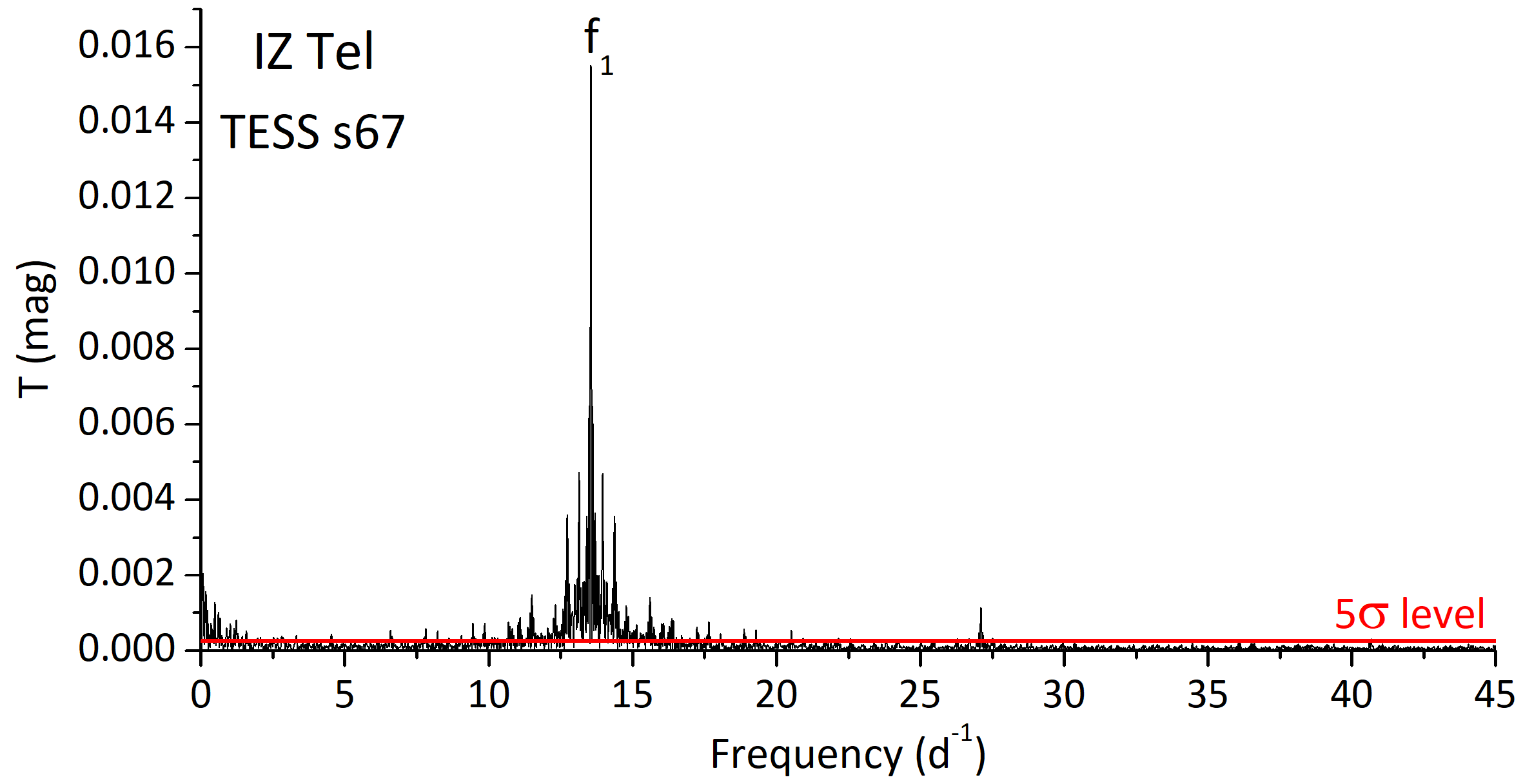}
\caption{Periodogram of the TESS data set of IZ~Tel.}
\label{fig:FSTEL}
\end{figure}

The pulsation analysis of the TESS LC resulted in only one independent frequency of $f_1\sim13.56$~d$^{-1}$ and 17 another frequencies that are harmonics of $f_1$ and $f_{\rm orb}$ (=0.205~d$^{-1}$) or combinations of them (Table~\ref{Tab:Telcombo}). The periodogram is plotted in Fig.~\ref{fig:FSTEL} and the Fourier fitting in Fig.~\ref{fig:FF}. The second strongest peak ($f_2 \sim 0.07$~d$^{-1}$) is likely an artefact, since neither harmonics nor combination frequencies involving it were identified. The respective analysis of $BVI$ LCs confirmed the value of $f_1$ (Table~\ref{tab:Indfreqs}) and also some of the frequencies detected in the TESS data such as $\sim27.1$~d$^{-1}$ and $\sim40.6$~d$^{-1}$ (i.e.~the second and third harmonics of $f_1$). However, in these data sets, low frequency values ranging between 0.16 and 2.9~d$^{-1}$ were found, but they are not verified by the far more accurate TESS LCs. Using the amplitude and phase ratios of $f_1$, as derived from the $BVI$ LCs pulsation analyses (Table~\ref{tab:Indfreqs}), and the absolute parameters of the primary component (Table~\ref{table:abs_par}) as inputs in the software FAMIAS \citep{ZIM08}, we identified that the dominant frequency is a non-radial pressure mode with $l$-degrees=1. Using the physical parameters, we also cross-checked this result with the models of \citet{FIT81} using the pulsation constant of the $\delta$~Sct star of IZ~Tel. Indeed, the latter models agree with the $l=1$ results for $f_1$ and, additionally, they also predict that this mode is the second overtone (radial order $n=2$).

\subsection{UW~Vir}
\label{Sec:Puls_UWVir}

\begin{figure}
\centering
\includegraphics[width=8cm]{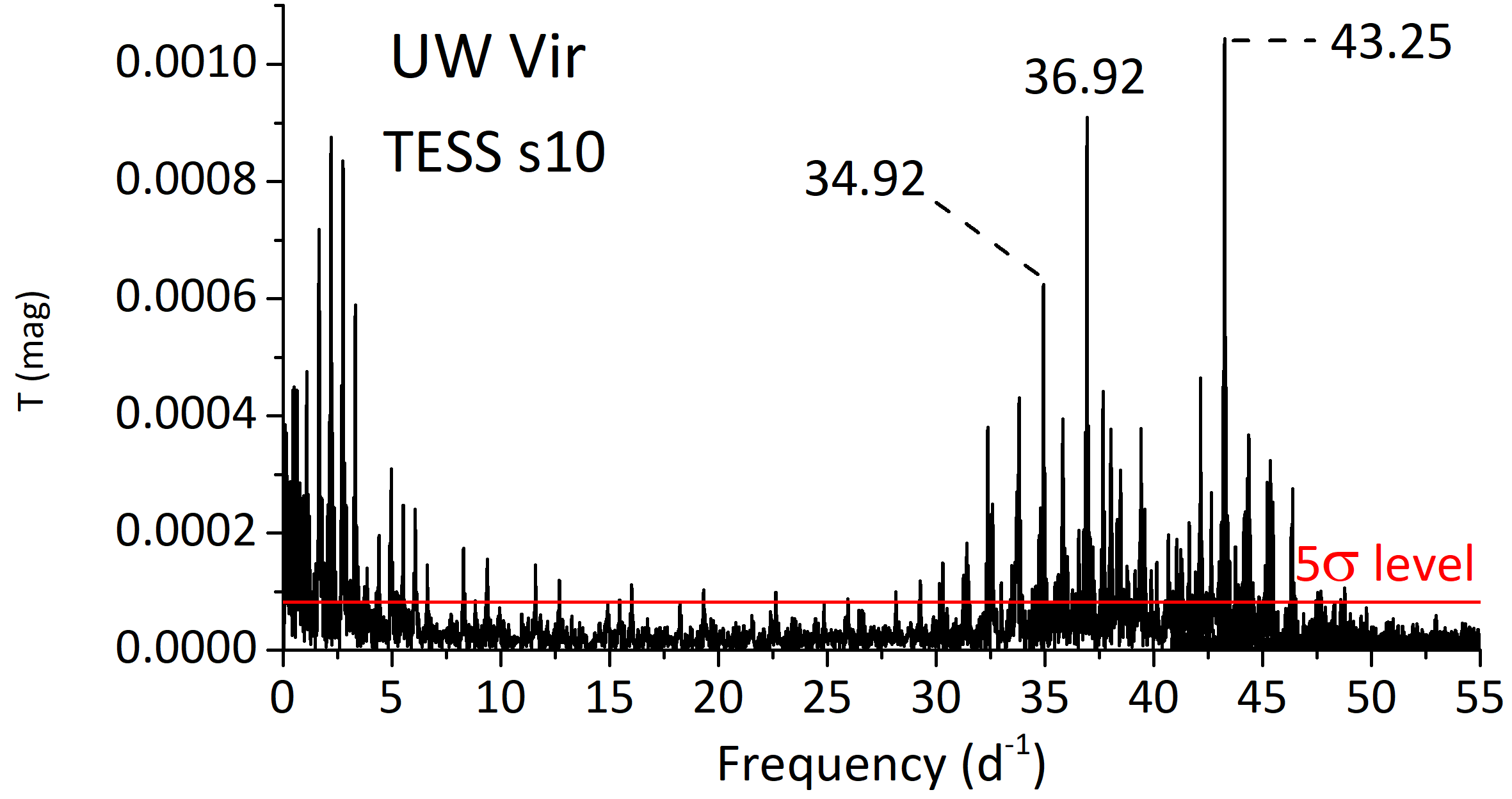}\\
\includegraphics[width=8cm]{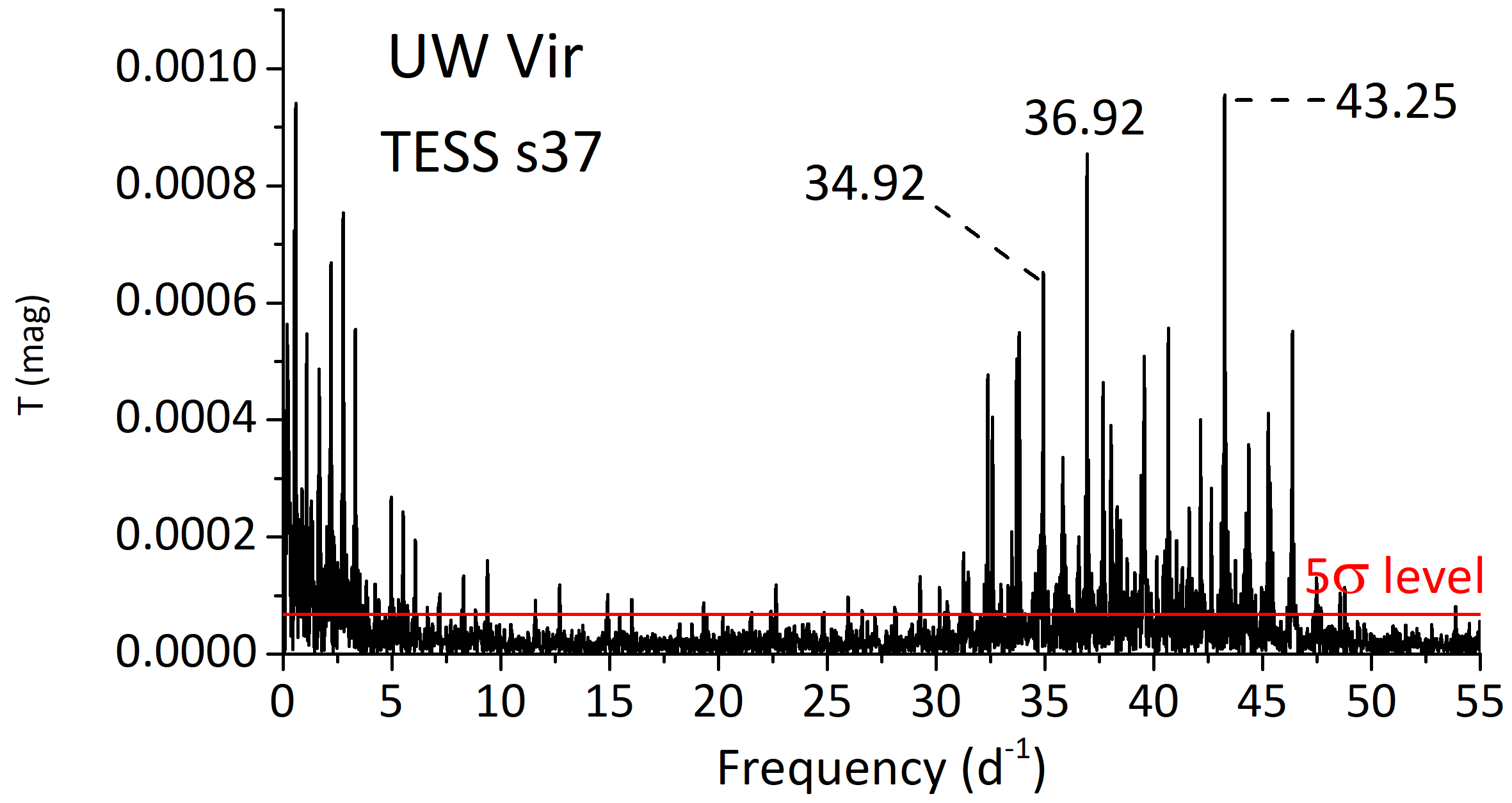}\\
\includegraphics[width=8cm]{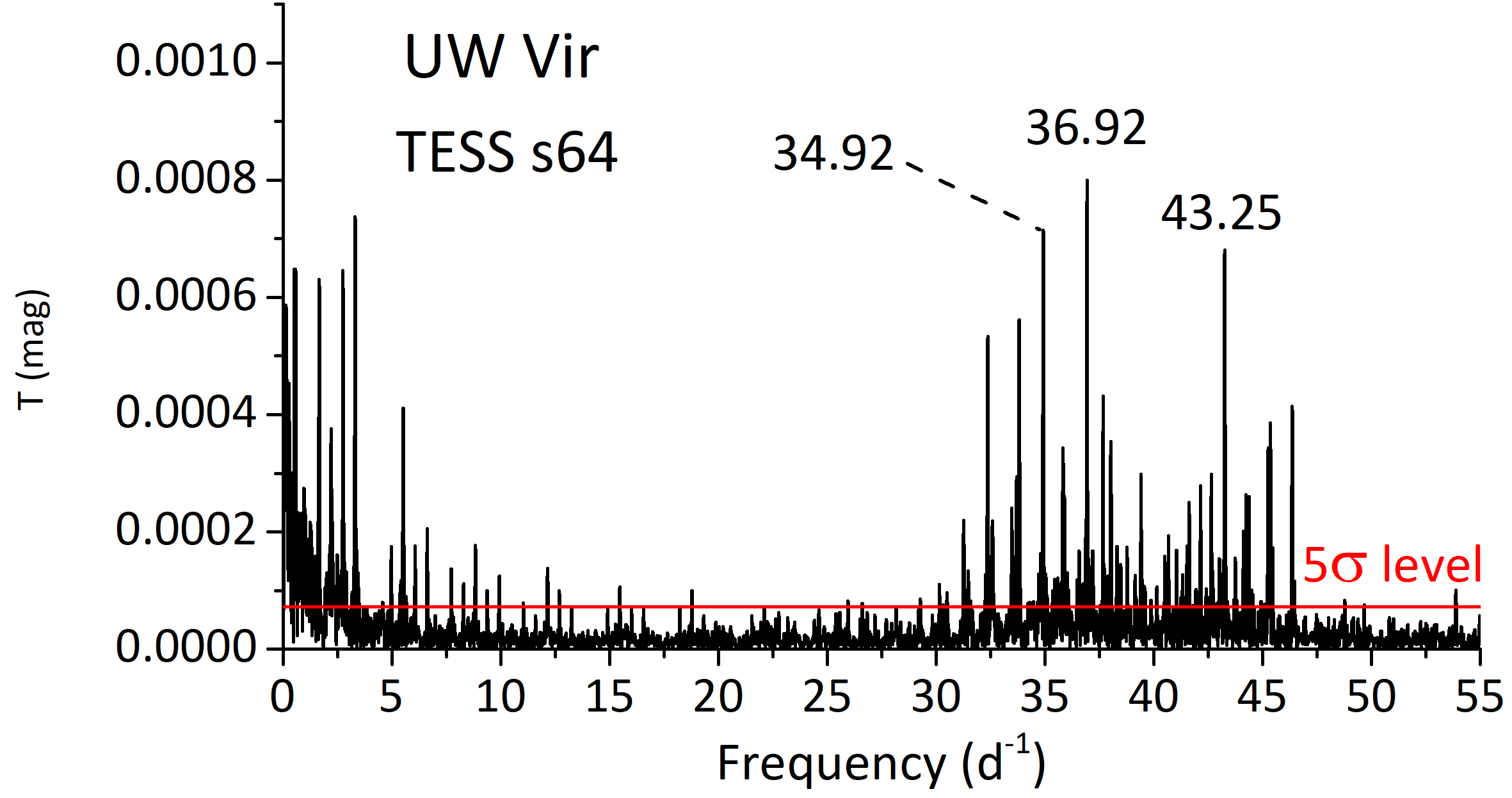}\\
\caption{Periodograms of the three analysed TESS sector data sets for UW~Vir.}
\label{fig:FSVIR}
\centering
\includegraphics[width=8cm]{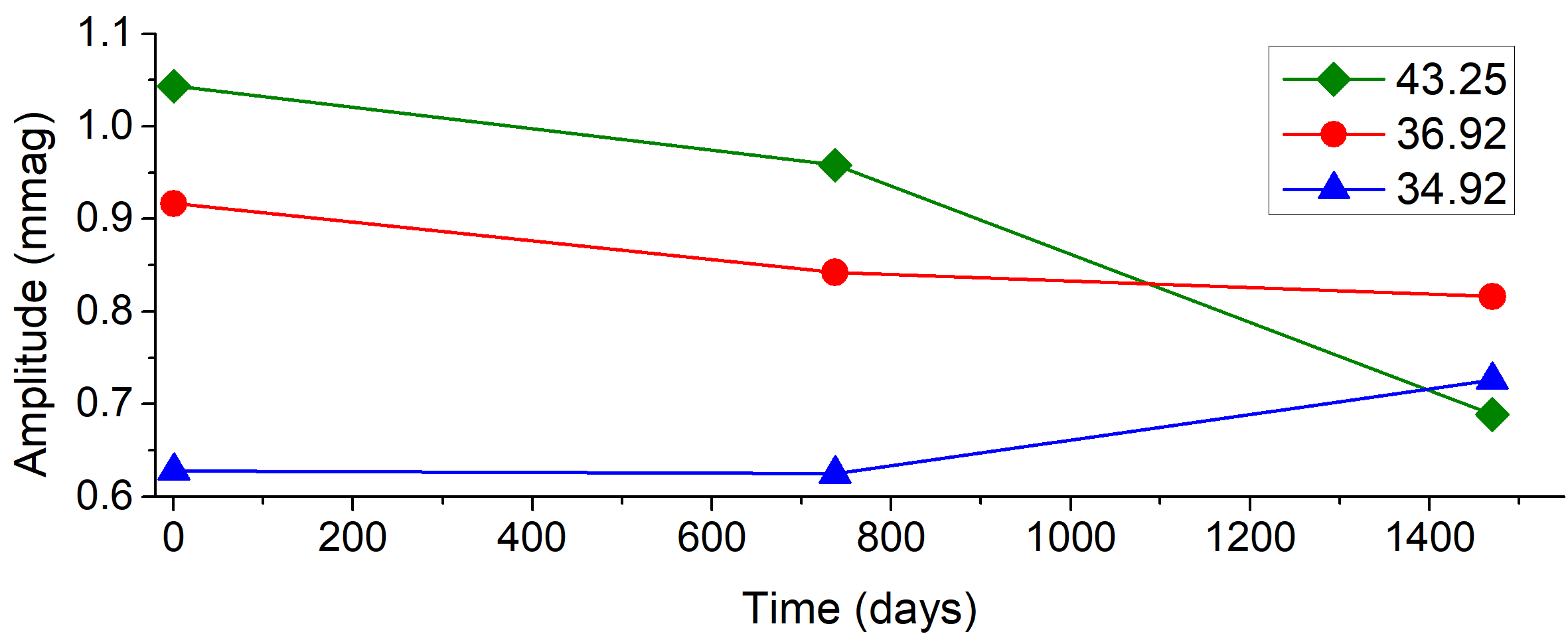}
\caption{Amplitude variation over time of the three independent frequencies of UW~Vir.}
\label{fig:AMPVIR}
\end{figure}

We did not have ground-based multi-filtered LCs of UW~Vir. However, TESS observed the system in three sectors s10, s37, and s64 in short-cadence mode. Each data set was analysed separately because of the long time gaps between them (i.e.~$\sim2$ years). In the analyses of these data sets, the strongest frequency ($f_1$) was not the same, but ranged between 0.0021 and 0.07~d$^{-1}$. We consider that these frequencies are not real and they probably originate from imperfect fit on the LCs. For the data sets of sectors 10 and 37, we detected three independent frequencies, namely 43.25~d$^{-1}$, 36.92~d$^{-1}$, and 34.92~d$^{-1}$ (in order of higher amplitude). Another 51 and 55 combination frequencies were also detected in the data sets of these two sectors, respectively. These two models have many common combination frequencies (Table~\ref{Tab:Vircombo1}) and their total number slightly differ because of different background noise of each set (i.e.~$\sim0.82~\mu$mag for s10 and $\sim0.68~\mu$mag for s37). Interestingly, the results for the data set of s64 regarding the amplitudes of the independent frequencies differ a lot (see Table~\ref{tab:Indfreqs} and also Table~\ref{Tab:Vircombo2} for the complete model). In particular, we found that the amplitude of the frequency 43.25~d$^{-1}$ decreased dramatically in comparison with its value in s10 data and become the weakest among these three. By taking the results of s10 as reference, this frequency decreased by $\sim8$\% after two years and reached $\sim34$\% decrease after four years. Similarly, a lower decrease was seen in the frequency 36.92~d$^{-1}$. It decreased by 8.7\% after two years and $\sim21$\% after four years. In contrast, the amplitude of the frequency 34.92~d$^{-1}$ remained almost the same for the first two years and increased rapidly by 17.7\% the second two years. The Fourier spectra for each data set are illustrated in Fig.~\ref{fig:FSVIR}, while the amplitude variations are shown in Fig.~\ref{fig:AMPVIR}. Furthermore, significant phase modulation of these frequencies was noticed. These variations can be explained either by mass transfer from the secondary component \citep{MIS21, KAH22} or by intrinsic mode coupling \citep{BAR15, LV21, NIU24}. Mass transfer from the companion, even at a relatively low rate, can deposit material in the star's outer layers, subtly altering the density and opacity in the He~II ionization zone where pulsations are driven. This can enhance the amplitude of some modes while damping others, without significantly affecting their frequencies. Intrinsic mode coupling provides an additional explanation: nonlinear interactions between pulsation modes can redistribute energy over time, causing certain modes to grow and others to decline, again without changing the oscillation frequencies. Together, these effects can naturally account for the observed mode-dependent amplitude variations over several years.

In the absence of multi-band photometry, the mode assignment of each detected frequency is determined only from the pulsation constant $Q$, the physical properties of the $\delta$~Sct star, and the models presented by \citet{FIT81}. All frequencies are identified as non-radial pressure modes. The frequencies 43.25~d$^{-1}$ and 36.92~d$^{-1}$ have $l$-degrees=2 and they are the fifth ($n=5$) and fourth overtones ($n=4$), respectively, whereas the frequency 34.92~d$^{-1}$ has $l=1$ and $n=4$. An example of Fourier fitting is given in Fig.~\ref{fig:FF}.

\section{Summary, discussion, and conclusions}
\label{Sec:Disc}
TESS photometry together with ground-based observations for the eclipsing systems IZ~Tel and UW~Vir were combined with medium-resolution spectroscopy to determine the physical characteristics of their components and to investigate their pulsational behaviour. Radial velocities for both members of each binary pair were obtained using broadening-function analyses in the $B$ and $V$ spectral bands. The latter were analysed together the ground-based and TESS LCs in order to calculate the absolute properties of the systems' components and to derive the LC residuals in the search for pulsations. Distances were calculated for both EBs and are in good agreement with the GAIA results. Both systems are classified as oEA stars, with their primary components exhibiting $\delta$~Sct–type pulsations. Using historical times of minima for UW~Vir, we determined the most likely modulating mechanisms of its orbital period.

The spectroscopy of IZ~Tel indicated mass transfer from the secondary to the primary star based on H$\alpha$ emission and broadened H$\alpha$ and Na~I~D absorption in the primary. Indeed, LC analyses revealed a semi-detached system with its secondary component filling its Roche lobe by more than 99\%. The primary component is now a main sequence star located close to the TAMS line and a little outside the red edge of the classical instability strip. It is the coolest star in the current sample of $\delta$~Sct stars in oEA systems and is among the five more evolved ones (Fig.~\ref{fig:evol}). The secondary has evolved to become a K2 subgiant. Currently, the primary component pulsates in a dominant non-radial pressure mode of $\sim13.56$~d$^{-1}$ and in another 16 combination frequencies. The result for the dominant mode is in perfect agreement with the findings of \citet{PIG07}.

Broadened H$\alpha$ and Na~I~D absorption in the UW~Vir primary component, together with episodic strengthening of the He~I~5876 and 6678~{\AA} lines, revealed active mass transfer from the secondary to the primary. These spectral features indicated that UW~Vir's secondary was undergoing an especially active transfer phase over the four months of our observations in 2021. The active state of mass transfer is supported by the presence of a hot spot indicated by the orbital modelling, as well as by the ETV analyses showing the mass transfer rate to be $1.9\times10^{-7}$~M$_\sun$~yr$^{-1}$ from the secondary to the primary component. The cyclic modulation of the system's $P_{\rm orb}$ is not clarified. However, both the photometric and spectroscopic analyses show no evidence for the presence of a third component; in contrast, the ETV analysis suggests that there is a third body with a minimum mass greater than that of the secondary component. This contradiction might be explained either by speculations on the nature of the third body or by the magnetic braking of the secondary component. The primary component, having gained a substantial amount of mass from the current secondary component is now a dwarf lying well inside the classical instability strip, although its spectral lines indicate that it has a luminosity class of V/IV. The secondary has evolved to the subgiant stage. The former is a $\delta$~Sct star that pulsates in more than 50 frequencies. However, only three of them were identified as parent frequencies and they range between 34.9-43.3~d$^{-1}$. These frequencies were found to be non-radial pressure modes but of different spherical degrees and radial orders. Moreover, the pulsation analyses of three different short cadence TESS data sets with a 2-yr interval revealed the amplitude decrease of the two out of three initially strongest modes and the amplitude increase of the initially weakest mode. Phase variation was also observed for these frequencies. These modulations are attributed probably to the mass accretion of the $\delta$~Sct star, as well as to intrinsic mode coupling. \citet{MKR17} reported two main pulsation frequencies (i.e.~28.8~d$^{-1}$ and 46.9~d$^{-1}$) based on the analysis of ground based data taken within 11 nights. Our results do not confirm the existence of a frequency of 28.8~d$^{-1}$, not even as a combination frequency. The other frequency (46.9~d$^{-1}$), although the authors mentioned that it had a marginally acceptable S/N, is relatively close to the 43.25~d$^{-1}$ found from our analysis.

\begin{figure}
\centering
\includegraphics[width=\columnwidth]{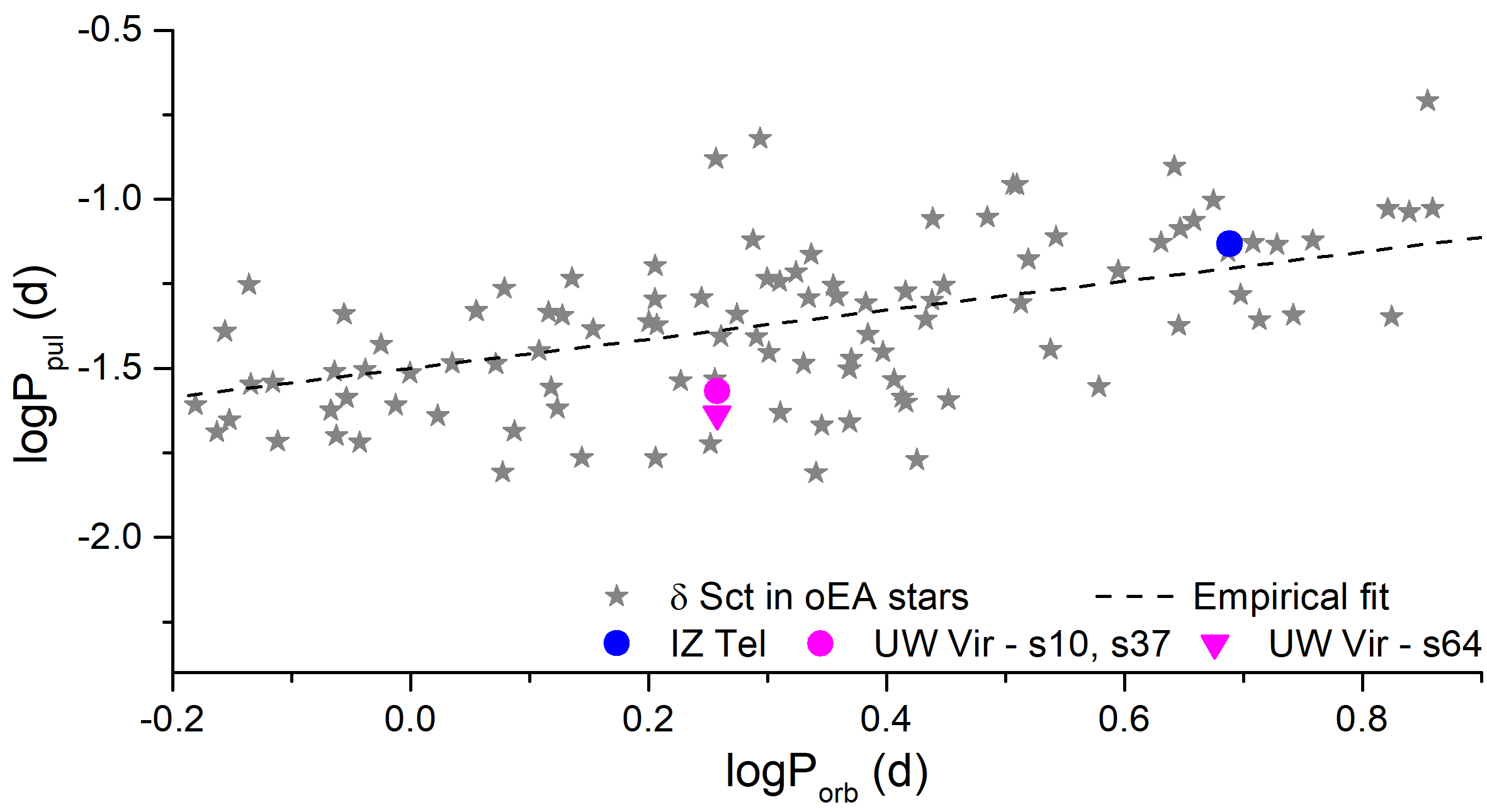}
\caption{$P_{\rm orb}-P_{\rm pul}$ correlation for $\delta$~Sct stars of 105 oEA systems (star symbols) and the locations of the pulsating (primary) components of IZ~Tel (blue) and UW~Vir (magenta). For the latter system, two locations are plotted due to the interchange of its dominant pulsation frequency between the TESS sectors 10, 37 (filled circle), and 64 (triangle). The black dashed line represents the linear fitting of \citet{LIA25}.}
\label{fig:PP}
\end{figure}

The locations of the $\delta$~Sct members of IZ~Tel and UW~Vir in the Mass–Radius and Hertzsprung–Russell diagrams are in the same range as other stars of the sample of oEA systems (Fig.~\ref{fig:evol}). The $P_{\rm orb}-P_{\rm pul}$ correlation for these two stars is in good agreement with pulsating stars in other oEA systems as shown in Fig.~\ref{fig:PP}. This plot incorporates the sample of 105 $\delta$~Sct components in oEA stars, along with the empirical fit provided by \citet{LIA25}. The primary of IZ~Tel is very close to the empirical curve, and that of UW~Vir deviates slightly from that of other similar stars.

Although this study expands the sample of oEA stars with well determined physical parameters by about 5\%, the overall number remains small, with only 43 systems currently available. Therefore, further studies of similar systems are highly encouraged. Enlarging the sample of oEA stars with accurately calculated absolute parameters will deepen our understanding of how interactions within close binary systems influence stellar pulsations.

\begin{acknowledgements}
It is a pleasure to express our appreciation of the high quality and ready availability, via the Mikulski Archive for Space Telescopes (MAST), of data collected by the TESS mission. Funding for the TESS mission is provided by the NASA Explorer Program. This research partly made use of the SIMBAD database, operated at CDS, Strasbourg, France, and of NASA's Astrophysics Data System Bibliographic Services. A.~L. acknowledges financial support from the NOA's internal fellowship `SPECIES' (No.~5094). D.~J.~W.~M. and J.~F.~W. acknowledge grants for time on the ANU 2.3~m telescope from the Edward Corbould Research Fund of the Astronomical Association of Queensland. D.~J.~W.~M. thanks Prof.~S.~Wyithe, Director, Research School of Astronomy and Astrophysics at ANU and Prof.~P.~Francis, Chair of the TAC of the ANU 2.3~m telescope for granting Discretionary Time to observe the primary eclipse of UW~Vir; I.~Price and B.~Martin for assistance with the proposal for the automated system and Dr~C.~Onken for modifying Python scripts to extract 1D spectra. He thanks R.~Jenkins, Variable Stars South and AAVSO, for observing a primary eclipse of UW~Virto confirm the revised ephemeris elements reported in this paper before time was requested on the ANU telescope. We also thank the anonymous reviewer for the helpful comments that improved the quality of this work.

\end{acknowledgements}

\section*{Data Availability}

The majority of data included in this article are available as listed in the paper or from the online supplementary material it refers to. The TESS data are available online from the MAST repository (\url{https://mast.stsci.edu/portal/Mashup/Clients/Mast/Portal.html}. 

\bibliographystyle{aa} 
\bibliography{references.bib} 

\clearpage
\onecolumn
\begin{appendix}

\section{Spectroscopy}
\label{Appendix:Spectroscopy}

\begin{table*} [h]
\begin{center}																	
\caption{Log of spectroscopic observations with the wide field spectrograph on the ANU 2.3~m telescope.}																	
\label{table:SPLOG}		
\begin{tabular}{l cccc cccc}																	
\hline \hline																	
System	&	Dates	&	Orbital phase	&	Number of spectra	&	Exp. time	&	Grating	&	S/N	&	Grating	&	S/N	\\
	&	(DD/MM/YY)	&		&		&	(s)	&		&		&		&		\\					
\hline																	
\multirow{5}{*}{IZ~Tel}	&	11/04/17	&	0	&	3	&	560	&	B3000	&	120	&	R3000	&	112	\\
	&	11/04/17	&	0	&	3	&	600	&	B7000	&	85	&	R7000	&	84	\\
	&	02/06/18	&	0.4	&	2	&	360	&	B3000	&	358	&	R7000	&	216	\\
	&	02/06/18-01/11/20	&	0.11--0.39	&	15	&	600-720	&	B7000	&	215	&	 R7000	&	212	\\
	&	12/06/17-03/11/20	&	0.65--0.75	&	12	&	180-600	&	B7000	&	118-215	&	 R7000	&	116-212	\\
 \hline																	
\multirow{4}{*}{UW~Vir}	&	27/02/21	&	0.33	&	2	&	60	&		&		&	 R7000 	&	377	\\
	&	28/05/21	&	0.5	&	6	&	25-30 	&	B3000	&	245	&	 R7000 	&	167	\\
	&	25/04/21-26/05/21	&	0.22--0.31	&	18	&	120-240	&	B7000	&	254-360	&	 R7000	&	254-360	\\
	&	28/03/21-26/04/21	&	0.70--0.80	&	21	&	180-240	&	B7000	&	240-440	&	 R7000	&	230-440	\\
    &	4/3/26	&	0	&	6	&	300	&	B3000	&	143	&	 R7000	&	132	\\
\hline																						
\end{tabular}	
\end{center}																	
\begin{center}
\caption{Radial velocity measurements of the components of IZ~Tel and UW~Vir.}
\label{Tab:RVs}
\scalebox{0.86}{
\begin{tabular}{cccccc|cccccc }
\hline																							
\hline																							
Time        & Phase     & RV$_{1}$  & (O-C)$_{1}$   &   RV$_{2}$    &   (O-C)$_{2}$ 											&	Time        & Phase     & RV$_{1}$  & (O-C)$_{1}$   &   RV$_{2}$    &   (O-C)$_{2}$ 											\\
(HJD-2450000)   &       & (km~s$^{-1}$) & (km~s$^{-1}$) & (km~s$^{-1}$) & (km~s$^{-1}$)											&	(HJD-2450000)   &       & (km~s$^{-1}$) & (km~s$^{-1}$) & (km~s$^{-1}$) & (km~s$^{-1}$)											\\
\hline																							
\multicolumn{6}{c}{IZ~Tel}											&	59358.9671	&	0.2577	&	$-32.51$	&	8.89	&	207.38	&	$-1.92$	\\
\cline{1-6}																							
59301.9679	&	0.7800	&	70.15	&	$-0.10$	&	$-174.41$	&	0.44	&	59358.9689	&	0.2586	&	$-33.05$	&	8.32	&	216.93	&	7.73	\\
59328.9788	&	0.6967	&	68.15	&	$-1.15$	&	$-174.62$	&	$-2.92$	&	59358.9706	&	0.2596	&	$-35.51$	&	5.82	&	210.57	&	1.52	\\
59328.9812	&	0.6980	&	68.15	&	$-1.30$	&	$-174.62$	&	$-2.47$	&	59358.9723	&	0.2606	&	$-41.16$	&	0.14	&	211.24	&	2.29	\\
59328.9837	&	0.6994	&	68.15	&	$-1.45$	&	$-174.62$	&	$-2.02$	&	59358.9741	&	0.2615	&	$-36.27$	&	4.99	&	205.38	&	$-3.42$	\\
59329.0334	&	0.7269	&	70.15	&	$-1.30$	&	$-176.95$	&	2	&	59358.9758	&	0.2625	&	$-36.04$	&	5.18	&	205.38	&	$-3.27$	\\
59329.0365	&	0.7286	&	70.15	&	$-1.35$	&	$-182.62$	&	$-3.47$	&	59360.8795	&	0.3138	&	$-33.08$	&	3.08	&	189.38	&	$-1.87$	\\
59329.0397	&	0.7303	&	70.15	&	$-1.40$	&	$-182.62$	&	$-3.27$	&	59360.8829	&	0.3157	&	$-33.67$	&	2.2	&	194.76	&	4.51	\\
\cline{7-12}																							
59329.0627	&	0.7430	&	70.15	&	$-1.60$	&	$-184.62$	&	$-4.62$	&	\multicolumn{6}{c}{UW~Vir}											\\
\cline{7-12}																							
59329.0659	&	0.7448	&	70.15	&	$-1.60$	&	$-167.81$	&	12.19	&	58329.2208	&	0.1095	&	$-39.45$	&	1.45	&	47.98	&	$-3.83$	\\
59329.0690	&	0.7465	&	70.15	&	$-1.55$	&	$-184.62$	&	$-4.67$	&	58329.2295	&	0.1113	&	$-40.65$	&	0.5	&	50.12	&	$-2.78$	\\
59329.1049	&	0.7663	&	70.15	&	$-1.00$	&	$-182.62$	&	$-4.67$	&	58329.2382	&	0.1131	&	$-44.22$	&	$-2.84$	&	50.95	&	$-3.05$	\\
59329.1080	&	0.7680	&	70.15	&	$-0.90$	&	$-182.62$	&	$-4.97$	&	58329.8944	&	0.2475	&	$-45.93$	&	5.72	&	102.53	&	1.53	\\
59329.1111	&	0.7698	&	70.15	&	$-0.80$	&	$-182.62$	&	$-5.32$	&	58329.9017	&	0.2490	&	$-50.93$	&	0.72	&	100.86	&	$-0.20$	\\
59329.1490	&	0.7907	&	68.15	&	$-1.10$	&	$-176.62$	&	$-5.22$	&	58329.9090	&	0.2505	&	$-53.79$	&	-2.08	&	97.41	&	$-3.69$	\\
59329.1515	&	0.7920	&	68.15	&	$-0.95$	&	$-176.62$	&	$-5.72$	&	58330.1322	&	0.2963	&	$-46.26$	&	4.54	&	100.29	&	3.18	\\
59329.1539	&	0.7934	&	68.15	&	$-0.80$	&	$-176.62$	&	$-6.22$	&	58330.1408	&	0.2980	&	$-50.19$	&	0.56	&	101.95	&	5.2	\\
59329.9175	&	0.2151	&	$-40.65$	&	0.09	&	203.38	&	$-3.67$	&	58330.1524	&	0.3004	&	$-53.00$	&	$-2.35$	&	97	&	0.75	\\
59329.9193	&	0.2161	&	$-39.84$	&	0.96	&	217.41	&	10.16	&	58272.0651	&	0.3980	&	$-38.33$	&	3.12	&	49.76	&	$-4.59$	\\
59329.9210	&	0.2170	&	$-41.47$	&	$-0.61$	&	211.15	&	3.7	&	58272.0944	&	0.4040	&	$-36.61$	&	4.02	&	51.91	&	1.31	\\
59329.9712	&	0.2447	&	$-45.66$	&	$-4.04$	&	207.38	&	$-2.72$	&	58272.1015	&	0.4055	&	$-43.00$	&	$-2.58$	&	48	&	$-1.66$	\\
59329.9743	&	0.2465	&	$-43.10$	&	$-1.49$	&	207.38	&	$-2.67$	&	58272.1060	&	0.4064	&	$-43.00$	&	$-2.71$	&	48	&	$-1.07$	\\
59329.9774	&	0.2482	&	$-43.44$	&	$-1.84$	&	207.38	&	$-2.62$	&	58239.1326	&	0.6500	&	$-4.05$	&	$-1.40$	&	$-129.76$	&	$-6.81$	\\
59330.0016	&	0.2615	&	$-43.97$	&	$-2.72$	&	206.84	&	$-1.96$	&	57917.1446	&	0.6732	&	$-2.00$	&	$-1.58$	&	$-135.00$	&	$-1.90$	\\
59330.0047	&	0.2632	&	$-45.87$	&	$-4.69$	&	209.16	&	0.61	&	57917.1470	&	0.6737	&	$-2.00$	&	$-1.62$	&	$-135.00$	&	$-1.70$	\\
59330.0078	&	0.2649	&	$-44.55$	&	$-3.44$	&	213.58	&	5.28	&	57917.1494	&	0.6741	&	$-2.00$	&	$-1.66$	&	$-135.00$	&	$-1.50$	\\
59330.0368	&	0.2810	&	$-47.11$	&	$-7.07$	&	203.38	&	$-1.27$	&	59156.9237	&	0.7097	&	4.22	&	2.16	&	$-142.15$	&	2.31	\\
59330.0400	&	0.2827	&	$-47.58$	&	$-7.68$	&	202.67	&	$-1.48$	&	59156.9310	&	0.7112	&	3.74	&	1.62	&	$-142.50$	&	2.25	\\
59330.0431	&	0.2844	&	$-48.84$	&	$-9.10$	&	201.38	&	$-2.22$	&	59156.9898	&	0.7232	&	2.6	&	$-0.03$	&	$-148.72$	&	$-1.67$	\\
59330.8843	&	0.7490	&	74.15	&	2.45	&	$-168.74$	&	11.11	&	59157.0057	&	0.7265	&	3.79	&	1.05	&	$-148.12$	&	$-0.57$	\\
59330.8874	&	0.7507	&	74.15	&	2.5	&	$-174.77$	&	5.03	&	59157.0137	&	0.7281	&	7.36	&	4.57	&	$-146.34$	&	1.42	\\
59330.9166	&	0.7668	&	74.15	&	3.05	&	$-172.57$	&	5.28	&	59157.0217	&	0.7298	&	5.95	&	3.12	&	$-142.74$	&	5.26	\\
59330.9197	&	0.7686	&	74.15	&	3.15	&	$-170.01$	&	7.54	&	57941.9169	&	0.7491	&	4	&	0.83	&	$-151.00$	&	$-1.45$	\\
59330.9229	&	0.7703	&	74.15	&	3.25	&	$-178.53$	&	$-1.33$	&	57941.9193	&	0.7496	&	4	&	0.83	&	$-151.00$	&	$-1.45$	\\
59330.9317	&	0.7752	&	74.15	&	3.55	&	$-167.20$	&	8.9	&		&		&		&		&		&		\\
\hline
\end{tabular}}
\end{center}
\begin{center}																	
\caption{Preference settings used in {\sc RAVESPAN}.}																	
\label{table:RAVESPAN}		
\begin{tabular}{lcc ccc}																	
\hline \hline																	
System	&	Wavelength range	&	$\log g$	&	Temperature & Metallicity & Resolution	\\
	& ({\AA})		&	(cm~s$^{-2}$)	&	(K)	& & (km~s$^{-1}$)\\						
\hline																	
IZ~Tel  & 4200--5270 & 3.0   & 5000&0& 2 \\
UW~Vir  & 4360--4840, 4880--5555  & 3.0   & 5250&0& 2  \\
\hline																	
\end{tabular}		
\end{center}																	
\end{table*}
\begin{table*}		
\begin{center}																	
\caption{Equivalent width of the He~I $\lambda$5876~{\AA}, He~I $\lambda$6678~{\AA}, Na~I~D$_1$,~D$_2$ lines (uncertainty $\pm0.02$), and the H$\alpha$ line (uncertainty $\pm0.04$) of UW~Vir in 2021.}	
\label{table:UWVir_ew}		
\scalebox{0.95}{
\begin{tabular}{lcccccccccc}															
\hline \hline																	
Date	&	Phase	&	$\lambda$5876 {\AA}	&	$\lambda$6678 {\AA}	&	S1D$_1$	&	S1D$_2$	&	S1D$_1$Cb$^a$	&	S1D$_2$Cb$^a$	&	S2D$_1$	&	S2D$_2$	&	H$\alpha$	\\
\hline
Feb-27th	&	0.33	&	0.18	&	0.14	&	0.24	&	0.33	&	0.24	&	0.43	&	0.16	&		&	5.57	\\
Mar-28th	&	0.78	&	0.09	&	0.10	&	0.30	&	0.33	&	0.04	&	0.12	&		&	0.20	&	4.60	\\
Apr-24th	&	0.70	&	0.02	&	0.06	&	0.37	&	0.50	&		&	0.03	&		&	0.23	&	4.23	\\
Apr-24th	&	0.75	&	0.08	&	0.07	&	0.36	&	0.52	&		&	0.06	&		&	0.23	&	4.39	\\
Apr-24th	&	0.77	&	0.09	&	0.07	&	0.38	&	0.57	&		&	0.03	&		&	0.24	&	4.17	\\
Apr-25th	&	0.22	&	0.05	&	0.06	&	0.41	&	0.38	&	 	&		&	0.20	&	 	    &	3.99	\\
Apr-26th	&	0.77	&	0.06	&	0.07	&	0.35	&	0.53	&		&	0.04	&		&	0.20	&	4.59	\\
May-24th	&	0.26	&	0.03	&	0.05	&	0.38	&	0.36	&		&		&	0.11	&		&	3.31	\\
May-26th	&	0.31	&	0.01	&	0.04	&	0.40	&	0.39	&		&		&	0.14	&		&	3.25	\\
\hline																	
\multicolumn{11}{l}{Note: S1=primary, S2=secondary, $^a$Gas stream around S1.}
\end{tabular}}															
\end{center}																	
\end{table*}	


\begin{figure}
\centering
\includegraphics[width=18.5cm]{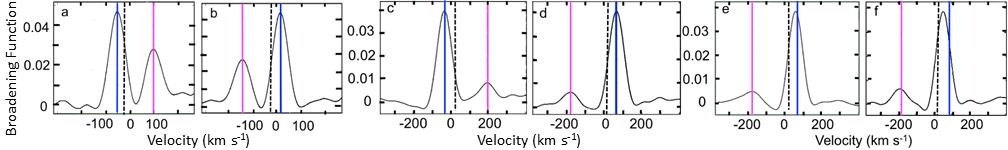}
\caption{Left: Examples of broadening function curves. The orbital velocity of the primary components is marked with a blue line and that of the secondary components with a magenta line. The systemic velocity is indicated by the dashed line. (a)~IZ~Tel phase 0.25, 2018~Jul.~30, exp.~time=600~s. (b)~IZ~Tel phase 0.73, 2020~Nov.~03, exp.~time=660~s. (c)~UW~Vir phase 0.25, 2021~Apr.~25, exp.~time=240~s. (d)~UW~Vir phase 0.75, 2021~Apr.~26, exp.~time=240~s. (e)~UW~Vir phase 0.78, 2021~Mar.~28, exp.~time=240~s. (f)~UW~Vir phase 0.75, 2021~Apr.~24, exp.~time=240~s.}
\label{Fig:BFs}
\centering
\includegraphics[width=5cm]{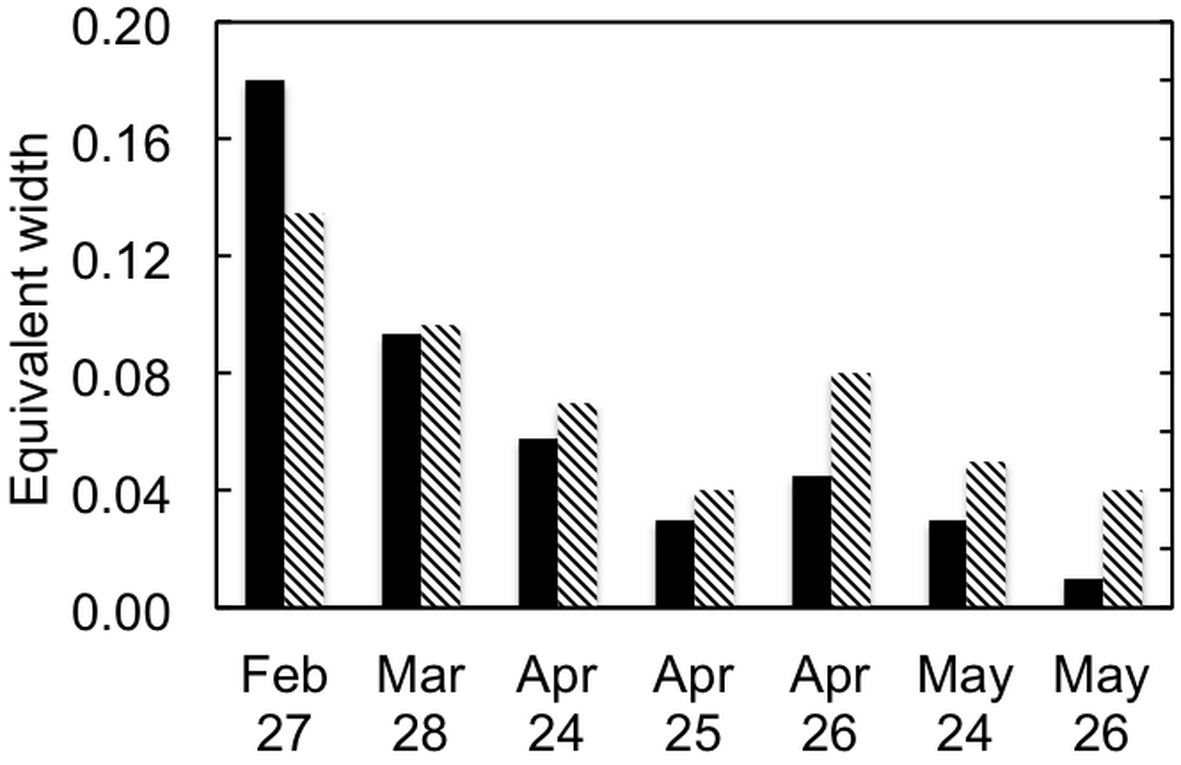}
\caption{
Equivalent widths of He~I $\lambda$5876~{\AA} (black) and $\lambda$6678~{\AA} (hatched) of UW~Vir between February and May~2021.}
\label{Fig:uwvir_ew}
\centering
\includegraphics[width=18cm]{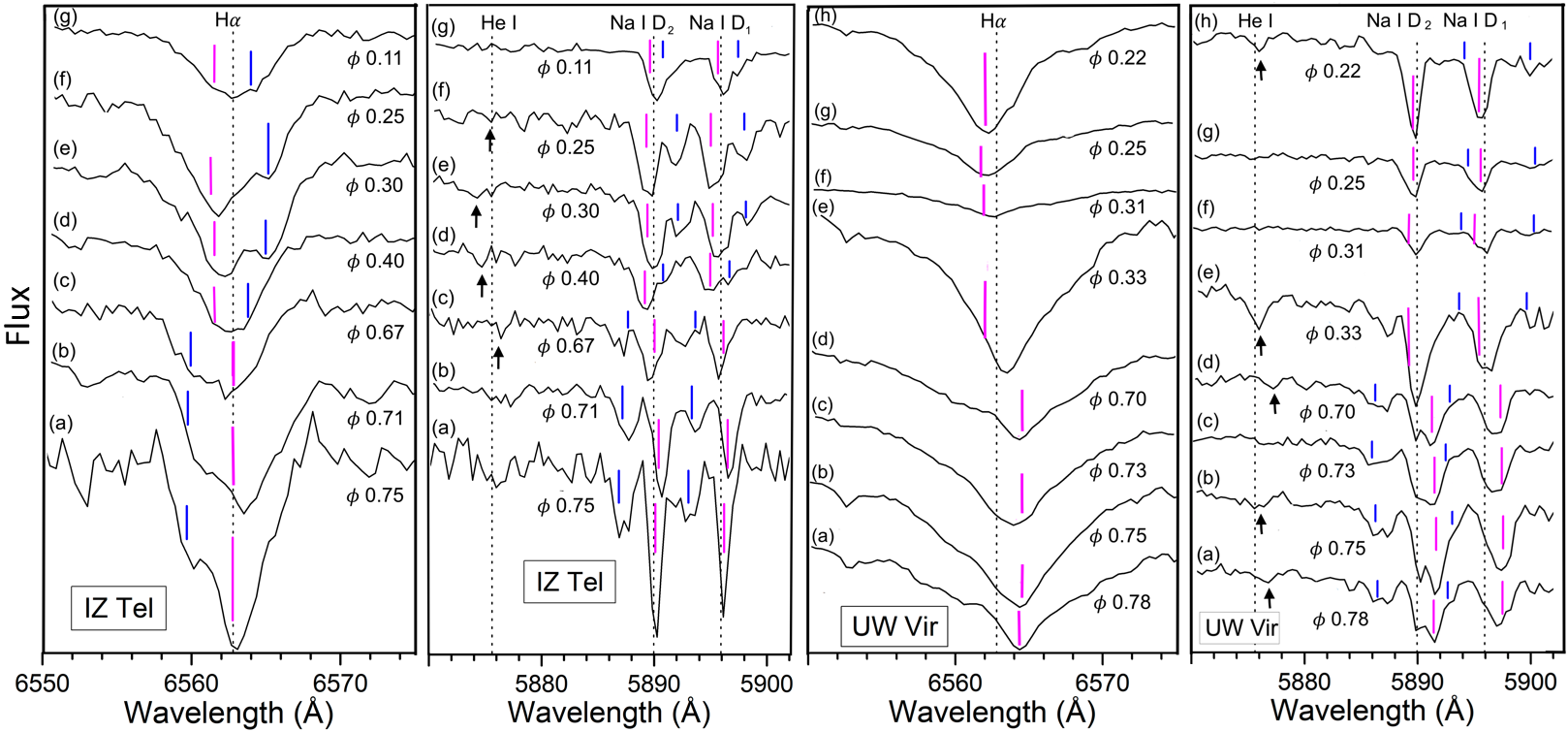}
\caption{Variation of H$\alpha$, Na~I~D and He~I spectra with orbital phase for IZ~Tel and UW~Vir. The expected line centres, calculated from their radial velocities, are marked with magenta bars (primary stars) and blue bars (secondary stars). Upward arrows indicate He~I lines that were above background variability. As the H$\alpha$ line of the UW~Vir secondary star was very weak, it is not evident at the scale on this figure.}
\label{Fig:PhSerSpec}
\centering
\includegraphics[width=13cm]{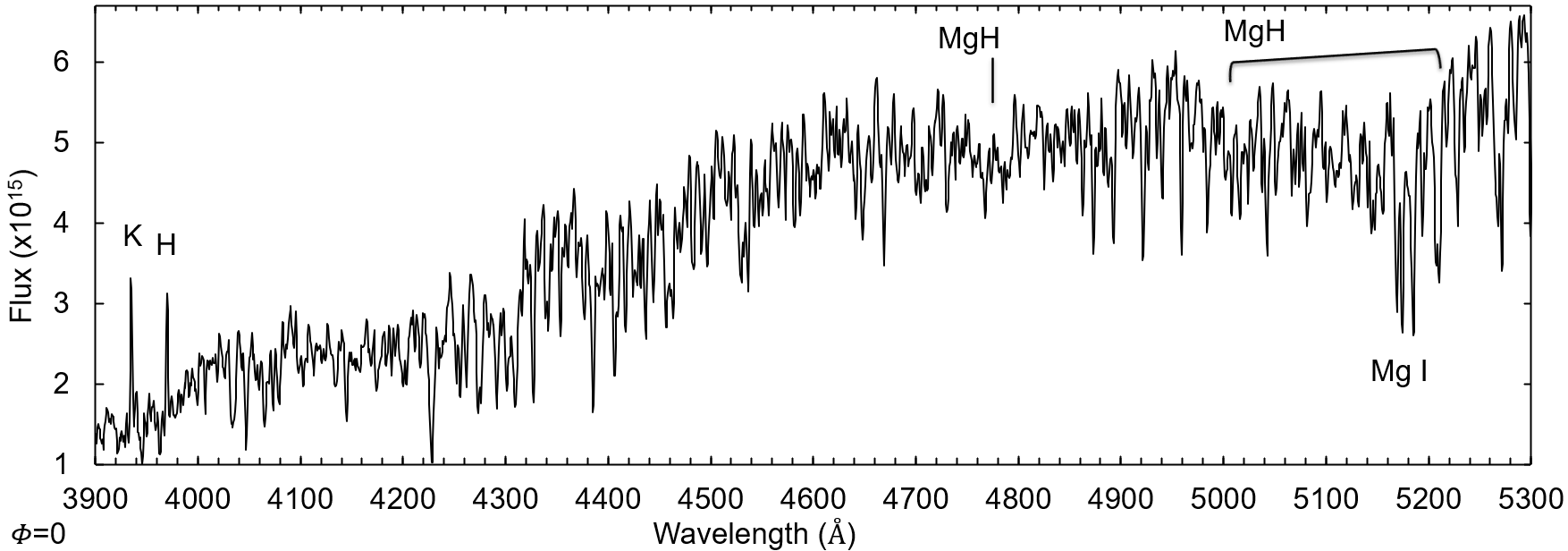}
\caption{Spectrum of the secondary component of UW~Vir in the blue region with R=3000 during a total primary eclipse.}
\label{Fig:uwvir_2rysp}
\end{figure}

\newpage

\section{Times of minima}
\label{Appendix:ETV}

The TESS ToM listed in Tables~\ref{Tab:ToM2} and \ref{Tab:ToM1} were calculated in this study.

\begin{table*}		[h]													
\begin{center}															
\caption{Historical times of minima of UW~Vir.}															
\label{Tab:ToM2}															
\scalebox{0.93}{
\begin{tabular}{cc cc cc cc}															
\hline\hline															
ToM (HJD)	&	type	&	ToM (HJD)	&	type	&	ToM (HJD)	&	type	&	ToM (HJD)	&	type	\\
\hline															
2415111.66300	&	pg	&	2443273.43200	&	vis	&	2457491.58500	&	ccd	&	2459325.00146	&	TESS	\\
2415886.60200	&	pg	&	2443510.64600	&	vis	&	2458571.71680	&	TESS	&	2459325.90760	&	TESS	\\
2416230.58900	&	pg	&	2443981.44700	&	vis	&	2458572.62230	&	TESS	&	2459326.81239	&	TESS	\\
2416250.56500	&	pg	&	2444010.42200	&	vis	&	2458573.52740	&	TESS	&	2459327.71840	&	TESS	\\
2417242.77400	&	pg	&	2444039.38900	&	vis	&	2458574.43310	&	TESS	&	2459328.62311	&	TESS	\\
2418468.62100	&	pg	&	2444267.54900	&	vis	&	2458575.33830	&	TESS	&	2459329.52946	&	TESS	\\
2419234.50400	&	pg	&	2444345.41300	&	vis	&	2458576.24390	&	TESS	&	2459330.43435	&	TESS	\\
2419460.83100	&	pg	&	2444354.46700	&	vis	&	2458577.14880	&	TESS	&	2459331.33990	&	TESS	\\
2419558.63100	&	pg	&	2444629.70000	&	vis	&	2458578.05470	&	TESS	&	2459332.24473	&	TESS	\\
2419942.51300	&	pg	&	2445022.64100	&	vis	&	2458578.95970	&	TESS	&	2459702.54900	&	ccd	\\
2419951.52700	&	pg	&	2445082.39900	&	vis	&	2458579.86540	&	TESS	&	2460042.07032	&	TESS	\\
2420954.69500	&	pg	&	2445111.38050	&	vis	&	2458580.77020	&	TESS	&	2460042.97583	&	TESS	\\
2421220.87300	&	pg	&	2445140.34900	&	vis	&	2458581.67620	&	TESS	&	2460043.88136	&	TESS	\\
2421336.72300	&	pg	&	2445357.63200	&	vis	&	2458585.28160	&	TESS	&	2460044.78670	&	TESS	\\
2421660.79900	&	pg	&	2445406.52500	&	vis	&	2458586.20310	&	TESS	&	2460045.69197	&	TESS	\\
2421749.53500	&	pg	&	2445464.47200	&	vis	&	2458588.01310	&	TESS	&	2460046.59694	&	TESS	\\
2422750.83500	&	pg	&	2446172.47700	&	vis	&	2458588.10780	&	TESS	&	2460047.50264	&	TESS	\\
2424941.78960	&	vis	&	2446938.43300	&	vis	&	2458588.91940	&	TESS	&	2460049.31334	&	TESS	\\
2424941.78990	&	vis	&	2447192.56600	&	vis	&	2458589.82400	&	TESS	&	2460050.21888	&	TESS	\\
2425327.45500	&	pg	&	2448001.36300	&	vis	&	2458590.73010	&	TESS	&	2460051.12395	&	TESS	\\
2427221.43800	&	vis	&	2448010.41200	&	vis	&	2458591.63540	&	TESS	&	2460052.02985	&	TESS	\\
2427576.33400	&	vis	&	2448314.62200	&	vis	&	2458592.54230	&	TESS	&	2460052.93510	&	TESS	\\
2427862.42200	&	vis	&	2449395.64400	&	vis	&	2458593.44610	&	TESS	&	2460053.84059	&	TESS	\\
2427929.41800	&	vis	&	2449779.52700	&	vis	&	2458594.35170	&	TESS	&	2460054.74592	&	TESS	\\
2431230.30100	&	vis	&	2449861.00700	&	vis	&	2458595.25730	&	TESS	&	2460055.65136	&	TESS	\\
2431554.41500	&	vis	&	2449866.44600	&	vis	&	2458657.72660	&	ccd	&	2460056.55643	&	TESS	\\
2431907.49900	&	vis	&	2449870.06600	&	ccd	&	2459308.70442	&	TESS	&	2460057.46222	&	TESS	\\
2433763.44700	&	vis	&	2450201.43000	&	vis	&	2459309.61051	&	TESS	&	2460058.36723	&	TESS	\\
2434480.49000	&	vis	&	2450525.56300	&	vis	&	2459310.51502	&	TESS	&	2460059.27295	&	TESS	\\
2441042.56900	&	vis	&	2451184.68000	&	vis	&	2459311.42140	&	TESS	&	2460060.17790	&	TESS	\\
2441062.48700	&	vis	&	2451275.21300	&	ccd	&	2459312.32588	&	TESS	&	2460061.08376	&	TESS	\\
2441062.49100	&	vis	&	2451606.58800	&	vis	&	2459313.23219	&	TESS	&	2460061.98892	&	TESS	\\
2441091.46000	&	vis	&	2452042.99200	&	vis	&	2459314.13692	&	TESS	&	2460062.89392	&	TESS	\\
2441397.47500	&	vis	&	2452323.65600	&	vis	&	2459315.04318	&	TESS	&	2460063.79934	&	TESS	\\
2442152.56500	&	vis	&	2452658.64500	&	vis	&	2459315.94747	&	TESS	&	2460064.70529	&	TESS	\\
2442545.48900	&	vis	&	2453060.63500	&	vis	&	2459316.85373	&	TESS	&	2460065.61036	&	TESS	\\
2442842.47200	&	vis	&	2453480.72790	&	ccd	&	2459317.75830	&	TESS	&	2460430.48200	&	ccd	\\
2442871.44600	&	vis	&	2453752.34600	&	vis	&	2459318.66451	&	TESS	&	2461075.11972	&	ccd	\\
2442900.40700	&	vis	&	2454165.20240	&	ccd	&	2459322.28605	&	TESS	&		&		\\
2442900.41200	&	vis	&	2455311.41530	&	ccd	&	2459323.19112	&	TESS	&		&		\\
2442900.41400	&	vis	&	2455979.59000	&	ccd	&	2459324.09683	&	TESS	&		&		\\
\hline																														
\end{tabular}	}
\newline
\textbf{Notes.} pg=photographic; vis=visual; The TESS ToM are in BJD units 													
\end{center}															
\end{table*}		

\begin{table*}
\begin{center}															
\caption{Historical times of minima of IZ~Tel.}															
\label{Tab:ToM1}															
\scalebox{0.94}{
\begin{tabular}{cc cc cc cc}															
\hline	\hline														
ToM (BJD)	&	type	&	ToM (BJD)	&	type	&	ToM (BJD)	&	type	&	ToM (BJD)	&	type	\\
\hline																													
2443173.05600	&	pg	&	2458665.43736	&	TESS	&	2459046.09554	&	TESS	&	2460132.00890	&	TESS	\\
2452086.87910	&	pg	&	2458670.31467	&	TESS	&	2459050.97558	&	TESS	&	2460134.39886	&	TESS	\\
2458314.05682$^a$	&	ccd	&	2458672.72556	&	TESS	&	2459053.38444	&	TESS	&	2460141.73756	&	TESS	\\
2458655.67363	&	TESS	&	2458677.63473	&	TESS	&	2459055.85590	&	TESS	&	2460144.15973	&	TESS	\\
2458658.11298	&	TESS	&	2458680.07337$^a$	&	ccd	&	2459058.31516	&	TESS	&	2460146.60667	&	TESS	\\
2458660.55416	&	TESS	&	2459038.77454	&	TESS	&	2460127.06843	&	TESS	&	2460149.04179	&	TESS	\\
2458662.98517	&	TESS	&	2459043.64438	&	TESS	&	2460129.50159	&	TESS	&		&		\\
\hline																												
\end{tabular}	}
\newline
\textbf{Notes.} pg=photographic; The pg and ccd ToM are in HJD units. $^a$Based on our $BVI$ LCs. 									
\end{center}															
\end{table*}

\newpage
\section{Combination frequencies}
\label{Appendix:Combofreqs}

\begin{table*}	[h]										
\begin{center}															
\caption{Combination pulsation frequencies of IZ~Tel.}															
\label{Tab:Telcombo}	
\begin{tabular}{ccc ccc ccc ccc}		
\hline\hline																							
$i$	&	  $f_{\rm i}$	&	$A$	&	  $\Phi$	&	S/N	&	Combination	&	$i$	&	  $f_{\rm i}$	&	$A$	&	  $\Phi$	&	S/N	&	Combination	\\
	&	     (d$^{-1}$)	&	(mmag)	&	(2$\pi$~rad)	&		&		&		&	     (d$^{-1}$)	&	(mmag)	&	(2$\pi$~rad)	&		&		\\
\hline																							
\multicolumn{6}{c}{TESS sector 67}											&	\multicolumn{6}{c}{$B$}											\\
\hline																							
2	&	0.071(1)	&	2.11(4)	&	0.539(3)	&	43.0	&	?	&	2	&	0.16(1)	&	14.9(7)	&	0.68(1)	&	19.4	&	?	\\
3	&	0.172(1)	&	1.60(4)	&	0.835(4)	&	32.5	&	$2f_2$	&	3	&	27.12(1)	&	6.7(7)	&	0.75(2)	&	8.7	&	$2f_1$	\\
4	&	27.116(1)	&	1.09(4)	&	0.09(1)	&	22.2	&	$2f_1$	&	4	&	2.08(2)	&	3.5(7)	&	0.25(3)	&	4.6	&	?	\\
5	&	0.340(1)	&	1.10(4)	&	0.57(1)	&	22.4	&	$4f_2$	&	5	&	40.62(6)	&	1.2(7)	&	0.62(9)	&	3.6	&	$3f_1$	\\
\cline{7-12}																							
6	&	0.613(1)	&	0.82(4)	&	0.19(1)	&	16.7	&	$3f_{\rm orb}$	&	\multicolumn{6}{c}{$V$}											\\
\cline{7-12}																							
7	&	1.228(2)	&	0.60(4)	&	0.93(1)	&	12.2	&	$6f_{\rm orb}$	&	2	&	0.22(1)	&	8.7(4)	&	0.58(1)	&	20.1	&	?	\\
8	&	0.040(2)	&	0.55(4)	&	0.17(1)	&	11.2	&	$\sim f_2$	&	3	&	27.12(1)	&	4.1(4)	&	0.76(2)	&	9.4	&	$2f_1$	\\
9	&	13.732(2)	&	0.49(4)	&	0.97(1)	&	9.9	&	$f_1+f_{\rm orb}$	&	4	&	2.30(2)	&	2.8(4)	&	0.13(3)	&	6.4	&	?	\\
10	&	1.644(2)	&	0.41(4)	&	0.39(2)	&	8.2	&	$8f_{\rm orb}$	&	5	&	2.93(2)	&	2.7(4)	&	0.71(3)	&	6.2	&	?	\\
11	&	13.174(2)	&	0.40(4)	&	0.96(2)	&	8.0	&	$f_1-2f_{\rm orb}$	&	6	&	40.65(2)	&	1.9(4)	&	0.38(4)	&	4.5	&	$3f_1$	\\
\cline{7-12}																							
12	&	12.362(2)	&	0.39(4)	&	0.60(2)	&	7.9	&	$f_1-6f_{\rm orb}$	&	\multicolumn{6}{c}{$I$}											\\
\cline{7-12}																							
13	&	40.674(3)	&	0.33(4)	&	0.82(2)	&	6.7	&	$3f_1$	&	2	&	0.27(1)	&	6.0(5)	&	0.35(1)	&	11.3	&	?	\\
14	&	0.772(3)	&	0.30(4)	&	0.77(2)	&	6.1	&	$\sim 4f_{\rm orb}$	&	3	&	2.19(1)	&	4.5(5)	&	0.68(2)	&	8.4	&	?	\\
15	&	13.057(3)	&	0.29(4)	&	0.15(2)	&	5.8	&	$f_1-2f_{\rm orb}-f_2$	&	4	&	1.54(1)	&	3.5(5)	&	0.10(2)	&	6.6	&	?	\\
16	&	1.087(3)	&	0.29(4)	&	0.98(2)	&	5.8	&	$6f_{\rm orb}-2f_2$	&	5	&	27.11(2)	&	2.7(5)	&	0.74(3)	&	5.1	&	$2f_1$	\\
17	&	16.897(3)	&	0.28(4)	&	0.99(2)	&	5.7	&	$f_1+16f_{\rm orb}$	&	6	&	40.72(3)	&	1.7(5)	&	0.01(5)	&	3.2	&	$3f_1$	\\
18	&	0.886(3)	&	0.27(4)	&	0.58(2)	&	5.5	&	$6f_{\rm orb}-4f_2$	&		&		&		&		&		&		\\
\hline																																													
\end{tabular}
\end{center}															
\end{table*}

\begin{table*}	
\begin{center}															
\caption{Combination pulsation frequencies of UW~Vir for TESS sectors 10 and 37.}													
\label{Tab:Vircombo1}	
\scalebox{0.9}{														
\begin{tabular}{ccc ccc ccc ccc}															
\hline\hline																						
\multicolumn{6}{c}{Sector 10}											&	\multicolumn{6}{c}{Sector 37}											\\
\hline																							
$i$	&	  $f_{\rm i}$	&	$A$	&	  $\Phi$	&	S/N	&	Combination	&	$i$	&	  $f_{\rm i}$	&	$A$	&	  $\Phi$	&	S/N	&	Combination	\\
	&	     (d$^{-1}$)	&	(mmag)	&	(2$\pi$~rad)	&		&		&		&	     (d$^{-1}$)	&	(mmag)	&	(2$\pi$~rad)	&		&		\\
\hline																				
1	&	0.0427(3)	&	1.04(1)	&	0.084(2)	&	62.9	&	?	&	1	&	0.0724(3)	&	0.85(1)	&	0.136(2)	&	62.5	&	?	\\
4	&	2.2063(3)	&	0.90(1)	&	0.197(2)	&	54.6	&	$4f_{\rm orb}$	&	3	&	0.5834(3)	&	0.77(1)	&	0.050(2)	&	56.7	&	$f_{\rm orb}$	\\
5	&	2.7645(4)	&	0.64(1)	&	0.244(3)	&	38.7	&	$5f_{\rm orb}$	&	5	&	2.7644(3)	&	0.74(1)	&	0.258(2)	&	54.0	&	$5f_{\rm orb}$	\\
7	&	1.6542(5)	&	0.58(1)	&	0.657(3)	&	35.3	&	$3f_{\rm orb}$	&	7	&	2.1989(4)	&	0.63(1)	&	0.339(3)	&	46.4	&	$4f_{\rm orb}$	\\
8	&	0.4855(6)	&	0.48(1)	&	0.641(4)	&	29.3	&	$f_2 + f_6 - 2f_3 - 7f_{\rm orb}$	&	8	&	0.1834(4)	&	0.60(1)	&	0.087(3)	&	44.0	&	$f_6 + 4f_{\rm orb} - f_4$	\\
9	&	37.6619(6)	&	0.44(1)	&	0.058(4)	&	26.8	&	$f_6 + 5f_{\rm orb}$	&	9	&	40.6566(5)	&	0.45(1)	&	0.508(4)	&	32.7	&	$2f_4 + 3f_{\rm orb} - f_6$	\\
10	&	0.1525(7)	&	0.40(1)	&	0.058(5)	&	24.6	&	$2f_3 - 2f_6 - 7f_{\rm orb}$	&	10	&	46.3604(5)	&	0.44(1)	&	0.745(4)	&	32.5	&	$f_2 + f_4 + 2f_{\rm orb} - f_6$	\\
11	&	39.4098(8)	&	0.37(1)	&	0.782(5)	&	22.5	&	$f_2 - 7f_{\rm orb}$	&	11	&	33.6888(4)	&	0.58(1)	&	0.602(3)	&	42.3	&	$f_4 - 6f_{\rm orb}$	\\
12	&	32.3656(8)	&	0.35(1)	&	0.965(6)	&	21.3	&	$2f_6 - f_3 - f_{\rm orb}$	&	12	&	32.3660(5)	&	0.44(1)	&	0.066(4)	&	32.5	&	$2f_6 - f_4 - f_{\rm orb}$	\\
13	&	0.5195(8)	&	0.34(1)	&	0.459(6)	&	20.6	&	$f_{\rm orb}$	&	13	&	37.6611(5)	&	0.47(1)	&	0.005(3)	&	34.4	&	$f_6 + 5f_{\rm orb}$	\\
14	&	45.3469(8)	&	0.33(1)	&	0.308(6)	&	20.3	&	$3f_3 + 8f_{\rm orb} - 2f_6$	&	14	&	0.2284(4)	&	0.55(1)	&	0.340(3)	&	40.6	&	$f_6 + 4f_{\rm orb} - f_4$	\\
15	&	3.3199(9)	&	0.31(1)	&	0.780(6)	&	18.9	&	$6f_{\rm orb}$	&	15	&	3.3192(6)	&	0.42(1)	&	0.834(4)	&	30.6	&	$6f_{\rm orb}$	\\
16	&	33.6887(9)	&	0.30(1)	&	0.029(6)	&	18.4	&	$3f_6 + 5f_{\rm orb} - 2f_3$	&	16	&	0.0326(6)	&	0.40(1)	&	0.106(4)	&	29.6	&	$2f_4 + 8f_{\rm orb} - f_2 - f_6$	\\
17	&	3.8689(9)	&	0.30(1)	&	0.434(7)	&	18.1	&	$7f_{\rm orb}$	&	17	&	39.5591(7)	&	0.35(1)	&	0.835(5)	&	25.7	&	$2f_4 + f_{\rm orb} - f_6$	\\
18	&	46.363(1)	&	0.29(1)	&	0.179(7)	&	17.7	&	$f_2 + f_3 + 2f_{\rm orb} - f_6$	&	18	&	3.8679(6)	&	0.39(1)	&	0.483(4)	&	28.9	&	$7f_{\rm orb}$	\\
19	&	0.764(1)	&	0.29(1)	&	0.498(7)	&	17.4	&	$f_6 + 5f_{\rm orb} - f_3$	&	19	&	33.8149(7)	&	0.31(1)	&	0.760(5)	&	23.0	&	$f_6 - 2f_{\rm orb}$	\\
20	&	38.470(1)	&	0.28(1)	&	0.418(7)	&	16.9	&	$f_2 + f_6 - f_3 - 5f_{\rm orb}$	&	20	&	34.7925(7)	&	0.34(1)	&	0.823(5)	&	25.2	&	$f_4 - 4f_{\rm orb}$	\\
21	&	42.639(1)	&	0.27(1)	&	0.773(7)	&	16.5	&	$f_2 - f_{\rm orb}$	&	21	&	1.6617(7)	&	0.35(1)	&	0.653(5)	&	25.6	&	$3f_{\rm orb}$	\\
22	&	44.125(1)	&	0.25(1)	&	0.152(8)	&	15.1	&	$f_2 + f_3 - f_6 - 2f_{\rm orb}$	&	22	&	39.4102(6)	&	0.39(1)	&	0.353(4)	&	28.6	&	$f_2 - 7f_{\rm orb}$	\\
23	&	4.422(1)	&	0.22(1)	&	0.012(9)	&	13.5	&	$8f_{\rm orb}$	&	23	&	45.3438(9)	&	0.21(1)	&	0.624(8)	&	15.3	&	$3f_4 + 8f_{\rm orb} - 2f_6$	\\
24	&	33.815(1)	&	0.20(1)	&	0.60(1)	&	12.4	&	$f_6 - 2f_{\rm orb}$	&	24	&	42.6379(9)	&	0.26(1)	&	0.989(6)	&	19.3	&	$f_2 - f_{\rm orb}$	\\
25	&	0.108(1)	&	0.20(1)	&	0.23(1)	&	12.3	&	$2f_3 - 2f_6 - 7f_{\rm orb}$	&	25	&	40.5145(8)	&	0.29(1)	&	0.561(5)	&	21.6	&	$f_2 - 5f_{\rm orb}$	\\
26	&	0.651(1)	&	0.20(1)	&	0.32(1)	&	12.3	&	$2f_3 - 2f_6 - 6f_{\rm orb}$	&	26	&	0.7718(8)	&	0.18(1)	&	0.460(9)	&	13.5	&	$f_6 + 5f_{\rm orb} - f_4$	\\
27	&	40.517(1)	&	0.20(1)	&	0.95(1)	&	12.2	&	$f_2 - 5f_{\rm orb}$	&	27	&	45.2324(8)	&	0.28(1)	&	0.492(6)	&	20.6	&	$f_2 + f_4 - f_6$	\\
28	&	1.118(1)	&	0.19(1)	&	0.76(1)	&	11.4	&	$2f_{\rm orb}$	&	28	&	0.3858(9)	&	0.21(1)	&	0.429(8)	&	15.2	&	$2f_6 + 8f_{\rm orb} - 2f_4$	\\
29	&	34.793(2)	&	0.18(1)	&	0.30(1)	&	11.2	&	$3f_6 + 7f_{\rm orb} - 2f_3$	&	29	&	0.282(1)	&	0.25(1)	&	0.499(7)	&	18.0	&	$f_4 - f_6 - 3f_{\rm orb}$	\\
30	&	45.218(2)	&	0.18(1)	&	0.54(1)	&	11.1	&	$f_2 + f_3 - f_6$	&	30	&	4.425(1)	&	0.21(1)	&	0.064(8)	&	15.3	&	$8f_{\rm orb}$	\\
31	&	33.715(2)	&	0.18(1)	&	0.92(1)	&	11.0	&	$3f_6 + 5f_{\rm orb} - 2f_3$	&	31	&	44.118(1)	&	0.26(1)	&	0.398(6)	&	18.7	&	$f_2 + f_4 - f_6 - 2f_{\rm orb}$	\\
32	&	41.288(2)	&	0.17(1)	&	0.85(1)	&	10.2	&	$f_3 + 8f_{\rm orb}$	&	32	&	33.724(1)	&	0.22(1)	&	0.337(7)	&	15.8	&	$3f_6 + 5f_{\rm orb} - 2f_4$	\\
33	&	0.427(2)	&	0.16(1)	&	0.43(1)	&	9.8	&	$2f_6 + 8f_{\rm orb} - 2f_3$	&	33	&	38.472(1)	&	0.16(1)	&	0.069(10)	&	12.0	&	$f_2 + f_6 - f_4 - 5f_{\rm orb}$	\\
34	&	0.369(2)	&	0.16(1)	&	0.37(1)	&	9.7	&	$f_3 - f_6 - 3f_{\rm orb}$	&	34	&	0.511(1)	&	0.17(1)	&	0.221(9)	&	12.5	&	$f_{\rm orb}$	\\
35	&	40.656(2)	&	0.16(1)	&	0.69(1)	&	9.6	&	$f_2 + f_6 - f_3 - f_{\rm orb}$	&	35	&	2.171(1)	&	0.19(1)	&	0.049(8)	&	14.2	&	$4f_{\rm orb}$	\\
36	&	30.301(2)	&	0.13(1)	&	0.69(1)	&	8.2	&	$3f_6 - 2f_3 - f_{\rm orb}$	&	36	&	1.100(1)	&	0.26(1)	&	0.857(6)	&	18.7	&	$2f_{\rm orb}$	\\
37	&	39.580(2)	&	0.13(1)	&	0.98(1)	&	8.0	&	$f_2 + f_6 - f_3 - 3f_{\rm orb}$	&	37	&	40.130(2)	&	0.15(1)	&	0.17(1)	&	10.7	&	$f_2 + f_6 - f_4 - 2f_{\rm orb}$	\\
38	&	1.068(2)	&	0.12(1)	&	0.06(2)	&	7.4	&	$2f_{\rm orb}$	&	38	&	44.244(1)	&	0.17(1)	&	0.86(1)	&	12.6	&	$3f_4 + 6f_{\rm orb} - 2f_6$	\\
39	&	44.242(2)	&	0.12(1)	&	0.60(2)	&	7.4	&	$3f_3 + 6f_{\rm orb} - 2f_6$	&	39	&	43.501(2)	&	0.13(1)	&	0.77(1)	&	9.7	&	$f_2 + f_6 + 4f_{\rm orb} - f_4$	\\
40	&	0.600(2)	&	0.12(1)	&	0.05(2)	&	7.3	&	$f_{\rm orb}$	&	40	&	0.976(2)	&	0.12(1)	&	0.10(1)	&	9.0	&	$f_4 - f_6 - 2f_{\rm orb}$	\\
41	&	40.131(2)	&	0.12(1)	&	0.25(2)	&	7.1	&	$f_2 + f_6 - f_3 - 2f_{\rm orb}$	&	41	&	4.983(1)	&	0.17(1)	&	0.59(1)	&	12.1	&	$f_2 - f_6 - 6f_{\rm orb}$	\\
42	&	1.845(2)	&	0.11(1)	&	0.12(2)	&	6.8	&	$f_6 + 7f_{\rm orb} - f_3$	&	42	&	36.482(2)	&	0.12(1)	&	0.99(1)	&	8.6	&	$3f_4 - 2f_6 - 8f_{\rm orb}$	\\
43	&	2.143(2)	&	0.11(1)	&	0.73(2)	&	6.8	&	$3f_3 - 3f_6 - 7f_{\rm orb}$	&	43	&	39.538(1)	&	0.16(1)	&	0.83(1)	&	12.0	&	$2f_4 + f_{\rm orb} - f_6$	\\
44	&	45.270(3)	&	0.11(1)	&	0.23(2)	&	6.7	&	$f_2 + f_3 - f_6$	&	44	&	0.640(2)	&	0.14(1)	&	0.56(1)	&	10.2	&	$2f_4 - 2f_6 - 6f_{\rm orb}$	\\
45	&	41.514(3)	&	0.11(1)	&	0.91(2)	&	6.4	&	$3f_3 + f_{\rm orb} - 2f_6$	&	45	&	45.256(1)	&	0.20(1)	&	0.97(1)	&	14.8	&	$f_2 + f_4 - f_6$	\\
46	&	34.926(3)	&	0.11(1)	&	0.75(2)	&	6.4	&	$f_6$	&	46	&	35.576(2)	&	0.11(1)	&	0.55(2)	&	7.8	&	$2f_4 - f_6 - 6f_{\rm orb}$	\\
47	&	32.990(3)	&	0.10(1)	&	0.65(2)	&	5.9	&	$f_3 + f_6 + 8f_{\rm orb} - f_2$	&	47	&	34.920(3)	&	0.09(1)	&	0.10(2)	&	6.6	&	$f_6$	\\
48	&	36.486(3)	&	0.10(1)	&	0.43(2)	&	5.9	&	$3f_3 - 2f_6 - 8f_{\rm orb}$	&	48	&	2.611(2)	&	0.10(1)	&	0.97(2)	&	7.4	&	$f_4 + f_{\rm orb} - f_6$	\\
49	&	45.107(3)	&	0.10(1)	&	0.10(2)	&	5.9	&	$2f_2 - f_3 - 8f_{\rm orb}$	&	49	&	57.556(2)	&	0.10(1)	&	0.52(2)	&	7.0	&	$3f_2 + 3f_{\rm orb} - 2f_4$	\\
50	&	2.419(3)	&	0.10(1)	&	0.58(2)	&	5.8	&	$f_6 + 8f_{\rm orb} - f_3$	&	50	&	0.857(2)	&	0.09(1)	&	0.93(2)	&	6.8	&	$f_4 - f_6 - 2f_{\rm orb}$	\\
51	&	0.233(3)	&	0.09(1)	&	0.05(2)	&	5.6	&	$f_6 + 4f_{\rm orb} - f_3$	&	51	&	32.986(3)	&	0.09(1)	&	0.32(2)	&	6.8	&	$f_4 - 7f_{\rm orb}$	\\
52	&	0.709(3)	&	0.09(1)	&	0.15(2)	&	5.5	&	$2f_3 - 2f_6 - 6f_{\rm orb}$	&	52	&	45.098(3)	&	0.09(1)	&	0.01(2)	&	6.4	&	$2f_2 - f_4 - 8f_{\rm orb}$	\\
53	&	43.023(3)	&	0.09(1)	&	0.31(2)	&	5.3	&	$f_2 + f_3 - f_6 - 4f_{\rm orb}$	&	53	&	53.858(3)	&	0.09(1)	&	0.07(2)	&	6.4	&	$3f_2 + f_6 - 3f_4$	\\
54	&	3.025(3)	&	0.09(1)	&	0.84(2)	&	5.2	&	$f_2 - f_3 - 6f_{\rm orb}$	&	54	&	32.513(3)	&	0.08(1)	&	0.56(2)	&	6.2	&	$f_4 - 8f_{\rm orb}$	\\
55	&	2.691(3)	&	0.08(1)	&	0.67(2)	&	5.1	&	$3f_3 - 3f_6 - 6f_{\rm orb}$	&	55	&	39.028(3)	&	0.08(1)	&	0.96(2)	&	6.1	&	$f_2 + f_6 - f_4 - 4f_{\rm orb}$	\\
	&		&		&		&		&		&	56	&	37.260(3)	&	0.08(1)	&	0.72(2)	&	5.8	&	$2f_4 - f_6 - 3f_{\rm orb}$	\\
	&		&		&		&		&		&	57	&	39.138(3)	&	0.08(1)	&	0.65(2)	&	5.7	&	$f_4 + 4f_{\rm orb}$	\\
	&		&		&		&		&		&	58	&	41.506(3)	&	0.07(1)	&	0.39(2)	&	5.2	&	$3f_4 + f_{\rm orb} - 2f_6$	\\
	&		&		&		&		&		&	59	&	41.301(3)	&	0.07(1)	&	0.69(2)	&	5.1	&	$f_4 + 8f_{\rm orb}$	\\
\hline	
\end{tabular}}													
\end{center}															
\end{table*}

\begin{table*}															
\begin{center}															
\caption{Combination pulsation frequencies of UW~Vir for TESS sector~64.}															
\label{Tab:Vircombo2}	
\scalebox{0.9}{														
\begin{tabular}{ccc ccc ccc ccc}
\hline\hline																							
\multicolumn{12}{c}{Sector 64}																							\\
\hline																							
$i$	&	  $f_{\rm i}$	&	$A$	&	  $\Phi$	&	S/N	&	Combination	&	$i$	&	  $f_{\rm i}$	&	$A$	&	  $\Phi$	&	S/N	&	Combination	\\
	&	     (d$^{-1}$)	&	(mmag)	&	(2$\pi$~rad)	&		&		&		&	     (d$^{-1}$)	&	(mmag)	&	(2$\pi$~rad)	&		&		\\
\hline																						
1	&	0.0021(1)	&	27.69(1)	&	0.808(1)	&	1926.5	&	?	&	32	&	2.225(2)	&	0.14(1)	&	0.99(1)	&	10.0	&	$4f_{\rm orb}$	\\
3	&	3.3118(3)	&	0.72(1)	&	0.709(2)	&	49.8	&	$6f_{\rm orb}$	&	33	&	0.709(1)	&	0.15(1)	&	0.63(1)	&	10.1	&	$2f_2 - 2f_4 - 6f_{\rm orb}$	\\
6	&	0.5670(3)	&	0.75(1)	&	0.435(2)	&	52.3	&	$f_{\rm orb}$	&	34	&	36.483(2)	&	0.14(1)	&	0.74(1)	&	9.8	&	$3f_2 - 2f_4 - 8f_{\rm orb}$	\\
7	&	2.7606(3)	&	0.69(1)	&	0.266(2)	&	48.3	&	$5f_{\rm orb}$	&	35	&	44.244(2)	&	0.14(1)	&	0.99(1)	&	9.5	&	$3f_2 + 6f_{\rm orb} - 2f_4$	\\
8	&	0.1464(3)	&	0.69(1)	&	0.389(2)	&	47.9	&	$2f_2 - 2f_4 - 7f_{\rm orb}$	&	36	&	35.590(2)	&	0.13(1)	&	0.15(1)	&	8.9	&	$2f_2 - f_4 - 6f_{\rm orb}$	\\
9	&	32.3644(4)	&	0.53(1)	&	0.715(3)	&	36.6	&	$2f_4 - f_2 - f_{\rm orb}$	&	37	&	0.620(2)	&	0.12(1)	&	0.67(1)	&	8.7	&	$f_{\rm orb}$	\\
10	&	0.2894(5)	&	0.47(1)	&	0.431(3)	&	32.8	&	$f_2 - f_4 - 3f_{\rm orb}$	&	38	&	41.308(2)	&	0.12(1)	&	0.01(1)	&	8.4	&	$f_2 + 8f_{\rm orb}$	\\
11	&	0.0908(5)	&	0.44(1)	&	0.464(4)	&	30.6	&	$2f_2 - 2f_4 - 7f_{\rm orb}$	&	39	&	2.788(2)	&	0.11(1)	&	0.83(1)	&	7.8	&	$5f_{\rm orb}$	\\
12	&	46.3483(5)	&	0.42(1)	&	0.809(4)	&	29.1	&	$f_2 + f_5 + 2f_{\rm orb} - f_4$	&	40	&	41.916(2)	&	0.11(1)	&	0.70(1)	&	7.7	&	$f_2 + f_5 - f_4 - 6f_{\rm orb}$	\\
13	&	37.6632(5)	&	0.41(1)	&	0.208(4)	&	28.7	&	$f_4 + 5f_{\rm orb}$	&	41	&	6.074(2)	&	0.11(1)	&	0.18(1)	&	7.5	&	$3f_2 - 3f_4$	\\
14	&	4.4175(6)	&	0.37(1)	&	0.525(4)	&	25.4	&	$8f_{\rm orb}$	&	42	&	33.684(2)	&	0.11(1)	&	0.28(1)	&	7.6	&	$3f_4 + 5f_{\rm orb} - 2f_2$	\\
15	&	45.3424(6)	&	0.36(1)	&	0.408(4)	&	25.3	&	$3f_2 + 8f_{\rm orb} - 2f_4$	&	43	&	0.872(2)	&	0.11(1)	&	0.48(1)	&	7.3	&	$f_2 - f_4 - 2f_{\rm orb}$	\\
16	&	0.1792(6)	&	0.36(1)	&	0.651(4)	&	24.7	&	$f_4 + 4f_{\rm orb} - f_2$	&	44	&	53.867(2)	&	0.10(1)	&	0.40(2)	&	7.1	&	$f_4 + 3f_5 - 3f_2$	\\
17	&	33.6870(7)	&	0.33(1)	&	0.275(5)	&	22.9	&	$3f_4 + 5f_{\rm orb} - 2f_2$	&	45	&	38.861(2)	&	0.10(1)	&	0.73(2)	&	7.0	&	$f_5 - 8f_{\rm orb}$	\\
18	&	42.6342(7)	&	0.32(1)	&	0.648(5)	&	22.1	&	$f_5 - f_{\rm orb}$	&	46	&	6.623(2)	&	0.10(1)	&	0.67(2)	&	6.9	&	$3f_2 + f_{\rm orb} - 3f_4$	\\
19	&	39.4104(7)	&	0.30(1)	&	0.11(1)	&	21.0	&	$f_5 - 7f_{\rm orb}$	&	47	&	40.640(2)	&	0.10(1)	&	0.63(2)	&	6.8	&	$2f_2 + 3f_{\rm orb} - f_4$	\\
20	&	0.5204(7)	&	0.30(1)	&	0.51(1)	&	20.8	&	$f_{\rm orb}$	&	48	&	40.129(2)	&	0.10(1)	&	0.31(2)	&	6.7	&	$f_4 + f_5 - f_2 - 2f_{\rm orb}$	\\
21	&	33.8136(8)	&	0.29(1)	&	0.64(1)	&	19.9	&	$f_4 - 2f_{\rm orb}$	&	49	&	38.473(2)	&	0.09(1)	&	0.69(2)	&	6.5	&	$f_4 + f_5 - f_2 - 5f_{\rm orb}$	\\
22	&	3.8683(8)	&	0.27(1)	&	0.99(1)	&	18.5	&	$7f_{\rm orb}$	&	50	&	1.547(2)	&	0.09(1)	&	0.35(2)	&	6.6	&	$3f_2 - 3f_4 - 8f_{\rm orb}$	\\
23	&	40.5123(8)	&	0.26(1)	&	0.01(1)	&	17.8	&	$f_5 - 5f_{\rm orb}$	&	51	&	7.176(2)	&	0.09(1)	&	0.96(2)	&	6.4	&	$f_5 - f_4 - 2f_{\rm orb}$	\\
24	&	33.7245(9)	&	0.24(1)	&	0.08(1)	&	16.8	&	$3f_4 + 5f_{\rm orb} - 2f_2$	&	52	&	0.801(2)	&	0.09(1)	&	0.54(2)	&	6.2	&	$f_4 + 5f_{\rm orb} - f_2$	\\
25	&	34.7930(9)	&	0.23(1)	&	0.07(1)	&	16.3	&	$3f_4 + 7f_{\rm orb} - 2f_2$	&	53	&	47.438(2)	&	0.09(1)	&	0.80(2)	&	6.1	&	$f_2 + f_5 + 4f_{\rm orb} - f_4$	\\
26	&	0.426(1)	&	0.23(1)	&	0.60(1)	&	15.9	&	$2f_4 + 8f_{\rm orb} - 2f_2$	&	54	&	34.912(2)	&	0.09(1)	&	0.46(2)	&	6.1	&	$f_4$	\\
27	&	43.019(1)	&	0.18(1)	&	0.62(1)	&	12.6	&	$f_2 + f_5 - f_4 - 4f_{\rm orb}$	&	55	&	1.291(3)	&	0.09(1)	&	0.74(2)	&	6.0	&	$f_4 + 6f_{\rm orb} - f_2$	\\
28	&	40.665(1)	&	0.17(1)	&	0.01(1)	&	12.1	&	$f_4 + f_5 - f_2 - f_{\rm orb}$	&	56	&	39.409(3)	&	0.08(1)	&	0.14(2)	&	5.7	&	$f_5 - 7f_{\rm orb}$	\\
29	&	3.317(1)	&	0.17(1)	&	0.73(1)	&	11.7	&	$6f_{\rm orb}$	&	57	&	0.082(3)	&	0.08(1)	&	0.66(2)	&	5.6	&	$2f_2 - 2f_4 - 7f_{\rm orb}$	\\
30	&	0.329(1)	&	0.17(1)	&	0.56(1)	&	11.5	&	$f_2 - f_4 - 3f_{\rm orb}$	&	58	&	39.134(3)	&	0.08(1)	&	0.01(2)	&	5.3	&	$f_2 + 4f_{\rm orb}$	\\
31	&	45.262(1)	&	0.15(1)	&	0.24(1)	&	10.4	&	$f_2 + f_5 - f_4$	&	59	&	8.275(3)	&	0.07(1)	&	0.77(2)	&	5.0	&	$f_5 - f_4$	\\
\hline	
\end{tabular}}													
\end{center}															
\end{table*}

\begin{figure*}[h]
\centering
\includegraphics[width=18cm]{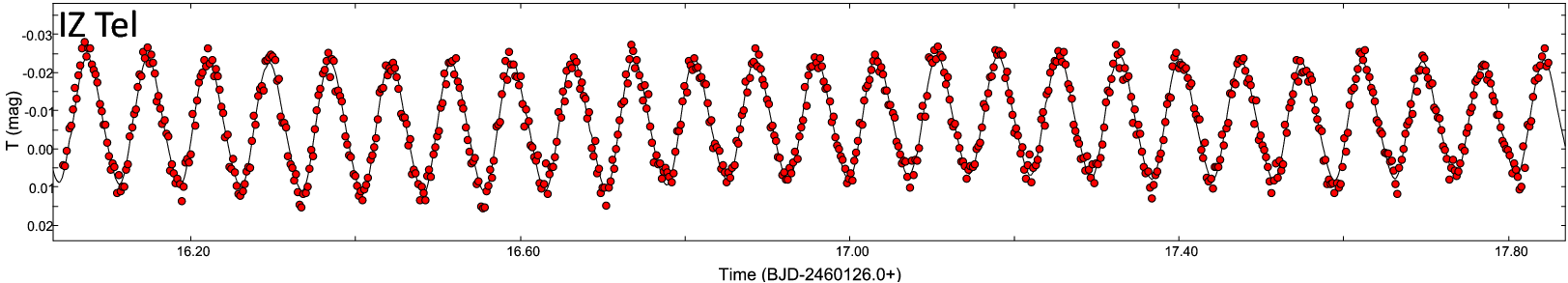}\\
\includegraphics[width=18cm]{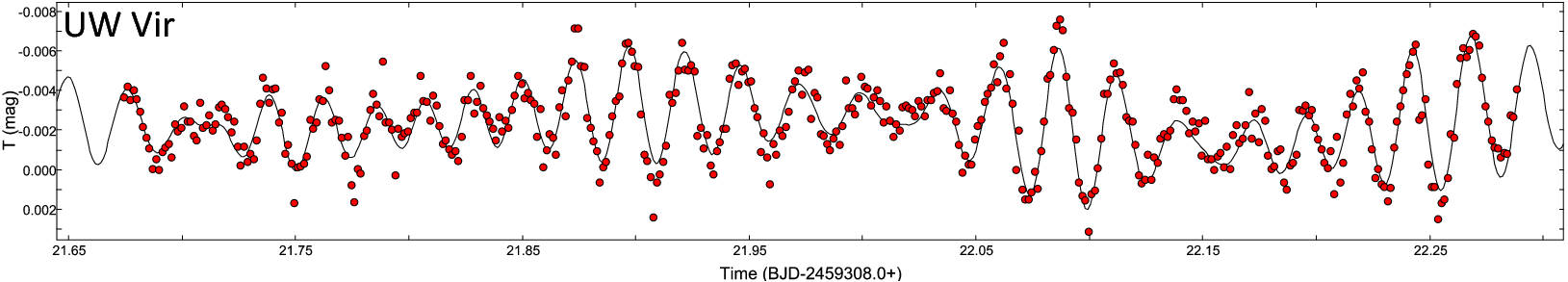}\\
\caption{Fourier fitting (solid lines) samples on various data points for both systems.}
\label{fig:FF}
\end{figure*}

\end{appendix}
\end{document}